\documentclass[lettersize,journal]{IEEEtran}
\usepackage{amsmath,amsfonts}
\usepackage{algorithmic}
\usepackage{array}
\pdfoutput=1
\usepackage[caption=false,font=normalsize,labelfont=sf,textfont=sf]{subfig}
\usepackage{textcomp}
\usepackage{stfloats}
\usepackage{url}
\usepackage{verbatim}
\usepackage{graphicx}
\usepackage{physics}
\usepackage{cleveref}
\usepackage{xcolor}
\usepackage{bbding}  
\usepackage{longtable} 
\usepackage{subfig}
\usepackage[figuresleft]{rotating} 
\usepackage{multirow} 
\usepackage{makecell}
\usepackage{adjustbox}
\usepackage{booktabs}
\usepackage{tabularx}

\usepackage{graphicx}
\usepackage{epstopdf}

\def\BibTeX{{\rm B\kern-.05em{\sc i\kern-.025em b}\kern-.08em
    T\kern-.1667em\lower.7ex\hbox{E}\kern-.125emX}}
\usepackage{balance}

\begin{document}

\title{Quantum Computing in Wireless Communications and Networking: A Tutorial-cum-Survey}

\author{Wei Zhao, \IEEEmembership{Member,~IEEE}, Tangjie Weng, Yue Ruan, \IEEEmembership{Member,~IEEE}, Zhi Liu, \IEEEmembership{Senior Member,~IEEE}, Xuangou Wu, Xiao Zheng, Nei Kato, \IEEEmembership{Fellow,~IEEE}
       
  \thanks{Wei Zhao, Tangjie Weng, Yue Ruan, Xuangou Wu and Xiao Zheng are with School of Computer Science and Technology, Anhui University of Technology,  China. Emails: \{zhaowei, wengtangjie, yue\_ruan, wuxgou, xzheng\}@ahut.edu.cn.}
    
  \thanks{Zhi Liu is with the University of Electro-Communications, Department of Computer and Network Engineering, Japan. Email: liuzhi@uec.ac.jp. }
  
  \thanks{Nei Kato is with Tohoku University, Sendai, Japan. Email: kato@it.is.tohoku.ac.jp.}

  \thanks{Wei Zhao, Tangjie Weng, Yue Ruan, Zhi Liu contributed equally to this work. The corresponding authors are Yue Ruan and Zhi Liu.}
  }



\markboth{Journal of \LaTeX\ Class Files,~Vol.~18, No.~9, September~2020}%
{How to Use the IEEEtran \LaTeX \ Templates}

\maketitle

\begin{abstract}
Owing to its outstanding parallel computing capabilities, quantum computing (QC) has been a subject of continuous attention. With the gradual maturation of QC platforms, it has increasingly played a significant role in various fields such as transportation, pharmaceuticals, and industrial manufacturing, achieving unprecedented milestones. In modern society, wireless communication stands as an indispensable infrastructure, with its essence lying in optimization. Although artificial intelligence (AI) algorithms such as reinforcement learning (RL) and mathematical optimization have greatly enhanced the performance of wireless communication, the rapid attainment of optimal solutions for wireless communication problems remains an unresolved challenge. QC, however, presents a new alternative. This paper aims to elucidate the fundamentals of QC and explore its applications in wireless communications and networking. First, we provide a tutorial on QC, covering its basics, characteristics, and popular QC algorithms. Next, we introduce the applications of QC in communications and networking, followed by its applications in miscellaneous areas such as security and privacy, localization and tracking, and video streaming. Finally, we discuss remaining open issues before concluding.

\end{abstract}


\section{Introduction} \label{Introduction}







\subsection{Background and Motivation}
QC \cite{o2007optical,ladd2010quantum}, a revolutionary approach to processing information based on quantum mechanics principles, shows great potential for transforming various industries. Unlike traditional computers that use bits in a binary state (0 or 1), quantum computers apply qubits that can exist in multiple states simultaneously, allowing for much more powerful computation. This technology is expected to greatly improve cryptography by breaking current encryption methods and enabling the development of more secure ones. Additionally, QC is highly effective in solving optimization problems, which can benefit applications like route planning and managing supply chains. It also has applications in fields such as drug discovery, materials science, and simulating quantum systems, offering faster insights into complex interactions at the molecular level.
As of December 2023, the global quantum market has reached an investment total of \$ 6.7 billion, with 261 active start-ups. The potential market size for quantum technology by 2040 is projected to reach \$ 173 billion \cite{marketsize}, driven by significant investments from governments and industries aiming to leverage the transformative capabilities of quantum technologies.

On the other hand, wireless communications and networking play a foundational role in modern connectivity, facilitating the smooth exchange of data across devices and supporting a wide range of applications, from mobile networks and WiFi to IoT devices and satellite communications \cite{lyu2023dynamic}. Optimization is crucial in wireless networks, especially as we move towards the era of 6G \cite{done-second-zuoqixue-9-ok}. With the increasing demand for higher data rates, lower latency, and greater device density, optimizing spectrum usage, reducing interference, and improving signal processing algorithms become even more critical. 6G, the next frontier in wireless technology, is expected to offer unprecedented speeds, ultra-low latency, and innovative applications \cite{tang2022roadmap}. Effectively optimizing these advanced wireless networks will be essential for fully realizing the potential of 6G, ensuring efficient and reliable communication in an increasingly interconnected world.

Additionally, the interaction among the aforementioned enabling technologies holds the potential for significant performance enhancements, encompassing increased capacity, enhanced energy efficiency, and improved reliability. 
For instance, incorporating edge intelligence for micro-service orchestration, as demonstrated in \cite{done-second-wucheng-19-ok}, can bolster network throughput and reduce service delay in next-generation networks. Leveraging QC algorithms for optimization tasks can accelerate this process. As wireless networks expand to encompass intelligent devices, such as autonomous vehicles, UAVs and satellites, the complexity of problems increases accordingly \cite{done-zhongrunhu-18-ok}.
In particular, optimization tasks in wireless communications and networking, like channel assignment, power allocation, edge computing, and video streaming demand substantial computational resources. While conventional solutions often rely on local servers, cloud infrastructure, or GPU servers, they still struggle to meet the growing research demands. QC techniques offer a promising approach to addressing these challenges by leveraging its ability to handle complex optimization scenarios efficiently, such as slicing  \cite{done-second-zuoqixue-4-ok} and ultra-reliable low latency communications (URLLC) \cite{done-second-yangdongling-8-ok} in 6G.

Given this landscape, our survey aims to explore the intersection of QC and wireless communications and networking, providing insights into the latest developments, challenges, and opportunities in this rapidly evolving field. We explore the application of QC techniques to address optimization challenges in wireless networks, offering a comprehensive overview of current research trends and future directions. 
Note that while quantum key distribution is indeed a crucial area in quantum communications, this survey focuses specifically on QC applications in wireless communications and networking.

\subsection{Related Surveys}
While previous surveys, primarily concentrated on quantum communications \cite{zuoqixue-6-wrong, li2023entanglement, third-shike-18-wrong}, have explored the applications and advantages of integrating QC into wireless networks from various applications \cite{botsinis2018quantum, nawaz2019quantum, 9870532, second-shike-11-wrong}, a comprehensive survey on QC-enabled communications and networking is still lacking. 
The first survey \cite{botsinis2018quantum} investigates the utilization of QC in addressing challenges within wireless communication systems. It provides an overview of QC fundamentals using linear algebra, showing its potential for enhancing wireless communication systems in existing literature. The survey extensively explores various optimization problems encountered in both the physical and network layers of wireless communications, offering comparisons between classical and quantum-assisted solutions. However, the survey from a few years back lacks coverage of AI-powered QC algorithms such as quantum neural networks (QNNs) and quantum deep reinforcement learning (QDRL). Moreover, it does not address particular concerns within wireless networks like edge computing and video streaming that are especially relevant in the context of 5G/6G networks.

\begin{table*}[!t]
\centering
\caption{Summary of related surveys on applications of QC in wireless networks}
\label{survey3}
\begin{tabular}{|c|c|ccc|ccc|ccc|ccc|}
\hline
\multicolumn{1}{|p{0.8cm}|}{\textbf{}} & \multicolumn{1}{p{0.8cm}|}{\textbf{}} & \multicolumn{3}{c|}{\textbf{Tutorial on QC}} & \multicolumn{3}{c|}{ \textbf{QC in Communication}} & \multicolumn{3}{c|}{\textbf{QC in Networking}} & \multicolumn{3}{c|}{\textbf{Others}} \\ \hline

\multicolumn{1}{|c|}{\textbf{Refs}} & \multicolumn{1}{c|}{\textbf{Year}} & \multicolumn{1}{p{0.8cm}|}{\centering\textbf{BQ}} & \multicolumn{1}{p{0.8cm}|}{\centering \textbf{CQ}} & \multicolumn{1}{p{0.8cm}|}{\centering\textbf{AQ}} & \multicolumn{1}{p{0.8cm}|}{\centering\textbf{PA}} & \multicolumn{1}{p{0.8cm}|}{\centering\textbf{CA}} & \multicolumn{1}{p{0.8cm}|}{\centering\textbf{UA}} & \multicolumn{1}{p{0.8cm}|}{\centering\textbf{R}} & \multicolumn{1}{p{0.8cm}|}{\centering\textbf{TO}} & \multicolumn{1}{p{0.8cm}|}{\centering\textbf{CC}} & \multicolumn{1}{p{0.8cm}|}{\centering\textbf{SP}} & \multicolumn{1}{p{0.8cm}|}{\centering\textbf{LT}} & \multicolumn{1}{p{0.8cm}|}{\centering\textbf{VS}} \\ \hline

\cite{botsinis2018quantum} & 2018 & \multicolumn{1}{c|}{\Checkmark} & \multicolumn{1}{c|}{PD} &  & \multicolumn{1}{c|}{} & \multicolumn{1}{c|}{\Checkmark} & \Checkmark & \multicolumn{1}{c|}{\Checkmark} & \multicolumn{1}{c|}{} &  & \multicolumn{1}{c|}{\Checkmark} & \multicolumn{1}{c|}{\Checkmark} &  \\ \hline

\cite{nawaz2019quantum} & 2019 & \multicolumn{1}{c|}{PD} & \multicolumn{1}{c|}{PD} & PD & \multicolumn{1}{c|}{PD} & \multicolumn{1}{c|}{PD} &  & \multicolumn{1}{c|}{PD} & \multicolumn{1}{c|}{PD} & PD & \multicolumn{1}{c|}{\Checkmark} & \multicolumn{1}{c|}{\Checkmark} &  \\ \hline

\cite{zuoqixue-6-wrong} & 2021 & \multicolumn{1}{c|}{\Checkmark} & \multicolumn{1}{c|}{} &  & \multicolumn{1}{c|}{} & \multicolumn{1}{c|}{} &  & \multicolumn{1}{c|}{} & \multicolumn{1}{c|}{} & & \multicolumn{1}{c|}{\Checkmark} & \multicolumn{1}{c|}{} &  \\ \hline

\cite{9870532} & 2022 & \multicolumn{1}{c|}{\Checkmark} & \multicolumn{1}{c|}{PD} & PD & \multicolumn{1}{c|}{PD} & \multicolumn{1}{c|}{PD} & & \multicolumn{1}{c|}{} & \multicolumn{1}{c|}{} &  & \multicolumn{1}{c|}{PD} & \multicolumn{1}{c|}{} &  \\ \hline

\cite{second-shike-11-wrong} & 2023 & \multicolumn{1}{c|}{\Checkmark} & \multicolumn{1}{c|}{\Checkmark} & \Checkmark & \multicolumn{1}{c|}{PD} & \multicolumn{1}{c|}{PD} &  & \multicolumn{1}{c|}{} & \multicolumn{1}{c|}{} & & \multicolumn{1}{c|}{} & \multicolumn{1}{c|}{} &  \\ \hline

\cite{li2023entanglement} & 2023 & \multicolumn{1}{c|}{\Checkmark} & \multicolumn{1}{c|}{} &  & \multicolumn{1}{c|}{} & \multicolumn{1}{c|}{} &  & \multicolumn{1}{c|}{} & \multicolumn{1}{c|}{} &  & \multicolumn{1}{c|}{} & \multicolumn{1}{c|}{} &  \\ \hline

\cite{third-shike-18-wrong} & 2023 & \multicolumn{1}{c|}{\Checkmark} & \multicolumn{1}{c|}{PD} &  & \multicolumn{1}{c|}{} & \multicolumn{1}{c|}{} &  & \multicolumn{1}{c|}{} & \multicolumn{1}{c|}{} &  & \multicolumn{1}{c|}{\Checkmark} & \multicolumn{1}{c|}{} &  \\ \hline

\cite{nguyen2024quantum} & 2024 & \multicolumn{1}{c|}{PD} & \multicolumn{1}{c|}{PD} & PD & \multicolumn{1}{c|}{} & \multicolumn{1}{c|}{} &  & \multicolumn{1}{c|}{} & \multicolumn{1}{c|}{PD} &  & \multicolumn{1}{c|}{PD} & \multicolumn{1}{c|}{} &  \\ \hline

ours & 2024 & \multicolumn{1}{c|}{\Checkmark} & \multicolumn{1}{c|}{\Checkmark} & \Checkmark & \multicolumn{1}{c|}{\Checkmark} & \multicolumn{1}{c|}{\Checkmark} & \Checkmark & \multicolumn{1}{c|}{\Checkmark} & \multicolumn{1}{c|}{\Checkmark} & \Checkmark & \multicolumn{1}{c|}{\Checkmark} & \multicolumn{1}{c|}{\Checkmark} & \Checkmark \\ \hline
\end{tabular}

\vspace{0.5em}{BQ: Basic of QC; 
CQ: Conventional QC;  AQ: AI-driven QC; PA: Power Allocation;  CA: Channel Assignment; UA: User Association; R: Routing; TO: Task Offloading; CC: Content Caching; SP: Security and Privacy; LT: Localization and Tracking; VS: Video Streaming; PD: Partially Addressed.}

\end{table*}

AI-driven QC algorithms have been explored in several studies, including \cite{nawaz2019quantum, 9870532, second-shike-11-wrong, nguyen2024quantum}. However, in \cite{nawaz2019quantum}, the discussion on conventional and AI-driven QC algorithms remains largely conceptual, with limited elaboration on specific algorithms like QDRL. Topics such as task offloading and content caching in wireless networks are also presented conceptually. Another brief survey on QC algorithms for 6G networks, focusing on quantum-inspired machine learning (ML) applications, can be found in \cite{9870532}. This survey provides a brief overview of both conventional and AI-driven algorithms, including the Shor algorithm, Grover algorithm, quantum reinforcement learning (QRL), and QNNs. It briefly discusses a few topics mainly related to security and resource allocation for 6G networks.
The fundamental principles of QC, as well as conventional and AI-driven QC algorithms, are relatively described in \cite{second-shike-11-wrong}. However, the paper provides only a limited introduction to their applications in wireless networks, focusing mainly on power allocation and channel assignment. 
In \cite{nguyen2024quantum}, the applications of QC are reviewed in detail, with a focus on quantum cloud computing. The paper discusses the latest advancements in quantum cloud technology, covering various cloud-based models, platforms, and recently developed technologies and software user cases. It also examines several applications of quantum cloud computing, such as resource management, quantum serverless computing, and addressing security and privacy concerns.
Table \ref{survey3} presents a comparative summary of the referenced surveys, underscoring that the most recent ones fail to provide a comprehensive overview of the QC applications in wireless networks. This observation highlights a gap in the existing literature.

\subsection{Contributions}
We take a broad approach aimed at a wide audience to show the QC advantages, offering the following contributions.
\begin{itemize}
    \item Our survey offers a comprehensive introduction to fundamental quantum mechanics concepts, such as quantum measurement, essential for developing a deep understanding of QC.
    \item  We thoroughly explain all aspects of QC, ensuring that even readers without a background in quantum mechanics can easily grasp the foundational principles of this emerging technology.
    \item We encompass all QC algorithms, ranging from classical approaches like quantum approximate optimization algorithm (QAOA), to quantum-inspired metaheuristic methods like quantum particle swarm optimization (QPSO), and even to AI-driven techniques such as QDRL.
    \item We examine recent advancements in the application of QC algorithms to address various communication and networking challenges in wireless networks, spanning power allocation, channel assignment, user association, routing, edge computing/task offloading, content caching,  security and privacy, localization and tracking, video streaming, and so forth. Additionally, we identify key challenges and open problems for future research to fully utilize the potential of QC. 
\end{itemize}

\subsection{Literature Categorization and Paper Organization}

\begin{figure} 
    \centering
    \includegraphics[width=0.94\linewidth]{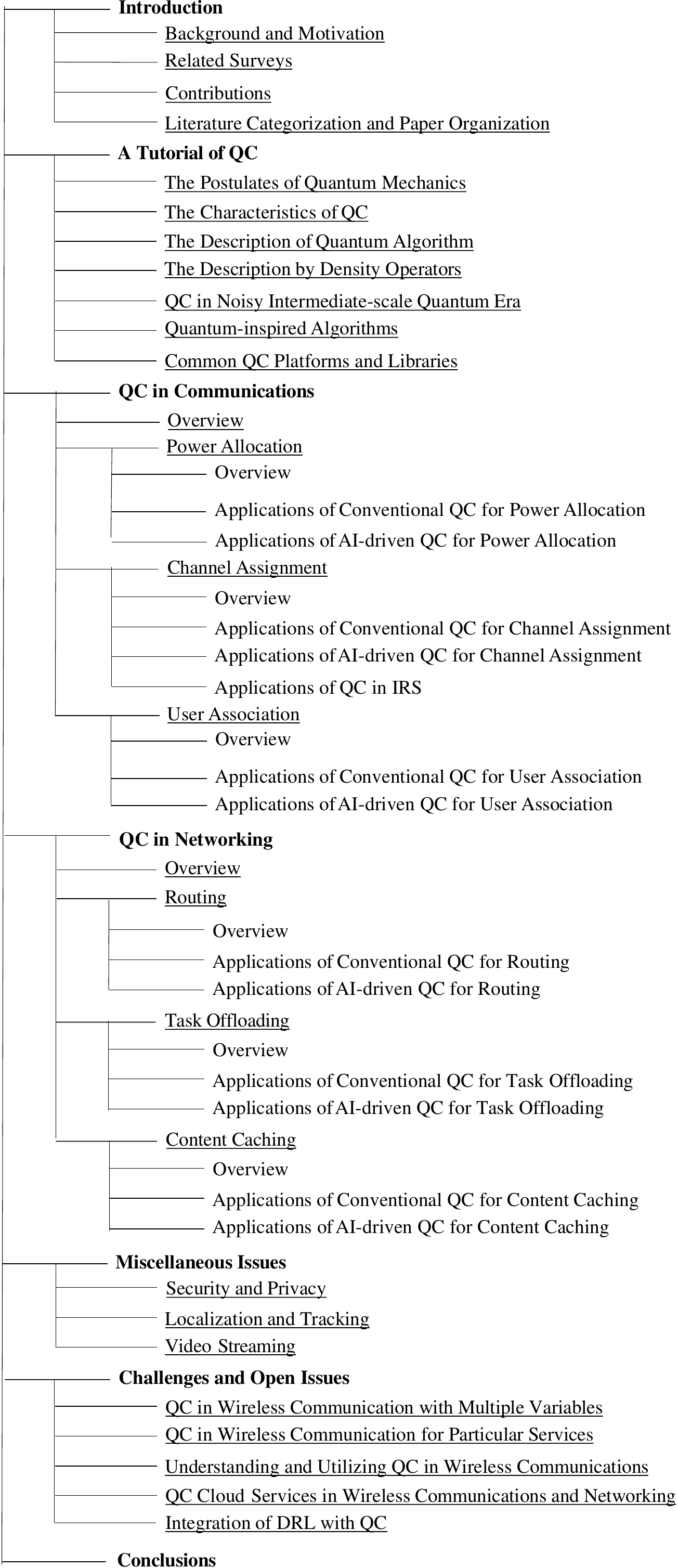}
    \caption{Paper organization.}
    \label{paper-outline}
\end{figure}

We have adopted the following methodology to ensure a comprehensive and in-depth review of applications of QC algorithms. We primarily organize the papers from two dimensions: the research issues and category of QC algorithms.
We divide the research issues in wireless networks into two parts: communications from the perspective of the physical layer, where issues typically arise within a single hop, and networking from the perspective of the routing layer, where issues typically arise within multiple hops. The former mainly considers topics including channel assignment, power allocation, and user association, while the latter mainly considers routing, task offloading, content caching, and among others. For each issue, we consider its network characteristics such as the presence of multiple cells in communications and mobility in networking.
Additionally, we divide QC algorithms into two categories. One is conventional QC algorithms, such as classical QAOA, quantum-inspired heuristics like QPSO, and quantum genetic algorithm (QGA). The other is AI-driven algorithms, such as QNNs and QDRL. The paper organization is described in Fig. \ref{paper-outline}.

\begin{table}
    \centering
    \caption{Acronyms and Abbreviations}\label{tab:my_label}
    \begin{tabular}{ll}
        \textbf{Acronym} & \textbf{Full form}\\
         QC & quantum computing\\
         AI & artificial intelligence\\
         URLLC & ultra-reliable low latency communications\\
         QNN & quantum neural network\\
         QDRL & quantum deep reinforcement learning\\
         QRL & quantum reinforcement learning\\
         QAOA & quantum approximate optimization algorithm\\
         QPSO & quantum particle swarm optimization\\
         QGA & quantum genetic algorithm\\
         NISQ & noisy intermediate-scale quantum\\
         VQE & variational quantum eigensolver\\
         QML & quantum machine learning\\
         PQC & parameterized quantum circuits \\
         QUBO & quadratic unconstrained binary optimization\\
         QGLM & quantum generative learning model\\
         BS & base station\\
         NOMA & non-orthogonal channel multiple access\\
         OFDMA & orthogonal frequency-division multiple access\\
         MIMO & multiple input multiple output antennas\\
         D2D & device to device\\
         QA & quantum annealing\\
         QMARL & quantum multiple agent reinforcement learning\\
         RSU & road-side unit\\
         MDP & Markov decision process\\
         IRS & intelligent reflective surface\\
         QEA & quantum evolution algorithm\\
         ILP & integer linear programming\\
         MEC & mobile edge computing

    \end{tabular}
\end{table}

\section{A Tutorial of QC} \label{An Overview of QC}
This section is organized into three parts. The first part introduces the fundamental principles and characteristics of quantum algorithms (Subsections A-D). The second part examines prominent quantum algorithms that are expected to achieve quantum supremacy during the noisy intermediate-scale quantum (NISQ) era (Subsections E and F). The final part covers widely used quantum hardware and software platforms. These sections are structured sequentially to support the discussion of quantum computing (QC) applications in wireless networks.



\subsection{The Postulates of Quantum Mechanics}
QC is grounded in the principles of quantum mechanics. In this section, we outline the essential principles of quantum mechanics, providing the requisite foundation for the subsequent discussions. This explanation is crucial in establishing a mathematical framework for comprehending QC, thereby facilitating a deeper understanding of this field for readers without a background in physics.

\vspace{5pt}

\emph{Postulate 1: Every enclosed physical system possesses a corresponding Hilbert space, wherein the system state can be precisely depicted by a unit vector.}\label{Postulate-1}

\vspace{5pt}

The Hilbert space, as described in Postulate 1, is a complete complex linear space characterized by a defined inner product, relevant to QC since it does not involve infinite spatial dimensions. A collection of $n$ linearly independent vectors is capable of spanning an $n$-dimensional linear space. When these vectors are both normalized and mutually orthogonal, they form orthonormal bases for the given space. A unique unit vector, referred to as a `state,' encapsulates the description of a specific physical system. This unit vector can be linearly decomposed using any complete set of orthonormal bases within the linear space, offering a mathematically precise representation of the physical system.

Physicist Dirac creates a set of symbols to describe system states \cite{dirac1981principles}. The symbol $\ket{\bullet}$ is termed a ket, corresponding to a column unit vector in the space. Thus, a specific system state, denoted by a quantum state $\ket{\phi}$  in eq. (\ref{Ket-represantation}), can be represented linearly by a set of orthonormal bases $\{\ket{a_1}, \ket{a_2}, \dots, \ket{a_n}\}$.
\begin{equation}\label{Ket-represantation}
\ket{\phi} = \sum_{i=1}^n c_i\ket{a_i}, \quad \text{s.t.} \sum_i |c_i|^2=1.
\end{equation}
The algebra reveals that the orthonormal bases of a Hilbert space are not unique; thereby, a quantum state can be described by different bases. This property is known as representation transformation in quantum mechanics, which is essentially equivalent to basis transformation.

The conjugate transpose of $\ket{\phi}$ is written as $\bra{\phi}$, corresponding to a row vector in the Hilbert space. The inner product of two vectors is defined by the dot product of the row vector $\bra{\psi}$ and the column vector $\ket{\phi}$, represented as $\bra{\psi}\ket{\phi}$.


\emph{Postulate 2: The evolution of a closed quantum system $\ket{\phi}$ can be described by the Schrodinger equation:
\begin{equation}
i\hbar \frac {d \ket{\phi (t)}}{dt} = H \ket{\phi (t)},
\end{equation} 
where $\hbar$ is reduced Plank constant, $H$ is the Hamiltonian of the system, mathematically expressed as a Hermite operator, i.e., $H^\dag=H$.}


In general, finding the Hamiltonian to characterize a particular physical system is difficult (requiring substantial experimental results), but fortunately this is not the primary concern for the involved theoretical framework. In the discussions of QC and quantum information, we always assume that $H$ is known. Thus, if the system in the state at time $t_0 $ is $\ket{\phi(t_0)}$, then the system at $t_1$ can be solved by the Schrodinger equation in eq. (\ref{Schrodinger}),
\begin{equation} \label{Schrodinger}
\ket{\phi(t_1)}=exp \left[ \frac{-iH(t_1-t_0)}\hbar \right]\ket{\phi(t_0)} = U(t)\ket{\phi(t_0)},
\end{equation}
where $U(t)$ is an evolutive operator from $t_0$ to $t_1$ in eq. (\ref{unitary-evolution}),
\begin{equation}\label{unitary-evolution}
U(t)=exp \left[ \frac{-iH(t_1-t_0)}\hbar \right].
\end{equation}
It is easy to verify that $U(t)$ is a unitary operator, i.e., $U(t)^\dag U(t)=I$. In many cases, for the convenience of discussion, \cref{unitary-evolution} can be simplified in eq. (\ref{sim_y}),
\begin{equation} \label{sim_y}
\ket{\phi_1} =U \ket{\phi_0}.
\end{equation}


\emph{Postulate 3:  A general quantum measurement is described by a family of linear operators $\{M_m\}$. These operators act on the state space of the system to be measured, satisfying the completeness condition $\sum_mM_m^\dag M_m=I$, and the indicator $m$ represents the possible measurement results in the experiment. }


Assuming that the state of the system before measurement is $\ket{\phi}$, the probability of observing the result $m$ is given by eq. (\ref{re_l}),
\begin{equation} \label{re_l}
p(m)=\bra{\phi}M_m^\dag M_m\ket{\phi}.
\end{equation}
The post-measurement state is given by eq. (\ref{st_1}),
\begin{equation} \label{st_1}
\frac{M_m\ket\phi}{\sqrt{\bra\phi M_m^\dag M_m \ket\phi}}.
\end{equation}


\emph{Postulate 4: The state space of a composite physical system is the tensor product of the state space of a sub-physical system.}

\vspace{5pt}

If the index numbers of the subsystems range from $1$ to $n$, and the state of subsystem-$i$ can be represented as $\ket{\phi_ i}$, then the state of the entire system is  $\ket{\phi_ 1} \otimes \cdots \otimes \ket{\phi_n}$. The notation $\otimes$ represents a tensor product, which combines smaller vector spaces to form a larger vector space. Eq. (\ref{tensor-product}) presents general rules for the tensor product, also known as the Kronecker product,
\begin{equation}\label{tensor-product}
  \mathbf{A\! \otimes \!B}\!\! =\!\!  \begin{array}{c@{\!\!\!}l}
   \overbrace{ \left(  \begin{array}[c]{ccccc}
    A_{11}B &  A_{12}B  & \cdots &  A_{1n}B \\
    A_{21}B &  A_{22}B  & \cdots &  A_{2n}B \\
    \vdots & \vdots && \vdots \\
     A_{m1}B & A_{m2}B  & \cdots & A_{mn}B \\
    \end{array}  \right) } ^\text{\normalsize{$nq$}}
&
 \begin{array}[c]{@{}l@{\,}l}
   \left. \begin{array}{c} \vphantom{0}  \\  \vphantom{0}  \\ \vphantom{\vdots}
   \\ \vphantom{0} \end{array} \right\}\!\!  & \! \text{$mp$} \\
\end{array}
\end{array},
 \end{equation}
where A and B are $m\times n$ and $p \times q$ matrix, respectively. Since a quantum state is essentially an $n \times 1$  matrix, eq. (\ref{tensor-product}) is applicable. If quantum states are described by density operators (see "D. The Description by Density Operators"), eq. (\ref{tensor-product}) can be directly applied.

\subsection{The Characteristics of QC}
The main advantage of QC arises from quantum parallelism, which is rooted in the unique properties of qubits.

1) Properties of Qubits

In classical computing, the bit, which can have a value of either $0$ or $1$, serves as the fundamental unit of information. This binary state system can represent a physical system with two clearly distinguishable states, such as the widely used high and low-level electrical signals.

The fundamental unit of information in QC is the quantum bit, or qubit. Unlike classical bits, which exist strictly in one of two states ($0$ or $1$), qubits can assume multiple physical forms. Examples include the ground and excited states of an electron in a hydrogen atom, the spin states of a proton ($+\frac 1 2$ and $-\frac 1 2$), or the clockwise and counterclockwise circular polarization of light. A key distinction of qubits is their ability to exist in a superposition, where they can simultaneously represent both $0$ and $1$, as well as any linear combination of these states. Mathematically, a qubit is represented as a unit vector in a two-dimensional Hilbert space, as described by Postulate 1 and shown in eq. (\ref{two-hilb}).

\begin{equation} \label{two-hilb}
\ket{\phi}=\alpha\ket 0 + \beta \ket1, \quad \text{s.t.} |\alpha|^2+|\beta|^2=1,
\end{equation}
where the complex numbers $\alpha$ and $\beta$ satisfy the normalization condition, and $\ket 0$ and $\ket 1$ are orthonormal basis $(0\ 1)^T$ and $(1\ 0)^T$, respectively. Thereby, $\ket\phi$ is represented as $(\alpha\ \beta)^T$.

There are several distinctive properties of qubits:

\textbullet\quad\emph{The amount of represented information increases exponentially with the number of qubits.}

A qubit is a unit vector in a two-dimensional Hilbert space. The representation of two qubits can be derived by  (Kronecker product \cref{tensor-product}) two single qubits. For example, $\ket{\phi_1} = \alpha_1\ket 0 + \beta_1 \ket 1$, $\ket{\phi_2} = \alpha_2\ket 0+ \beta_2 \ket 1$, then two qubits:

\begin{equation}
\label{two-qubits}
\begin{split}
\!\ket{\phi_1\phi_2} &=\! \ket{\phi_1} \otimes \ket{\phi_2} \\
&=\!  (\alpha_1\ket 0 + \beta_1 \ket 1) \otimes (\alpha_2 \ket 0 + \beta_2 \ket 1)\\
&=\!\alpha_1\alpha_2 \ket{00} \!+\! \alpha_1\beta_2 \ket{01} \!+\! \beta_1\alpha_2 \ket{10} \!+ \!\beta_1\beta_2 \ket{11}
\end{split},
\end{equation}
where $\ket{00} = (1\ 0\ 0\ 0)^T$, $\ket{01} = (0\ 1\ 0\ 0)^T$, $\ket{10} = (0\ 0\ 1\ 0)^T$, and $\ket{11} = (0\ 0\ 0\ 1)^T$, form a set of orthonormal bases for a four-dimensional Hilbert space. Obviously, $|\alpha_1\alpha_2|^2+|\alpha_1\beta_2|^2+|\beta_1\alpha_2|^2+|\beta_1\beta_2|^2=1$, and $\ket{\phi_1\phi_2}$ is a normalized vector in a four-dimensional Hilbert space.
Likewise, the extension of $n$ qubits $\ket{\phi_1\phi_2\cdots\phi_n}$ is expressed in eq. (\ref{superposition}),
\begin{equation}\label{superposition}
\ket{\phi_1\phi_2\cdots\phi_n} = \sum_{i=0}^{2^n-1}c_i \ket{i}, \qquad \text{s.t.} \sum|c_i|^2=1.
\end{equation}

Suppose a quantum register in a quantum computer stores $n$ qubits. According to eq. (\ref{two-qubits}), the quantum register stores every number from $0$ to $2^n-1$ simultaneously with the probability of $|c_i|^2$. In other words, an $n$-qubit quantum register holds $2^n$ binary numbers.

\textbullet\quad\emph{Probability statistical properties of measurement}

While $n$ superposed qubits can concurrently represent $2^n$ binary numbers, there is generally no method to extract each binary number concealed within the superposition state simultaneously. More specifically, it is impossible to accurately determine the probability amplitude $c_i$ in \cref{superposition} of each individual binary number for an unknown $n$-qubit. The primary tool for probing an unknown quantum state is through measurement or observation. However, each observation only yields a fragment of information about the original state since the state collapses upon measurement, leading to significant data loss.

To illustrate, consider a beam of light emitting a mixture of horizontally and vertically polarized photons $\ket{\phi} = \alpha \ket{\uparrow} +\beta\ket{\rightarrow}$. Once this beam passes through a horizontal polarization filter, only vertically polarized photons $\ket{\uparrow}$ can be observed, resulting in $\ket{\phi} \Rightarrow \ket{\uparrow}$, with the frequency of observing vertically polarized light (or vertically polarized photons) approximating $|\alpha|^2$. Conversely, if this light is subjected to a vertical polarization filter, the observable light is horizontally polarized with a probability of $|\beta|^2$.

The methodology of this aforementioned observation characterizes the process of projective measurement, a specific case of the more general measurement process described by Postulate 3. This type of measurement, also known as Von Neumann measurement, is extensively employed across various studies. Regardless of physical circumstances, this can be abstracted as an observable or a Hermitian operator $M$, presented in the spectral decomposition form in \cref{spe-dec},

\begin{equation} 
\label{spe-dec}
M=\sum_m mP_m,
\end{equation}
where the projective operator $P_m$ is the projection onto the subspace of $M$ corresponding to the eigenvalue $m$. It satisfies $P_mP_{m'}=\delta_{mm'}P_m$ and the completeness condition $\sum_m P_m = I $. 

The measurement outputs are eigenvalues $m$. When measuring the state $\ket{\phi}$, the probability of getting a certain $m$ is given by:
\begin{equation} \label{get-cer}
p(m)=\bra\phi P_m\ket\phi.
\end{equation}

After the measurement, $\ket{\phi}$ collapses to:
\begin{equation} \label{mea-coll}
\frac{P_m\ket\phi}{\sqrt {p(m)}},
\end{equation}
If there is no ambiguity in the context, it is said that the measurement is performed in basis $\ket{m}$; thereby, the projective operator $P_m$ can be written as $P_m = \ket m \bra m $. 

From the preceding discussion, it is clear that, in general, post-measurement qubits $\ket\phi$ collapse to a particular basis state $\ket m$ with probability $p(m)$, unless the measurement basis aligns with $\ket\phi$ and its orthogonal complement $\ket{\phi^\bot}$. This behavior contrasts with classical bits, which retain their initial state regardless of the form or frequency of observations (measurements). For a quantum state such as $\ket\phi = \sum_{i = 0}^{2^n - 1} c_i \ket{i}$, performing a projective measurement $P_i = \ket i \bra i$ will yield the outcome $\ket{i}$ with a probability of $|c_i|^2$.

\textbullet\quad\emph{Quantum states can not be cloned.}

A key property of quantum states is that only limited information can be extracted from a single measurement. It is fundamentally impossible to fully determine a quantum state by repeatedly cloning and measuring it, due to the no-cloning theorem. A simplified proof of this principle is presented here, with further details available in \cite{wootters1982single}.

Assume a unitary operation $U$ is able to replicate $\ket\phi$ and substitute resource $\ket s$ with $\ket\phi$:

\begin{equation} \label{noclone} \ket\phi\otimes \ket s \xrightarrow{\ U\ } \ket\phi \otimes \ket\phi . \end{equation}

Accordingly, for $\ket\phi$ and $\ket\psi$: 
\begin{equation} \label{r_x_t}
\begin{split} U(\ket\phi\otimes \ket s) = \ket\phi \otimes \ket\phi, \ U(\ket\psi\otimes \ket s) = \ket\psi \otimes \ket\psi. 
\end{split} 
\end{equation}

Apply the inner product on both sides of eqs. (\ref{noclone}) and (\ref{r_x_t}), we obtain:
\begin{equation}
\bra{\phi} \ket{\psi} = (\bra{\phi} \ket{\psi})^2.
\end{equation}

The resulting equation suggests $x = x^2$, where $x = 0$ or $x = 1$ ($\bra{\phi} \ket{\psi}=x$). This implies that either $\ket\phi = \ket\psi$ or $\ket\phi$ and $\ket\psi$ is orthogonal. Consequently, a quantum cloning device can only clone mutually orthogonal states, thereby proving that a universal quantum cloning machine cannot exist. In essence, a unitary operation that perfectly replicates an unknown quantum state does not exist unless the qubit has effectively collapsed into a classical bit.

2) Quantum Parallelism

The defining feature of QC is its ability to perform parallel computation, enabled by the principles of superposition and entanglement of quantum states. According to Postulate 2, QC operates on the unitary evolution of quantum states. Mathematically, this evolution is represented by a unitary matrix $U_f$, which satisfies the condition in eq. (\ref{unitary-m}),

\begin{equation} \label{unitary-m}
U_f U_f^\dag=I,
\end{equation}
where $U_f^\dag$ signifies the conjugate transpose of $U_f$ and identity matrix $I$.

Consider a classical computational function $f: x \rightarrow f(x)$. This function $f$ can be transformed into a corresponding QC function $U_f$ with polynomial computational cost \cite{kitaev2002classical}. The transformed function $U_f$ will then operate as $U_f: \ket x \rightarrow \ket{f(x)}$. Assume an $n$-qubit register holds qubits with initial values $\ket{00\cdots 0}$. Initially, the Hadamard transform $H^{\otimes n}$ is applied, as shown in eq. (\ref{h-gate-opera}), resulting in a superposition state of all binary numbers between $0$ and $2^n-1$, each occurring with equal probability of $\frac{1}{2^n}$.

\begin{equation}\label{h-gate-opera}
H^{\otimes n}\ket{00\cdots0} =\frac 1 {\sqrt {2^n}}\sum_{x=0}^{2^n-1}\ket{x}.
\end{equation}
Subsequently, $f(x)$ is computed by $U_f$ in eq. \ref{uf-eq},
\begin{equation}\label{uf-eq}
\!U_f \!\! \left(\!\!\frac 1 {\sqrt {2^n}}\!\sum_{x=0}^{2^n-1}\ket{x} \!\!\right)\!\!=\!\!\frac 1 {\sqrt {2^n}}\sum_{x=0}^{2^n-1}U_f\ket{x}\!\!=\!\!\frac 1 {\sqrt {2^n}} \! \sum_{x=0}^{2^n \! - \! 1}\ket{f(x)}.
\end{equation}
Eq. (\ref{uf-eq}) demonstrates that $U_f$ performs all computations of $\ket{f(x)}$ for inputs ranging from $\ket{00 \cdots 0}$ to $\ket{11 \cdots 1}$. In contrast, on a classical computer, calculating $f(x)$ for every $x$ from $00 \cdots 0$ to $11 \cdots 1$ requires either $2^n$ sequential loops or $2^n$ parallel processors.

This capability of QC, referred as quantum parallelism, strengthens its advantage over classical computation. Although quantum parallel operations can generate $2^n$ results simultaneously, reading all $\ket{f(x)}$ states in one step presents a significant challenge. Information from quantum states can only be extracted via measurement, but each measurement reveals only partial information and collapses the quantum state. To gather the full spectrum of information on the final outcome, multiple measurements in different bases (often exponentially large in number) are needed, which requires substantial computational resources.

3) Entanglement

Entanglement can be viewed as a specialized form of superposition, exhibiting properties of quantum parallelism, since quantum parallelism originates from the superposition of quantum states. For example, consider the Bell state $\frac {1} {\sqrt 2} (\ket{00} + \ket{11})$, which represents a superposition of the states $\ket{00}$ and $\ket{11}$. When inputs are $\ket{00}$ and $\ket{11}$, a suitable unitary operation can process these states simultaneously, demonstrating quantum parallelism.

However, entanglement possesses unique characteristics that distinguish it from general superposition states. For instance, the uniform superposition state $\frac {1} {2} (\ket{00}+\ket{01}+\ket{10}+\ket{11})$ is not entangled, as it can be decomposed into $\frac {1} {\sqrt 2} (\ket{0} + \ket{1}) \otimes \frac {1} {\sqrt 2} (\ket{0} + \ket{1})$. In contrast, the Bell state $\frac {1} {\sqrt 2} (\ket{00} + \ket{11})$ can not be factored into a separable tensor product, indicating that it is an entangled state.

Similarly, the state $\frac {1} {\sqrt 3} (\ket{00} + \ket{10} + \ket{11})$ is also entangled since it can not be expressed as a separable tensor product. However, its entanglement is weaker than that of the Bell state $\frac {1} {\sqrt 2} (\ket{00} + \ket{11})$. This difference in entanglement can be quantitatively measured using entanglement entropy, as defined in eq. (\ref{srho}),

\begin{equation}\label{srho}
S(\rho)=-tr(\rho^A \log \rho^A)=-\sum_i \lambda_i \log \lambda_i,
\end{equation}
where $\rho^A$ is reduced density operator of subsystem A of a composite system $\rho^{AB}$. $\rho^A$ is defined as a partial trace over system B in eq. (\ref{trB}):
\begin{equation}\label{trB}
tr_B(\ket{a_1}\bra{a_2}\otimes \ket{b_1}\bra{b_2})=\ket{a_1}\bra{a_2}tr(\ket{b_1}\bra{b_2}),
\end{equation}
where $\ket {a_1}$ and $\ket {a_2}$ are vectors of subsystem A, and $\ket {b_1}$ and $\ket {b_2}$ are vectors of subsystem B.

Simple calculations reveal that the entanglement entropy of the Bell state is 1, while the state $\frac {1} {\sqrt 3} (\ket{00}+\ket{10}+\ket{11})$ has an entanglement entropy of approximately 0.17, significantly lower than that of the Bell state.

Entanglement plays a crucial role in quantum communication protocols, such as super-dense coding and quantum teleportation. However, in QML, the effect of entangled data on prediction error presents a dual nature, contingent upon the number of allowed measurements. When a sufficient number of measurements are available, increasing the entanglement of the training data consistently reduces prediction error or lowers the required size of the training set to achieve a comparable error rate. On the other hand, when the number of measurements is limited, highly entangled data may result in increased prediction errors \cite{wang2024transition}.

\subsection{The Description of Quantum Algorithm}
1) Overview of Quantum Algorithm Capabilities

The preceding discussion on quantum parallelism presents two contrasting views. On one hand, quantum parallelism has the potential to significantly enhance computational efficiency. On the other hand, obtaining complete and precise solutions can be resource-intensive. This leads to a fundamental question: do quantum algorithms truly provide an advantage?

To address the problem, we need to examine quantum algorithms from the perspective of computational theory and basic problem structures. How can one determine if a problem is suitable for a quantum algorithm? How a quantum algorithm is described? And what methods can be used to evaluate its computational efficiency and resource usage?

This exploration began with Benioff's foundational work \cite{benioff1980computer}, where he introduced the quantum Turing machine model. This model was later expanded by Deutsch \cite{ deutsch1989quantum} and Yao \cite{yao1993quantum}, who introduced the quantum circuit model, showing that it is equivalent to the Turing machine model. Further work by Bernstein \cite{bernstein1993quantum} and Knill \cite{knill2001quantum} established the theoretical foundation of quantum computational complexity, linking it to classical complexity theory.
These classic studies provide a key insight:

\textbullet\quad\emph{A problem unsolvable by classical computation also remains unsolvable by quantum algorithms.} 

According to the Church-Turing thesis, any deterministic computational process, no matter how complex, can be represented by a Turing machine. This means that all computations can be effectively simulated by a Turing machine, whether performed by an abacus or a supercomputer. The solvability of a problem, therefore, depends on whether the input is countable and whether the Turing machine halts. Aside from the well-known halting problem, several problems remain unsolvable \cite{myasnikov2008generic}, such as determining whether two topological spaces are homeomorphic.

Are these theoretically solvable but practically are unsolvable problems solved on a quantum computer? The answer is no. A quantum Turing machine can not solve problems that a classical Turing machine can not solve, meaning that in terms of computability, both models are equivalent. Therefore, problems that are unsolvable by classical means remain unsolvable by quantum algorithms as well.

\textbullet\quad\emph{Quantum algorithms can solve problems that can not be solved efficiently in classical computation.}

The distinction between solving problems and solving them efficiently is critical. A problem that can be efficiently solved implies that there exists an algorithm capable of addressing the problem with a cost that grows polynomially with the size of the input. It is generally acknowledged that quantum algorithms can efficiently solve certain problems that classical computation can not handle efficiently. Although this claim has not been definitively proven, there is substantial evidence supporting its validity.

Shor’s algorithm \cite{shor1994algorithms} serves as a notable example, as it efficiently solves problems like large integer factorization and discrete logarithms in $O(\text{poly}(n))$, while no classical algorithms can achieve this efficiently. The key difference lies in the nature of the computational models. Quantum Turing machines manipulate quantum states (vectors), whereas classical Turing machines operate on character strings. Moreover, quantum systems have a greater variety of allowed operations due to the larger set of valid quantum gates compared to classical logic gates.

Consequently, it is widely accepted that $P \subseteq BQP$, where $BQP$ represents the class of problems that can be efficiently solved by quantum algorithms. This inclusion indicates that quantum computation can address at least some problems more efficiently than classical methods.

\textbullet\quad\emph{Quantum circuits serve as an efficient tool for the depiction and analysis of quantum algorithms.}

The quantum Turing machine, while useful for understanding the theoretical relationship between quantum and classical algorithms, is generally different for practical algorithm description. Like high-level applications abstracted away from assembly language, quantum circuits offer a more practical framework for analyzing and describing quantum algorithms. Yao \cite{yao1993quantum}  demonstrates the equivalence between quantum circuits and quantum Turing machines \cite{yao1993quantum}. \cite{barenco1995elementary} provides foundational methods for analyzing quantum circuits and identifying their elementary gates. These elementary gates form a universal set for constructing and analyzing quantum algorithms. Consequently, any quantum algorithm, regardless of complexity, can be represented and evaluated in terms of circuit depth and gate count.

In addition to quantum circuits, there are two higher-level methods for describing quantum algorithms. The first method involves a sequence of unitary operators, while the second method builds upon the first by incorporating an oracle (a black box that performs specific functions). The use of an oracle is often justified because it might represent a trivial operation or because an efficient procedure should exist for those operations theoretically, which is not the primary concern here. Quantum query complexity \cite{ambainis2018understanding} is another approach to analyzing the complexity of a quantum algorithm based on the number of oracle queries.

While various methods for describing quantum algorithms are discussed in the literature, this survey does not strictly differentiate among them due to their inherent equivalency. However, constructing quantum circuits from unitary operators is not straightforward. For example, popular superconducting quantum hardware features sparse near-neighbor connectivity, requiring the use of swap gates to exchange non-adjacent qubits. The efficient layout of the circuit to minimize the use of swap gates is crucial for performance. Additionally, the native gates supported by different hardware platforms vary significantly, impacting performance based on which gates are used to build quantum circuits. This problem, known as ``quantum circuit compilation," has been proven NP-hard \cite{siraichi2018qubit}.

Given that quantum circuits and gates are essential for both describing and executing quantum algorithms, the next section will focus on discussing the most widely used elementary quantum gates.

2) Quantum Circuits and Elementary Quantum Gates

Quantum algorithms can be conceptualized as unitary evolutions starting from a predetermined initial state within a quantum system. Since unitary operations are essentially quantum circuits, decomposing a quantum circuit into elementary gates is equivalent to breaking down complex unitary evolutions into fundamental operations. A standardized evaluation criterion, using a uniform set of elementary gates, facilitates the comparison and assessment of the efficiency and costs associated with different quantum algorithms. Typically, the elementary gates of quantum circuits include single and two-qubit gates. It is important to note that the set of elementary gates is not fixed. It varies significantly depending on the hardware support. The following section examines some of the widely recognized elementary gates.

\textbullet\quad\emph{Single qubit gates}

\textcircled{1} Identity gate

$I=|0\rangle\langle0|+|1\rangle\langle1|$, i.e., the identity matrix. $I=\begin{pmatrix} 1 & 0\\ 0 & 1  \end{pmatrix}$.

\textcircled{2} Bit flip gate

$X=|0\rangle\langle1|+|1\rangle\langle0|$, $X=\begin{pmatrix} 0 & 1\\ 1 & 0  \end{pmatrix}$.

When acting on $|0\rangle$, $|1\rangle$, and $|\phi\rangle=\alpha|0\rangle+\beta|1\rangle$:
\begin{equation}
\begin{split}
X&|0\rangle=|1\rangle\\ 
X&|1\rangle=|0\rangle\\
X&|\phi\rangle=X(\alpha|0\rangle+\beta|1\rangle)=\alpha|1\rangle+\beta|0\rangle
\end{split}.
\end{equation}

\textcircled{3} Phase flip gate

$Z=|0\rangle\langle0|-|1\rangle\langle1|$, $Z=\begin{pmatrix} 1 & 0\\ 0 & -1  \end{pmatrix}$.

When acting on $|0\rangle$, $|1\rangle$, and $|\phi\rangle=\alpha|0\rangle+\beta|1\rangle$:
\begin{equation}
\begin{split}
Z&|0\rangle=|0\rangle\\
Z&|1\rangle=-|1\rangle\\
Z&|\phi\rangle=Z(\alpha|0\rangle+\beta|1\rangle)=\alpha|0\rangle-\beta|1\rangle
\end{split}.
\end{equation}

\textcircled{4} Bit and phase flip gate

$Y=iXZ=i(|1\rangle\langle0|-|0\rangle\langle1|)$, $Y=\begin{pmatrix} 0 & -i\\ i & 0  \end{pmatrix}$.

When acting on $|0\rangle$, $|1\rangle$, and $|\phi\rangle=\alpha|0\rangle+\beta|1\rangle$:
\begin{equation}
\begin{split}
Y&|0\rangle=i|1\rangle\\
Y&|1\rangle=-i|0\rangle\\
Y&|\phi\rangle=Y(\alpha|0\rangle+\beta|1\rangle)=i(\alpha|1\rangle-\beta|0\rangle)
\end{split}.
\end{equation}

$X$, $Y$, and $Z$ gates are fundamental single-qubit gates, frequently represented by the Pauli matrices: $\sigma^x$, $\sigma^y$, and $\sigma^z$. These matrices are not only unitary but also Hermitian, with their eigenvectors defining the coordinate axes on the Bloch sphere of a single qubit. These gates are essential for constructing $X$-, $Y$-, and $Z$-axis rotation gates, which serves as the building blocks for parameterized quantum circuits.

\textcircled{5} Rotation gates
\begin{equation}
\begin{split}
R_x(\theta)&=e^{-i\theta X/2}=cos\frac \theta 2 I -i sin \frac \theta 2 X\\
R_y(\theta)&=e^{-i\theta Y/2}=cos\frac \theta 2 I -i sin \frac \theta 2 Y\\
R_z(\theta)&=e^{-i\theta Z/2}=cos\frac \theta 2 I -i sin \frac \theta 2 Z
\end{split}.
\end{equation}

\textcircled{6} H gate

$H=\frac1{\sqrt{2}}\left((|0\rangle+|1\rangle)\langle0|+(|0\rangle-|1\rangle)\langle1|\right)$. Its matrix representation $H=\frac1{\sqrt{2}}\begin{pmatrix} 1 & 1\\ 1 & -1 \end{pmatrix}$. When acting on $|0\rangle$, $|1\rangle$, and $|\phi\rangle=\alpha|0\rangle+\beta|1\rangle$:
\begin{equation}
\begin{split}
H|0\rangle&=\frac1{\sqrt{2}}(|0\rangle+|1\rangle)=|+\rangle\\
H|1\rangle&=\frac1{\sqrt{2}}(|0\rangle-|1\rangle)=|-\rangle\\
H|\phi\rangle&\!=\!H(\alpha|0\rangle \! +\!\beta|1\rangle)\!\!=\!\!\frac1{\sqrt{2}}(\alpha\!+\!\beta)|0\rangle\!+\!\frac1{\sqrt{2}}(\alpha\!-\!\beta)|1\rangle
\end{split}.
\end{equation}

\begin{figure}[!htbp]
     \centering
    \subfloat[]{  	
    \includegraphics[width=2cm]{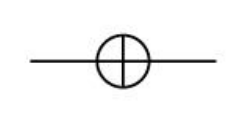}
    \label{not}
    }  
\quad  
    \subfloat[]{  \label{x}
    \includegraphics[width=1.8cm]{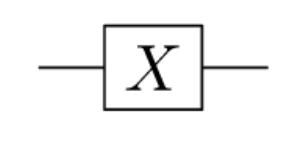}
    }  
     \caption{The bit flip gate.}
\end{figure}

Hadamard gate ($H$) is widely used in generating quantum circuits, such as creating the initial state $\ket{00 \cdots 0}$ for many quantum algorithms. It is important to note that the bit-flip gate ($X$) in Figs. (\ref{not}) and (\ref{x}) can be depicted differently in various sources. Actually, these representations are equivalent.

\textbullet\quad\emph{Double qubit gates}

Double qubit gates perform unitary operations involving two qubits. Among these, the controlled unitary gates are notable, where one qubit acts as the control qubit and the other as the target (controlled) qubit. The state of the control qubit determines whether the unitary operation is applied to the target qubit.

Here, we focus on the widely used CNOT gate in Fig. (\ref{cnot-gate-1-2}), also known as the quantum XOR gate. This gate operates on two qubits: $\ket{x}$ as the control qubit and $\ket{y}$ as the target qubit. The CNOT gate operates as follows:

\vspace{6pt}
\textbf{IF} {$\ket x = \ket 0$} \textbf{THEN} $\ket y$ unchanged, perform I operation;
\vspace{3pt}

\textbf{IF} {$\ket x = \ket 1$} \textbf{THEN} $\ket y$ flipped, perform X operation;
\vspace{6pt}

Thereby, $CNOT: \ket{x, y} \rightarrow \ket{x, x\oplus y}$, the definition is:
\begin{equation}
CNOT=|0\rangle\langle0|\otimes I+|1\rangle\langle1|\otimes X.
\end{equation}

Acting effect:
\begin{equation}
\begin{split}
CNOT|00\rangle=|00\rangle\qquad CNOT|01\rangle=|01\rangle\\ 
CNOT|10\rangle=|11\rangle\qquad CNOT|11\rangle=|10\rangle
\end{split}.
\end{equation}

Matrix representation:
\begin{equation}
\begin{pmatrix} 
1 & 0 & 0 & 0\\ 
0 & 1 & 0 & 0\\
0 & 0 & 0 & 1\\
0 & 0 & 1 & 0\\
\end{pmatrix}.
\end{equation}

Gate representation:
\begin{figure}[!h] 
   \centerline{\includegraphics[scale=0.4]{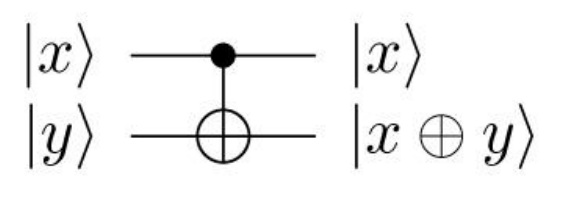}}
   \caption{CNOT gate.}\label{cnot-gate-1-2}
\end{figure}

\textbullet\quad\emph{Treble qubit gate}

It has been established that any unitary matrix within a $d$-dimensional Hilbert space can be decomposed into a product of two-level unitary matrices. Furthermore, it has been shown that a combination of single-qubit gates and a CNOT gate can facilitate any two-level unitary operation \cite{reck1994experimental}. Consequently, any three-qubit gate can be simplified using single-qubit gates and two-qubit gates. However, a specific three-qubit gate, known as the Toffoli gate, widely features as a fundamental module within quantum circuits and usually does not undergo further simplification:

The Toffoli gate has three inputs: $\ket{x}$, $\ket{y}$, and $\ket{z}$, where $\ket{x}$ and $\ket{y}$ are control qubits and $\ket{z}$ is the target qubit. The semantics of the Toffoli gate are as follows:

\vspace{6pt}
\textbf{IF}  $\ket x = \ket 1$ \textbf{AND} $\ket y = \ket 1$ \quad \textbf{THEN}

\qquad $\ket z$ flipped, perform X operation;

\vspace{3pt}
\textbf{ELSE}

\qquad $\ket z$ unchanged, perform I operation;
\vspace{6pt}

Therefore, $\text{Toffli}: |x,y,z\rangle \rightarrow |x,y,(x\land y)\oplus z\rangle$, the definition is:
\begin{equation}
\text{Toffli}=|0\rangle\langle0|\otimes I \otimes I+|1\rangle\langle1|\otimes CNOT.
\end{equation}

Acting effect:
\begin{equation}
\begin{split}
\text{Toffli}|000\rangle=|000\rangle\qquad \text{Toffli}|001\rangle=|001\rangle\\ 
\text{Toffli}|010\rangle=|010\rangle\qquad \text{Toffli}|011\rangle=|011\rangle\\ 
\text{Toffli}|100\rangle=|100\rangle\qquad \text{Toffli}|101\rangle=|101\rangle\\ 
\text{Toffli}|110\rangle=|111\rangle\qquad \text{Toffli}|111\rangle=|110\rangle 
\end{split}.
\end{equation}

The matrix representation:
\begin{equation}
\begin{pmatrix} 
1 & 0 & 0 & 0 & 0 & 0 & 0 & 0\\ 
0 & 1 & 0 & 0 & 0 & 0 & 0 & 0\\
0 & 0 & 1 & 0 & 0 & 0 & 0 & 0\\
0 & 0 & 0 & 1 & 0 & 0 & 0 & 0\\
0 & 0 & 0 & 0 & 1 & 0 & 0 & 0\\
0 & 0 & 0 & 0 & 0 & 1 & 0 & 0\\
0 & 0 & 0 & 0 & 0 & 0 & 0 & 1\\
0 & 0 & 0 & 0 & 0 & 0 & 1 & 0
\end{pmatrix}.
\end{equation}

\begin{figure}[!htbp]
     \centering
    \subfloat[]{  
    \includegraphics[width=4.2cm]{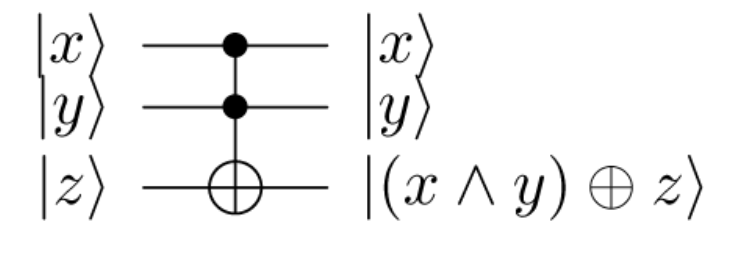}
    \label{toffoli}
    }  
\quad  
    \subfloat[]{  
    \includegraphics[width=3cm]{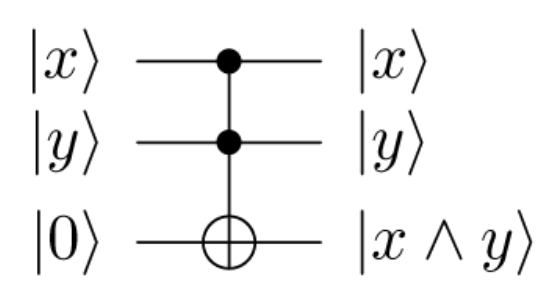}
    \label{AND}
    }  
     \caption{Toffli gate.}
\end{figure}

Toffoli gate representation is shown in Fig. (\ref{toffoli}). In particular, when $|z\rangle=|0\rangle$, the Toffoli gate can be exploited as a quantum `AND' gate in Fig. (\ref{AND}).

3) Measurement

Measurement (observation) is the only method to determine and extract the result of a quantum algorithm. Observing or measuring implies an interaction between macroscopic or classical objects and a quantum system, resulting in correlations and entanglements \cite{durr1998origin}.
Assume the state of a system under observation is $\ket{\phi} = \sum_i c_i \ket{i}$. When the system interacts with a measuring instrument, the combined wave function of the system and the instrument is given by \cref{instru},

\begin{equation} \label{instru} \ket{\Psi} = \sum_i c_i \ket{i} \otimes \ket{e_i}, \end{equation}
where ${ \ket{e_i} }$ represents the set of wave functions of the measurement instrument. This correlation signifies a form of entanglement: once the instrument is observed to be in the state $\ket{e_i}$, the entire wave function collapses to $\ket{i} \otimes \ket{e_i}$, allowing the system state $\ket{i}$ to be inferred.

The physical realization of this process involves significant experimental preparation, including tasks such as calibrating the parameterized Hamiltonian and determining the ground state of the entire system. However, in this context, we focus only on the algorithmic (circuit) level. We use a simple example to illustrate the generation of entanglement and the determination of the remaining segment when a partial state of an entangled system is observed.

\begin{figure}[!h]
   \centerline{\includegraphics[scale=1]{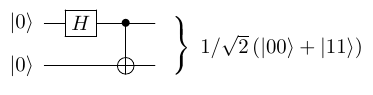}}
   \caption{The circuit to generate Bell state.}\label{Bell}
\end{figure}

Fig. \ref{Bell} shows the circuit used to generate one of the Bell states: $\frac{1}{\sqrt{2}} \left(\ket{00} + \ket{11}\right)$. The process starts with the first qubit $\ket{0}$, which is transformed into a superposition state $\frac{1}{\sqrt{2}} \left(\ket{0} + \ket{1}\right)$ after passing through gate $H$. Next, gate CNOT is applied, where $\ket{1}$ in the first qubit flips $\ket{0}$ in the second qubit, creating the entangled state in \cref{eq: Bell}. After generating entanglement, these qubits cannot be expressed as a $\otimes$ product of independent subsystems.
When observing the second qubit, if it is found in state $\ket{0}$, the first qubit collapses into $\ket{0}$. Conversely, if the second qubit is in state $\ket{1}$, the first qubit collapses into $\ket{1}$. In this scenario, the first qubit can be viewed as the quantum system being observed, while the second qubit serves as the measuring instrument.

\begin{equation}
\ket{00} \stackrel{H^1}{\longrightarrow}\frac {1} {\sqrt{2}} \left(\ket{0}+\ket{1}\right)\otimes\ket{0}\stackrel{CNOT}{\longrightarrow}\frac {1} {\sqrt{2}} \left(\ket{00}+\ket{11}\right)\label{eq: Bell}.
\end{equation}

In the preceding discussion, the description of an observer identifying a qubit in a specific state is imprecise. In quantum mechanics, a quantum state can not be observed directly; rather, it requires an observable for examination. Mathematically, an observable is represented by a Hermitian matrix.
The simplest and most widely used observable is the Pauli $Z$ operator, defined as $Z = \ket{0} \bra{0} - \ket{1} \bra{1}$. To observe an individual qubit in the state $\ket{\phi} = \alpha \ket{0} + \beta \ket{1}$, we proceed as follows:

\begin{equation}
\begin{split}
&\bra\phi Z \ket\phi \\
&= (\alpha^* \bra 0 + \beta^* \bra 1)(\ket 0 \bra 0 - \ket 1 \bra 1) (\alpha \ket 0 + \beta \ket 1)\\
&=|\alpha|^2(1)+|\beta|^2(-1)\label{eq: observation}
\end{split}.
\end{equation}
There are two possible outcomes when measuring the Pauli $Z$ observable: $+1$ and $-1$, corresponding to its eigenvalues. The probability of obtaining the result $+1$ is $|\alpha|^2$, while the probability of obtaining $-1$ is $|\beta|^2$. Thus, if every measurement yields $+1$, it can be inferred that the qubit is in the state $\ket{0}$, and similarly, a result of $-1$ indicates the qubit is in the state $\ket{1}$.

For a partial measurement of state Bell, where the observable is $I \otimes Z$, `observed by $Z$' refers to measurements in bases $\ket{0}$ and $\ket{1}$ (the computational bases). The choice of measurement basis is crucial. For instance, for the state $\frac{1}{\sqrt{2}} (\ket{0} + \ket{1})$, measuring in the $\ket{0}$ and $\ket{1}$ basis yields a probability of $\frac{1}{2}$ for either outcome, reflecting the uncertainty of the result. However, if the measurement is performed using the $\ket{+}$ and $\ket{-}$ basis (i.e., measuring with the Pauli $X$ observable), the outcome $\ket{+}$ is certain.

\begin{figure}[!h]
   \centerline{\includegraphics[scale=0.3]{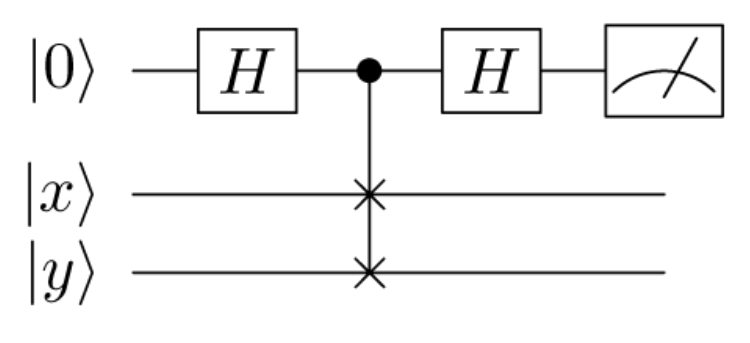}}
   \caption{Swap test to evaluate the similarity of two states.}\label{swaptest}
\end{figure}

Being able to deterministically obtain the result of a quantum algorithm implies that the chosen measurement bases can reliably differentiate the final state. Although only orthogonal quantum states can be distinguished with certainty, the ultimate states of a quantum algorithm are sometimes non-orthogonal. In such cases, the goal is to measure the similarity between states rather than distinguishing them with absolute certainty. This can be effectively achieved through probability estimation using partial measurements.
Fig. \ref{swaptest} illustrates this technique, where an initial $\ket{0}$ acts as an auxiliary qubit and transitions into superposition state $\frac{1}{\sqrt{2}}(\ket{0} + \ket{1})$ after passing through gate $H$. The circuit then uses state $\ket{1}$ as control to perform a swap operation between $\ket{x}$ and $\ket{y}$, resulting in the final state given by \cref{eq:swaptest}\footnote{The swap gate can be implemented using 3 CNOT gates.}:

\begin{equation}
\!\!\!\!\ket{0_{anc}} \otimes  \ket{xy} \!\! \rightarrow \!\! \frac 1 2 \! \ket{0_{anc}}\! \left(\! \ket{xy} \!+\! \ket{yx} \! \right)  +   \frac 1 2 \! \ket{1_{anc}} \! \left( \! \ket{xy} \! - \! \ket{yx} \! \right) \!\label{eq:swaptest}.
\end{equation}

If measuring the auxiliary qubit alone in bases $\ket{0}$ and $\ket{1}$, the probability of obtaining post-measurement state $\ket{0_{anx}}$ is given by:

\begin{equation} P(\ket{0_{anx}}) = \frac{1}{2} + \frac{1}{2} |\bra{x} \ket{y}|^2, \end{equation}
where $|\bra{x} \ket{y}|$ is known as \emph{fidelity} in quantum information theory, analogous to the classical \emph{cosine distance}. This probability equals 1 if $\ket{x}$ and $\ket{y}$ are identical, and $\frac{1}{2}$ if $\ket{x}$ and $\ket{y}$ are orthogonal. It is crucial to note that this probability estimation is independent of the dimension of the state space; larger dimensions enhance the effectiveness of this quantum technique. Additionally, the general similarity measure can be computed as $Euclidean\ distance = \sqrt{2 - 2|\bra{x} \ket{y}|}$.
In practice, this technique underpins many QML algorithms, including supervised and unsupervised ML, kernel methods in QML \cite{schuld2019quantum}, quantum support vector machines (QSVM) \cite{rebentrost2014quantum}, and quantum nearest-neighbor algorithms \cite{wiebe2015quantum}.

Lastly, it is worth mentioning weak measurement, which can be considered a generalization of strong (projective) measurement \cite{tamir2013introduction}. Unlike strong measurement, weak measurement does not directly observe the original system. Instead, it couples the system to an auxiliary quantum system, where partial information about the original system is extracted through a strong measurement of the auxiliary system, referred to as the probe. The wave function of the probe is typically a Gaussian wave packet centered at 0 with variance $\sigma^2$.
Consider a simple example of a two-qubit coupling, where the wave function is described in eq. (\ref{wave}):

\begin{equation}\label{wave}
\left(e^{-\frac{(x_0-\hbar/2)^2}{4\sigma^2}}\alpha\ket 0 + e^{-\frac{(x_0+\hbar/2)^2}{4\sigma^2}}\beta\ket 1\right) \otimes \ket{x_0},
\end{equation}
where $\ket{x_0}$ represents the post-measurement state of the probe system. When $x_0 \approx 1$, the probability amplitude of the subsequent state vector $\ket{0}$ is significantly high, and the converse is true when $x_0$ deviates from 1. If $\sigma \gg 1$, the measurement is considered very weak because the Gaussian term $e^{-\frac{(x_0-\hbar/2)^2}{4\sigma^2}}\approx 1$, indicating that the measurement induces minimal disturbance to the original system. In practice, the width of the probe's wave packet, $\sigma$, is tunable; increasing $\sigma$ reduces the impact of the measurement on the original system. In essence, the state vector of the system gradually shifts in the direction of the probe's collapse with a direct correspondence. 

Weak measurement allows for information extraction from quantum systems while preserving quantum properties. Leveraging this capability, \cite{ding2020retrieving} introduced a technique that utilizes weak measurements to label additional samples, thus improving classification accuracy in active learning frameworks.

\subsection{The Description by Density Operators}

The density operator provides a more powerful mathematical framework for describing quantum systems compared to state vectors, which is limited to representing pure states. In contrast, the density operator unifies the description of both pure and mixed states, making it a more powerful tool in quantum mechanics. This operator is especially useful in quantum information theory, such as in analyzing the relationship between entanglement entropy and algorithmic accuracy \cite{dupont2022entanglement}.

The density operator for a quantum system is defined as follows: suppose a system is in a pure state $\ket {\psi_i}$ with a probability $p_i$, where $i$ indexes a set of states ${\ket {\psi_i}}$. The quantum system is then described as an ensemble of pure states, with its density operator given by:

\begin{equation} \rho = \sum_i p_i \ket{\psi_i} \bra{\psi_i}. \end{equation}

The density operator is also referred to as the density matrix, and the terms are generally used interchangeably. For a density matrix $\rho$ to represent a valid quantum ensemble ${p_i, \ket{\psi_i}}$, it must satisfy the following conditions: (1) $tr (\rho) = 1$, and (2) $\rho$ is a positive semi-definite matrix. These properties ensure that $\rho$ is a legitimate description of a quantum system.

Using the density operator, we can effectively distinguish between pure and mixed quantum states. A pure state refers to a deterministic quantum system where $\rho = \ket{\psi}\bra{\psi}$, meaning the system is in state $\ket\psi$ with probability 1. In contrast, a mixed state describes a quantum system that is in a probabilistic combination of different states. The criterion to differentiate between pure and mixed states is based on the trace of $\rho^2$: if $tr(\rho^2) = 1$, then $\rho$ is a pure state; if $tr(\rho^2) < 1$, the system is in a mixed state.

In terms of density operators, the quantum algorithmic core steps can be described as: 

\noindent 1) unitary evolution
\begin{equation}
\rho_2 = U \rho_1 U^\dag,
\end{equation}
where $\rho_1$ is the state in time $t_1$ and $\rho_2$ is the state in time $t_2$.
2) measurement
\begin{equation}
\begin{split}
p(m)=&tr(M_m M_m^\dag \rho) \\
\rho_m=&\frac{M_m \rho M_m^\dag}{tr(M_m M_m^\dag \rho)}
\end{split},
\end{equation}
where $\{M_m\}$ is a set of measurement operators, the index $m$ refers to the possible measurement outcomes. $p(m)$ is the the probability of obtaining the outcome $m$ and $\rho_m$ is the corresponding post-measurement state. 
\subsection{QC in Noisy Intermediate-scale Quantum Era}

In this section, we do not cover widely known algorithms like Shor’s algorithm \cite{shor1994algorithms} and Grover’s algorithm \cite{grover1996fast}, as they have been extensively explored in various materials \cite{botsinis2018quantum}. Additionally, the circuit depths required to implement these algorithms far exceed what is currently feasible with NISQ devices. Instead, we focus on promising applications suited for NISQ devices, such as VQE \cite{peruzzo2014variational}, QAOA \cite{farhi2014quantum}, and QML algorithms, which have demonstrated potential quantum superiority over classical computing \cite{wu2021strong}. These algorithms share a common feature: a hybrid classical and quantum iterations to calibrate parameterized ansatzes. 

1) VQE

VQE is mainly applied in computational chemistry. With the development of modern quantum chemistry, computational chemistry has become an important tool for drug synthesis, catalyst preparation, etc, which involves a huge computation burden.
VQE simulates chemical computation on a NISQ quantum computer. It not only ensures the coherence of the quantum state but also achieves chemical accuracy for finding the ground state of a quantum chemical system. Its pipeline involves 3 major steps: 

\emph{Step 1 -- Pre-process}

Firstly, a quantum chemical system (electron system of a molecular) is represented as a Hamiltonian by Born-Oppenheimer approximation:

\begin{equation}
H=-\frac{h^2}{2m_e}\sum_i \nabla_i^2-\sum_i\frac{Ze^2}{r_i}+\sum_i\sum_j\frac{e^2}{r_{ij}}\label{first-quantization},
\end{equation}
where $\frac{h^2}{2m_e}\sum_i \nabla_i^2$ is the kinetic energy term of the electron, $\sum_i\frac{Ze^2}{r_i}$ is the gravitational potential energy term between nucleus and electrons, and $\sum_i\sum_j\frac{e^2}{r_{ij}}$ is repulsive energy term among electrons. This process is called the first quantization, and then $H$ in \cref{first-quantization} is transformed to the structure represented by the fermion operators,
\begin{equation}
H=\sum_{pq}h_{pq}a_p^\dagger a_q + \frac 1 2 \sum_{pqrs}a_p^\dagger a_q^\dagger a_r a_s,
\end{equation}
where $a_j^\dagger$ and $a_j$ are known as creation and annihilation operators, respectively. 
This process is called second quantization. After this process, the number of qubits to characterize $H$ drops from scale $\mathcal O(m log(n))$ to $\mathcal O(n)$ \cite{berry2018improved}, where $m$ and $n$ are the number of electrons and basis functions, respectively.

Finally, it needs to translate fermionic Hamiltonian into Pauli operators. Fermionic Hamiltonian follows the antisymmetry relationship, thereby the linear combination of translated Pauli strings should maintain this property. Exploiting Jordan-Wigner transformation \cite{jordan1993paulische}, Parity transformation \cite{seeley2012bravyi}, or other methods \cite{jiang2020optimal}, $H$ composed by Pauli operators satisfying this criterion can be obtained in eq. (\ref{VQE-Hamiltonian}),

\begin{equation}
H=\sum_j  w_j H_j\label{VQE-Hamiltonian},
\end{equation}
where $w_j$ are the weights of the Pauli strings in Hamiltonian.

\emph{Step 2 -- Core-process}

\begin{figure}[!h]
   \centerline{\includegraphics[scale=0.3]{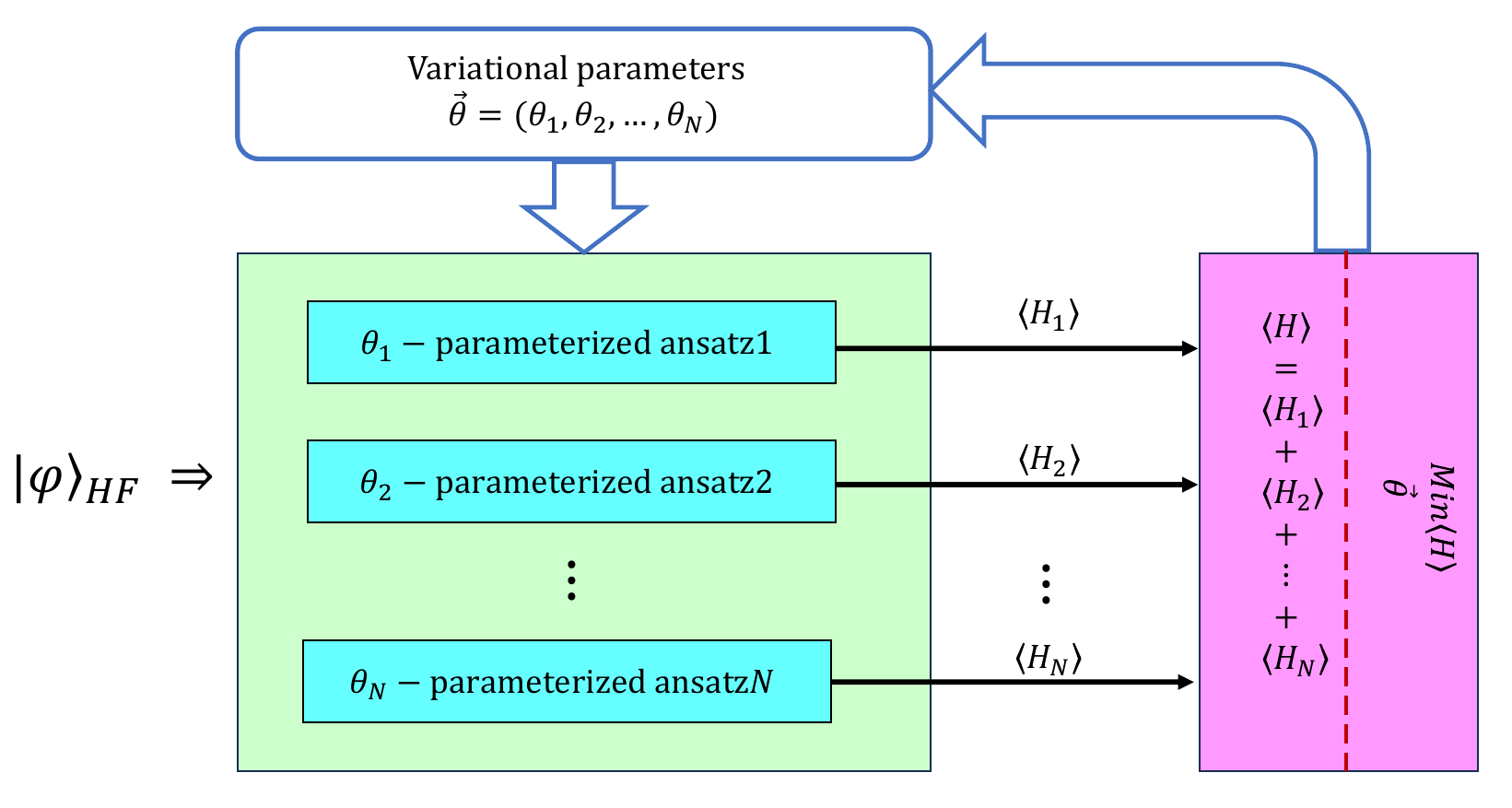}}
   \caption{The schematic diagram of VQE core-process.}\label{VQE}
\end{figure}

The core step is sketched in Fig. \ref{VQE}. First, the trail state is initialized as Hartree-Fock wave function. Simply speaking, for the molecular system with electronic spin-orbits, the Hartree-Fock state can be represented as $\ket{0\dots 0 1\dots 1}$, where $\ket{0}$ denotes empty orbit and $\ket{1}$ denotes occupied orbit. Then, this state goes into the parameterized devices (circuits) for evolving. All the parameters $\vec{\theta}$  of the devices is initialized randomly or by some specific methods to alleviate the suffering from barren plateaus in Fig. \ref{VQE}. When trial wave function $\ket{\varphi({\vec{\theta}})}$ is obtained, it is rotated into basis Z (computational basis) for measurements. In terms of the expectation value on each component in \cref{VQE-expectation}, we can get the total expectation $\langle{H}\rangle$ by weighted summation \cref{VQE-Hamiltonian}. Based on $\langle{H}\rangle$ obtained, we update the device parameters using a certain optimization strategy and begin a new iteration until obtaining $\min\limits_{{{\vec{\theta}}}}\langle{H}\rangle$,
\begin{equation}
\langle{H_j}\rangle=\bra{\varphi({\vec{\theta}})}U_j^\dagger(Z^{\otimes n})U_j\ket{\varphi({\vec{\theta}})}\label{VQE-expectation}.
\end{equation}

\emph{Step 3 -- Post-process}

The post-process, i.e., error mitigation, is an additional layer of processing measurement results aimed at reducing the impact of quantum noise for more accurate results. Since it is involved in all practical QC programs, we present a unified discussion after introducing other NISQ algorithms.

\vspace{5pt}

2) QAOA

QAOA \cite{farhi2014quantum} is mainly used to solve combinatorial optimization problems. Due to the ubiquitousness of optimization in various scientific and technological domains, this algorithm has aroused broad research interests. 
QAOA is approximate to quantum adiabatic algorithm (QAA) \cite{farhi2001quantum}. By applying Trotter approximation, time-dependent Hamiltonians of QAA are broken into a sequence of $2p$ time-independent Hamiltonians. Two parameterized unitary operators $\hat{U}_B(\beta)=e^{-i\beta \hat{B}}$ and $\hat{U}_C(\gamma)=e^{-i\gamma \hat{C}}$ is induced, where $\hat{B}$ and $\hat{C}$ are initial Hamiltonian and final Hamiltonian corresponding to ``mixing Hamiltonian'' and ``problem Hamiltonian'' of QAOA, respectively. Acting two unitary operators with parameters $\beta$ and $\gamma$ on the initial state iteratively,  final state $\ket{\beta, \gamma}$ approximates optimal solutions for optimization problems in eq. (\ref{eq:QAOA}),
\begin{equation}\label{eq:QAOA}
\ket{\beta, \gamma}=\hat{U}_B(\beta_p) \hat{U}_C(\gamma_p) \dots \hat{U}_B(\beta_1) \hat{U}_C(\gamma_1) \ket{s}.
\end{equation}

QAOA schematic diagram involves 3 steps in Fig. \ref{QAOA}:
\begin{figure}[!h]
   \centerline{\includegraphics[scale=0.6]{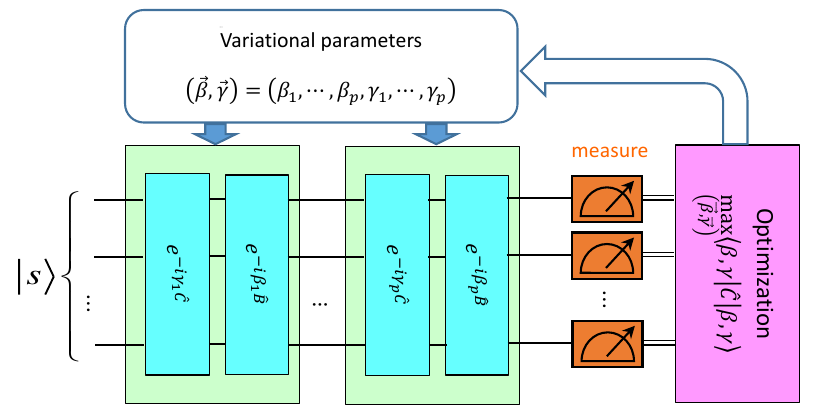}}
   \caption{The schematic diagram of QAOA.}\label{QAOA}
\end{figure}

\emph{Step 1 -- Prepare initial state}

In the vanilla QAOA applied to solve an unconstrained problem such as MAXCUT, the initial state is typically prepared as $\ket{++\dots+}$, which is the uniform superposition over all possible states. For constrained problems, the straightforward approach is to incorporate penalty terms into the problem Hamiltonian. This ensures that infeasible solutions violating the constraints are inferior to optimal feasible solutions in terms of the objective function. This method, often referred to as quadratic unconstrained binary optimization (QUBO), does not change the operator framework, allowing the initial state to remain as $\ket{++\dots+}$.

Nevertheless, numerous QAOA variants have demonstrated that initial state choices need not be restricted to $\ket{++\dots+}$. A simple approach is to employ a pre-processing technique that yields a biased initial solution (state) closer to the optimal solution. This so-called "warm-starting" approach accelerates convergence towards the final state. Various methods can be employed for warm-starting, including variable relaxation, positive semi-definite programming, graph neural networks, matrix product states, and more.

Another method for initial state preparation arises in an alternative approach to handling constraints, which focuses on restricting evolution to the feasible subspace. In this case, $\ket{++\dots+}$, which includes infeasible states, is not suitable. Instead, the initial state has to be prepared as a superposition composed of only feasible solutions that satisfy the constraints, excluding any infeasible solutions. For general constrained optimization problems, such as the maximum independent set, preparing a uniform superposition of all feasible solutions is computationally expensive \cite{volya2023state}. However, this approach is particularly effective for a class of NP-hard problems, including graph partition \cite{li2022large}, maximum k-vertex cover \cite{cook2020quantum}, and max-k colorable subgraph \cite{wang2020x}. These problems have feasible subspaces spanned by permutations of bit-strings with specific Hamming weights (the number of 1s in a bit-string equals $k$). As a result, their search space can be represented by a specific form of Dicke states, $\ket{D^n_k}=\frac{1}{\sqrt{\tbinom{n}{k}}}\sum_{x\in{0,1}^n,wt(x)=k}\ket{x}$, which can be efficiently prepared \cite{bartschi2019deterministic}.

\emph{Step 2 -- Construct mixing Hamiltonian and problem Hamiltonian}

The mixing Hamiltonian $\hat{B}$ is designed to perform "mixing" operations. When $\beta$ is treated as a time-dependent parameter, $e^{-i \beta \hat{B}}$ describes a continuous quantum walk on the graph defined by $\hat{B}$. This allows the quantum state to evolve from the initial state to a superposition of other feasible states. In simpler terms, the mixing operation represents a specific unitary transformation that drives the system from its initial configuration into a mixed superposition of possible solutions.
In the case of unconstrained problems, or QUBO-based constrained problems, the mixing Hamiltonian is typically defined as $\sum_i \sigma^x_i$. In physics, this operator is referred to as the transverse field operator, and its corresponding graph is a hyper-graph where nodes are connected pairwise by a Hamming distance of 1.

For constrained optimization problems, designing the mixing Hamiltonian becomes more intricate. By eliminating penalty terms and restricting the quantum evolution to the feasible subspace, one can achieve faster convergence at a lower computational cost. QAOA introduces specialized mixers such as XY-mixer and partial mixers for different types of constraints \cite{hadfield2019quantum}. Recent works have introduced further variants of the mixing Hamiltonian \cite{chandarana2022digitized}, focusing on reducing the required number of iterations $p$ through enhanced mixing capabilities. A smaller $p$ corresponds to a shallower quantum circuit, which is beneficial for maintaining quantum coherence and reducing gate errors.

Problem Hamiltonian $\hat{C}$ encodes an objective function to be optimized, where $\bra{x} \hat{C} \ket{x} = C(x)$, and $x$ is a bit string that indicates the inclusion of combinatorial variables in the solution. $C(x) \in \mathbb{R}$ is the value of the objective function for a given bit string $x$. Depending on specific optimization problems, the highest or lowest energy eigenstate $\ket{x^*}$ of the Ising Hamiltonian $\hat{C}$ represents the optimal solution. For Ansatz-based schemes, each component of $x$ is formulated using $\frac{I - \sigma_i^z}{2}$, known as the spin-to-binary conversion, which is used to generate matrix representation of $\hat{C}$. In QUBO-based schemes, additional penalty terms have to be constructed within $\hat{C}$ \cite{lucas2014ising}.

\emph{Step 3 -- Generate the corresponding circuit on specific hardware and run}

In terms of the unitary operators $\hat{U}_B$ and $\hat{U}_C$, they are translated into quantum circuits for execution on a quantum computer. The initial state is prepared and evolved through the circuit, producing a final state $\ket{x}$. A measurement is performed to evaluate $\bra{x}\hat{C}\ket{x}$, after which parameters $\vec{\beta}$ and $\vec{\gamma}$ are adjusted for the next iteration. This process continues until final state $\ket{x}$ converges to the optimal solution $\ket{x^*}$. Similar to VQE, this procedure involves the selection of classical optimization strategies and error mitigation techniques, which will be discussed in subsequent sections.

It is important to note that implementing QAOA-based algorithms on real quantum hardware presents additional challenges. The operator forms of different QAOA variants are often complex, making the quantum circuit deeper and more prone to errors. Furthermore, current quantum devices vary significantly in qubit connectivity topology, native gate sets, and operational fidelity. Compiling software provided by vendors is not yet capable of abstracting these hardware-specific characteristics efficiently.

\vspace{5pt}

3) QML

QML, a fascinating cross-research field, holds a solid foundation since the matrix manipulating techniques used in AI are well-studied in quantum theory. Over the past two decades, research in this area has achieved significant milestones. For example, QSVM exhibits a logarithmic runtime in both training and classification phases, offering an exponential speedup over its classical counterpart \cite{rebentrost2014quantum}. However, this advancement remains theoretical, since a key subroutine of QSVM, quantum principal component analysis \cite{lloyd2014quantum}, relies on the quantum Fourier transform, where the decomposition of controlled unitary gates results in quantum circuits too deep for near-term devices in the NISQ era. Thus, QML, along with other quantum algorithms, must adapt to these constraints and operate efficiently within the limited capabilities of present QC resources. This adaptation has led to the hybrid workflow model, where  PQC runs on quantum computers while classical computers optimize the parameters. PQC has effectively become synonymous with QNNs, and algorithms like VQE and QAOA can be considered special cases of PQC \cite{du2021learnability}. Given extensive QML literature, this discussion focuses on a few key areas, particularly classification, generative AI, and RL.

\vspace{5pt}

\textbullet\quad\emph{Classification}

\begin{figure}[!h]
   \centerline{\includegraphics[scale=0.55]{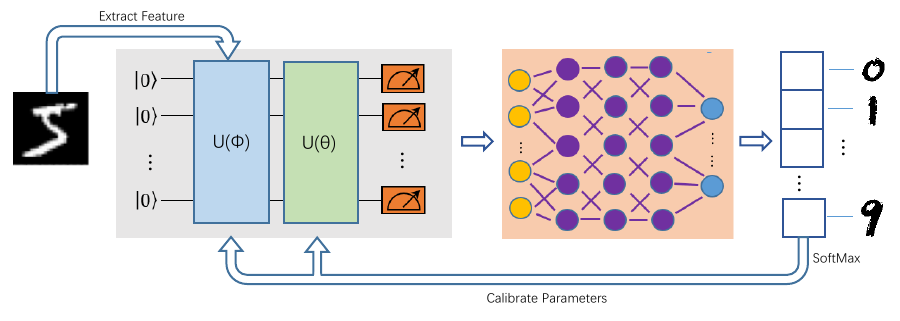}}
   \caption{The schematic diagram of typical hybrid quantum convolutional neural network architecture.}\label{classification}
\end{figure}

Classification, including clustering, is one of the most extensive studies in ML. In PQC-based frameworks, quantum subroutines replace one or more steps in a classical neural network, with the aim of enhancing the network expressibility, trainability, and generalization capabilities \cite{du2021learnability, lloyd2020quantum}. Fig. \ref{classification} illustrates a typical architecture of such frameworks. The leftmost block represents a pre-processing step, say feature extraction, on a classical computer. This step may be straightforward. For example, in the case of an image, the gray value of each pixel is embedded as the rotation angle of a quantum rotation gate in eq. (\ref{rot-g}) for subsequent quantum processing \cite{liu2021hybrid}. Here, $a_i \in [0, 255]$ represents the gray value of a pixel:

\begin{equation}\label{rot-g} R_y(\theta_i), \quad \theta_i = \frac{a_i}{255}*\pi/2. \end{equation}

The process could be more complex. For instance, \cite{srikumar2021clustering} utilizes quantum auto-encoders to extract key features from quantum states into a classical latent space. This enables clustering and classification using reduced quantum data.

The abstract notations $U(\phi)$ and $U(\theta)$ in the middle quantum processing block denote unitary operations for mapping classical data to the quantum domain and generating entanglement, respectively. The specific forms of $U(\phi)$ and $U(\theta)$ can vary depending on scenarios.
In quantum convolutional neural networks (QCNNs), $U(\phi)$ and $U(\theta)$ are combined to create a quantum convolution layer, typically realized by a series of rotation gate $R_y(\theta)$ and CNOT. Another variant of QCNNs incorporates multi-scale entanglement renormalization ansatz to implement a pooling layer in addition to the quantum convolutional layer.
It is important to highlight that entanglement plays a crucial role in achieving a quantum advantage with hybrid neural networks. In quantum transformer architectures, entanglement is generated using gate RBS, which can be further decomposed into four Hadamard gates, two rotation gates $R_y$, and two two-qubit gates $CZ$ \cite{kerenidis2024quantum}.

After quantum data are remeasured, they are converted back into classical data to proceed with the remaining operations in classical neural networks. Depending on specific problems, various objective or loss functions are computed. The results are then used to adjust the variational parameters $\boldsymbol{\phi}$ and $\boldsymbol{\theta}$ until convergence is achieved.
This iterative training process also involves addressing noise and gate errors, challenges that are common to all PQC-based algorithms. A detailed discussion on these aspects is provided in a subsequent section.

\vspace{5pt}

\textbullet\quad\emph{Generative AI}

In recent years, generative AI has emerged as a leading technology, capable of generating new samples by modeling the underlying probability distribution of datasets such as images and videos. This includes a range of methods from the Boltzmann machine \cite{zoufal2019quantum} to contemporary approaches like generative adversarial networks (GANs) \cite{romero2021variational} and denoising diffusion probabilistic models (DDPM) \cite{zhang2024generative}. Each of these techniques has its quantum counterpart.

Quantum generative learning models (QGLMs) have demonstrated greater versatility compared to classical models. They have been applied to quantum state estimation \cite{hu2019quantum}, continuous distribution generation \cite{anand2021noise}, and distribution learning \cite{zoufal2019quantum}, showing their capability to tackle problems that classical methods struggle with.

\begin{figure}[!h]
   \centerline{\includegraphics[scale=1]{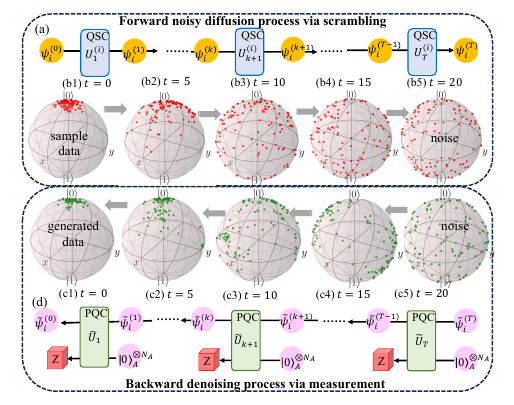}}
   \caption{The schematic diagram of QuDDPM \cite{zhang2024generative}.}\label{QuDDPM}
\end{figure}

QGLMs generally follow the same workflow as their classical equivalents. For instance, the quantum version of the denoising diffusion probabilistic model (QuDDPM) \cite{zhang2024generative}, depicted in Fig. \ref{QuDDPM}, consists of both forward noisy and backward denoising processes.
In the forward noisy diffusion process, a quantum scrambling circuit is utilized to evolve sample data towards a random ensemble of pure states. It is achieved by applying a series of random unitary gates independently. Conversely, the backward denoising process involves gradually reducing noise to obtain the final generated data. This is done by performing projective measurements after applying unitaries on ancilla qubits. This approach ensures that the generated data remains pure and closely resembles the sample data.
The integration of QC techniques with diffusion models provides significant advantages, not only for efficient training and learning of quantum data but also for enhancing performance in classical tasks. For example, \cite{kolle2024quantum} demonstrates that two quantum diffusion models, Q-Dense and QU-Net, outperform their classical counterparts.

The primary challenge in QGLMs is assessing their learnability, particularly in terms of generalization, expressivity, and trainability. Compared to quantum discriminative learning models, the generalization and expressivity of QGLMs have been less extensively studied. The concept of generalization in the context of quantum generative learning remains ambiguous across different model settings.
Trainability is another significant challenge for QGLMs, due to in part the non-convex nature of loss landscapes and the presence of barren plateaus, where gradients become small and hinder effective training.
Furthermore, the performance of QGLMs is heavily influenced by quantum hardware imperfections, including gate noise and measurement errors. When noise levels exceed a certain threshold, near-term quantum computers may perform worse than classical counterparts. Additionally, quantum system noise can lead to divergent optimization processes in QGLMs, further complicating their practical implementation.

\vspace{5pt}

\textbullet\quad\emph{RL}


Given that QC can enhance computational power, combining QC and RL has been studied for more than a decade.
QRL can be roughly divided into three categories according to the proportion of `quantum': the first is so-called inspired RL, which basically belongs to the category of classical heuristics. The second is PQC-based RL, which can run on NISQ devices and is the focus of our subsequent discussion. The last one is full-quantum RL. As the name suggests, algorithms in this category only work well on fault-tolerated quantum computers. 

\begin{figure}[!h]
   \centerline{\includegraphics[scale=0.56]{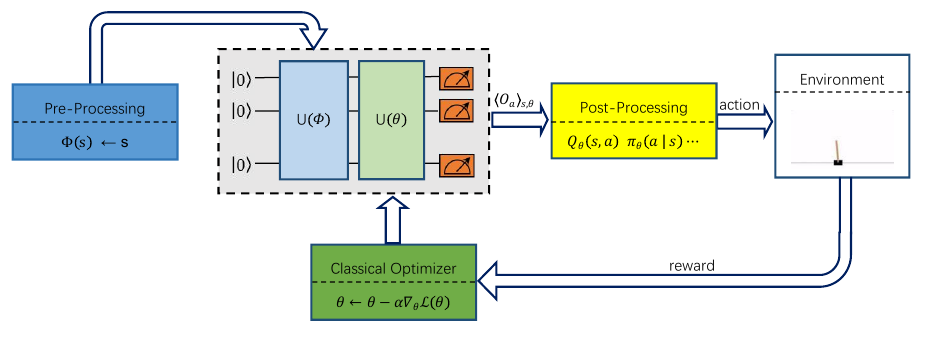}}
   \caption{The schematic diagram of PQC-based RL.}\label{QRL}
\end{figure}

PQC-based RL can be further classified in terms of the role of PQC in the pipeline, one is the value-function approximation approaches; one is policy approximation approaches, and the combined approximation for value and policy, i.e., actor-critic (AC) approaches.
A typical PQC-based QRL pipeline is summarized in Fig. \ref{QRL}. The hybrid agent observes the initial environment and applies pre-processing to encode the observation into the feature map $U(\phi)$. A $\theta$-parameterized unitary operation $U(\theta)$ then generates the current state $\ket{s_t} = U(\theta) U(\phi) \ket{0}^{\otimes n}$. Based on the measurements of the action-selection observable $\bra{s_t}O_a\ket{s_t}$, either the state-action value function $Q_\theta(s_t, a_t)$ or the policy $\pi_\theta(a_t|s_t)$ determines the next favorable action $a_{t+1}$. The cumulative reward $r_t$ obtained from interacting with the environment is subsequently fed into a classical optimizer for iterative parameter updates and optimization until convergence. This iterative pipeline resembles classical DRL, with the distinction that state changes occur in the quantum domain, where the variant can be viewed as a form of QNNs for approximation.

Applying quantum approaches to RL presents several challenges. One key challenge is the limited exploration of the true potential of near-term quantum methods in RL, as existing research has struggled to solve classical benchmarking tasks using PQC. Additionally, achieving sufficient performance in RL environments with PQCs as value-function or policy approximators has proven difficult.
The complexity and resource-intensive nature of quantum computations further complicate their scalability and practical implementation in RL tasks. Quantum computations require substantial resources, which can hinder their feasibility for real-world applications.
Moreover, the interpretability of quantum learning models is a concern. The quantum nature of these computations makes them difficult to understand compared to classical models, complicating efforts to improve models.
Finally, the accessibility of quantum computers for training QRL models could exacerbate existing inequalities in AI development and usage, potentially creating disparities in the ability to leverage quantum-enhanced technologies.

\vspace{5pt}

4) Challenges and Current Solutions

PQC-based algorithms, or rather hybrid variational quantum algorithms (VQA), all have a subroutine that employs classical optimizers to enable the training of parametrized quantum circuits to find the optimality of the objective function. However, barren plateaus become a core issue that confines the training efficiency. In the following section, we will summarize the related work regarding the circumstances where barren plateaus may arise and the solutions to avoid/alleviate them. On the other hand, due to the underlying discrepancy of QC infrastructure and various types of noise, running quantum algorithms on a genuine NISQ device is far more complex than running them on a simulator, which is another challenge that will be discussed afterward.

PQC-based algorithms, or hybrid variational quantum algorithms, typically involve a classical optimization subroutine to train parametrized quantum circuits and identify the optimal values for the objective function. A significant challenge in this context is the presence of barren plateaus, which can severely limit training efficiency. In the following section, we review related work addressing the conditions under which barren plateaus arise and explore strategies to avoid or mitigate them.
Additionally, the practical implementation of quantum algorithms on NISQ devices presents its own set of challenges. Due to variations in QC infrastructure and the presence of various types of noise, executing quantum algorithms on actual hardware is substantially more complex than running them on simulators. This challenge is discussed in detail in the subsequent section.

\textbf{\textbullet\quad\emph{Barren plateaus}}: 

Barren plateaus (BP) are characterized by an exponential vanishing of gradients across the optimization landscape. These plateaus not only impede gradient-based optimizers but also pose challenges for gradient-free algorithms like Nelder-Mead and COBYLA, as the flat landscape tends to exponentially dampen the cost function \cite{arrasmith2021effect}. The emergence of BP limits the potential of PQCs to achieve quantum advantages. This is because the exponential decline of gradients necessitates exponential resources for optimization, whereas quantum algorithms typically aim for polynomial scaling of the system.

BP occurs in varied cases \cite{du2022quantum} like random initialization in parametric circuit, objective function defined in terms of global observables, entanglement-induced, noise-induced, and PQC structure-induced. Layerwise learning, transfer learning, and Bayesian learning have been used to find a favorable initial set of parameters for mitigating BP \cite{liu2023mitigating}. \cite{haug2022generalized} proposes an algorithm called generalized quantum assisted simulator, which does not require a classical-quantum feedback loop. \cite{holmes2022connecting} establishes a fundamental connection between expressibility and trainability by extending the concept of BP to arbitrary ansatz, which provides insights into designing strategies for problem-inspired ansatz to avoid BP. \cite{larocca2023theory} show that over-parametrization in QNNs can mitigate BP.

Note that although there has been a wealth of research on the BP problem, little is known about the optimization landscape of a generalized PQC-based algorithm. To the best of our knowledge, the only one that can be sure to avoid BP is the work \cite{pesah2021absence}. The paper rigorously analyzes gradient scaling for parameters in QCNN architecture. It finds that the variance of the gradient vanishes no faster than polynomials, indicating that QCNNs do not exhibit BP. This result is significant as it provides an analytical guarantee for the trainability of randomly initialized QCNNs.

\textbf{\textbullet\quad\emph{Run quantum algorithms in real quantum computers}}:

The first step in implementing a quantum algorithm involves decomposing unitary operators (matrices) into smaller components—typically single-qubit and two-qubit gates. For a generalized unitary matrix $e^{-i H}$, it is currently known that the Hamiltonian $H$ must be sparse or low-rank to allow for a polynomial decomposition \cite{lloyd2014quantum}. Over the past few decades, significant efforts have been dedicated to discovering more efficient methods for such decompositions.
Once the unitary operators are decomposed, the next step is translating logical gates into native gates supported by the specific quantum hardware. Since QC platforms vary widely in underlying technologies (superconducting qubits, ion traps, optical photons, etc.), and even within the same technology there are differences at lower levels, considerable effort is required for optimization \cite{blinov2021comparison}. For instance, while both IBM and Rigetti machines are based on superconducting technology, their native gate sets differ. IBM natively supports the CNOT gate, whereas Rigetti requires the translation of a $CNOT(A, B)$ gate into a series of rotation and CZ operations. Vendor-provided compilers, such as IBM's Qiskit, are capable of translating multi-controlled gates into native gates that are optimized for specific hardware architectures. 

Finally, programmers must address various types of errors inherent in quantum systems. In the NISQ era, quantum hardware lacks the resources to fully support intensive demands of quantum error correction codes. Thus, quantum error mitigation, which reduces errors through classical postprocessing techniques, is currently the most practical solution for handling errors in NISQ devices \cite{bharti2022noisy}.
One common method requires knowledge of the noise model beforehand. It estimates parameters for the expected value of the noise-free model by analyzing the behavior of the noise model under different noise levels. Another technique is probabilistic error cancellation, which does not rely on precise knowledge of physical noise models in advance. It is based on gate set tomography and uses a linearly independent basis of operations to systematically remove the effects of localized Markovian errors \cite{kwon2020hybrid}. By measuring the impact of these errors, probabilistic error cancellation has been successfully implemented in both trapped-ion and superconducting systems.

\subsection{Quantum-inspired Algorithms}

The origin of quantum-inspired algorithms stems from the incorporation of QC concepts into classical algorithms. These algorithms are designed to run on classical computers while emulating certain quantum properties, such as superposition or entanglement, to enhance computational performance. By leveraging quantum principles, quantum-inspired algorithms offer a way to achieve improvements in efficiency and scalability, even without the use of quantum hardware.

A subclass of quantum-inspired algorithms considers their implementation on quantum hardware. For example, in QRL algorithms, state transitions are represented through measurements of an observable \cite{dong2008quantum}. After the measurement, the quantum state collapses to one of the observable’s eigenstates, corresponding to the next state, and the action is probabilistically selected by the agent. While theoretically plausible, practical challenges arise in more complex scenarios. Physically implementing an observable at each step is non-trivial, and the no-cloning theorem introduces complications regarding how to generate sufficient independent copies of quantum states to perform statistical analysis on measurement outcomes.

The second subclass of quantum-inspired algorithms explicitly operates on classical hardware. A notable example is Tang's work \cite{arrazola2020quantum}, where a binary tree-like data structure is introduced to meet quantum state preparation assumptions. This structure allows a classical algorithm to generate an $\mathcal{l}^2$-norm sample from a rank-$k$ approximation of a matrix in nearly polynomial time. It demonstrates how certain quantum advantages can be imitated classically under specific conditions, without requiring actual quantum computation. 

The third category is referred to as quantum-inspired metaheuristic algorithms. As the name suggests, these algorithms incorporate quantum principles to modify steps in classical metaheuristic algorithms, aiming to enhance performance \cite{gharehchopogh2023quantum}. Numerous studies have explored this approach, including quantum ant colony algorithms \cite{das2023quantum}, quantum evolutionary algorithms \cite{szwarcman2022quantum}, QGA \cite{konar2017improved}, and QPSO \cite{agrawal2021quantum}. These algorithms have been applied to solve classical NP-hard problems such as the knapsack and traveling salesman problems, as well as practical issues like clustering and project scheduling.

Modifications to quantum-inspired algorithms based on QC principles typically follow a consistent pattern. It involves representing original data, such as chromosomes in genetic algorithms, using qubits or n-dimensional normalized vectors. Rotation gates or matrices are then employed to evolve qubits for enhanced exploration. A random number generator is often used to generate new feasible solutions, simulating the collapse of quantum measurements to ground states. Additionally, entangled qubit states are integrated into the representation, ensuring that when a new population is created by measuring the qubit string (solution), the entangled qubits introduce diversity, which helps in avoiding local optima.

Quantum-inspired metaheuristic algorithms improve optimization by leveraging QC concepts to accelerate convergence, enhance exploration and exploitation, and find optimal solutions efficiently. However, much of the existing work still lacks rigorous complexity analysis to validate these performance improvements. Another concern is the challenge of embedding QC concepts into the optimization workflow in a way that aligns with metaheuristic algorithms, without compromising the algorithm's computational efficiency.

\subsection{Common QC Platforms and Libraries}

1) \emph{Hardware}

In the development of QC hardware, multiple technologies are being explored to realize universal quantum computers. For clarity, the key advantages and disadvantages of these hardware architectures are summarized in Table \ref{Comparsion-hardware}.

\begin{table*}[!t]
    \centering
    \caption{Comparison of varied QC hardware}
    \label{Comparsion-hardware}
    \begin{tabular}{|>{\centering\arraybackslash}m{1.5cm}|>{\centering\arraybackslash}m{3cm}|>{\centering\arraybackslash}m{3cm}|>{\centering\arraybackslash}m{3cm}|>{\centering\arraybackslash}m{3cm}|>{\centering\arraybackslash}m{1.5cm}|}
    \hline
     \textbf{Hardware architecture}&\textbf{technical route} & \textbf{Pros} & \textbf{Cons} & \textbf{Challenges} & \textbf{Typical vendors} \\ \hline
     Superconducting & Use superconducting circuits to generate and manipulate qubits, employ microwave pulses to control quantum gates. & Fast response, scalable, and seamlessly integrable with existing microelectronics technology. & Be susceptible to environmental perturbations, including thermal noise, electromagnetic noise, etc., resulting in shorter coherent time. &  Maintain high fidelity while scaling. & IBM, Google, Rigetti \\ \hline
     Ion Trap & Use light and magnetic fields, manipulate them with lasers. & High stability, long coherence times, and high quantum gate fidelity. & Cumbersome physical manipulation of ion traps, high optical stability requirements, and complex laser operating systems. & Complexity in scaling up the system. & IonQ \\ \hline
     Optical QC& Use photons as information carriers. & Less affected by environmental noise, with relatively mature techniques for generating, manipulating, and detecting single photons. & Require precise optical operations, reliable single-photon sources, and efficient single-photon detectors. & Implement quantum logic gates with high efficiency and fidelity, as well as large-scale optical QC. & Xanadu   \\ \hline
     Topological QC & Rely on emergent quasiparticles called Majorana zero-modes, which is robust against perturbations. & More stable qubits that are resistant to many types of errors and noise. & Manufacture and controll Majorana zero-modes is complex and technically challenging. &  Reliable fabrication and detection of Majorana zero-modes in experiments have not yet been achieved. & Microsoft  \\ \hline
    \end{tabular}
\end{table*}

2) \emph{Software}

\textbf{\textbullet\quad\emph{Qiskit \cite{cross2018ibm}}}: 
Qiskit, an open-source software development kit (SDK) by IBM, provides robust tools for programming quantum computers or simulators using Python. It enables access to IBM quantum systems and offers extensive documentation with tutorials and case studies, serving as a valuable resource for both learning and research in QC.
Qiskit has evolved rapidly, incorporating cutting-edge advancements such as Qiskit Machine Learning, which facilitates the creation and training of QML models. These models are capable of addressing various tasks, including classification, regression, and RL, making Qiskit a key platform for researchers and practitioners in the field.

\textbf{\textbullet\quad\emph{Cirq \& Tensorflow Quantum \cite{broughton2020tensorflow}}}:

Cirq, developed by Google's AI quantum team, is a powerful QC framework that supports both simulation and programming for specific quantum processors. Cirq offers two distinctive components: OpenFermion, designed primarily for quantum chemistry, and TensorFlow quantum (TFQ), an open-source ML library built on TensorFlow.
Cirq introduces two key data types: quantum circuit batch, which represents quantum circuits of varying sizes, and Pauli sum batch, which captures linear combinations of tensor products of Pauli operators. Using these data types, TFQ can efficiently handle tasks such as sampling the output distribution of circuit batches, computing the expected values of Pauli sums and circuit batches, calculating gradients, and simulating quantum circuits and quantum state batches.
By integrating TFQ with TensorFlow, practitioners can leverage TensorFlow's extensive capabilities, including automatic differentiation and ML tools, while also taking advantage of quantum circuit simulators and quantum hardware access. However, TFQ’s strong dependence on TensorFlow and Cirq presents compatibility challenges with other QC libraries and frameworks.

\textbf{\textbullet\quad\emph{PennyLane \cite{bergholm2018pennylane}}}:

PennyLane is a python-based QML library that integrates classical ML algorithms with QC using automatic differentiation. It operates independently of specific underlying frameworks, allowing seamless interaction with both classical ML libraries like Pytorch and TensorFlow and QC frameworks such as Qiskit and Cirq. PennyLane uses a plugin system to connect to various quantum hardware and simulators.
The library offers quantum-specific operations for evaluating and optimizing quantum circuits, encapsulates quantum circuits in quantum nodes, and supports hybrid quantum-classical ML models, including neural networks. PennyLane’s automatic differentiation quantum nodes facilitate the training of ML models and provide user-friendly interfaces for quantum algorithms and applications. However, its performance relies on the stability and quality of the plugin system, and its smaller community may result in fewer supporting resources compared to other tools integrated with traditional ML frameworks.

\vspace{2pt}

\textbf{\textbullet\quad\emph{Torch Quantum (QuantumNAS) \cite{wang2022quantumnas}}}:

Torch quantum, developed by a consortium including MIT, Yale, and the University of Texas at Austin, integrates seamlessly with Pytorch, adding several vital components: a quantum state vector simulator optimized for GPU acceleration, quantum circuit modules enabling the definition and manipulation of PQCs similar to classical neural networks, quantum optimization facilitating parameter optimization using classical algorithms like gradient descent, and an interface to real quantum devices. This extension of Pytorch allows developers to construct quantum programs using familiar tools and interfaces while benefiting from automatic differentiation for training QNNs. Additionally, the high-performance computing state vector simulator supports large-scale simulations on GPUs. However, its performance might not match specialized QC platforms like IBM quantum and Rigetti due to its reliance on Pytorch optimization.

\section{QC in Communications} 
\label{Communicating}

\subsection{Overview}
\begin{figure}
    \centering
    \includegraphics[width=1\linewidth]{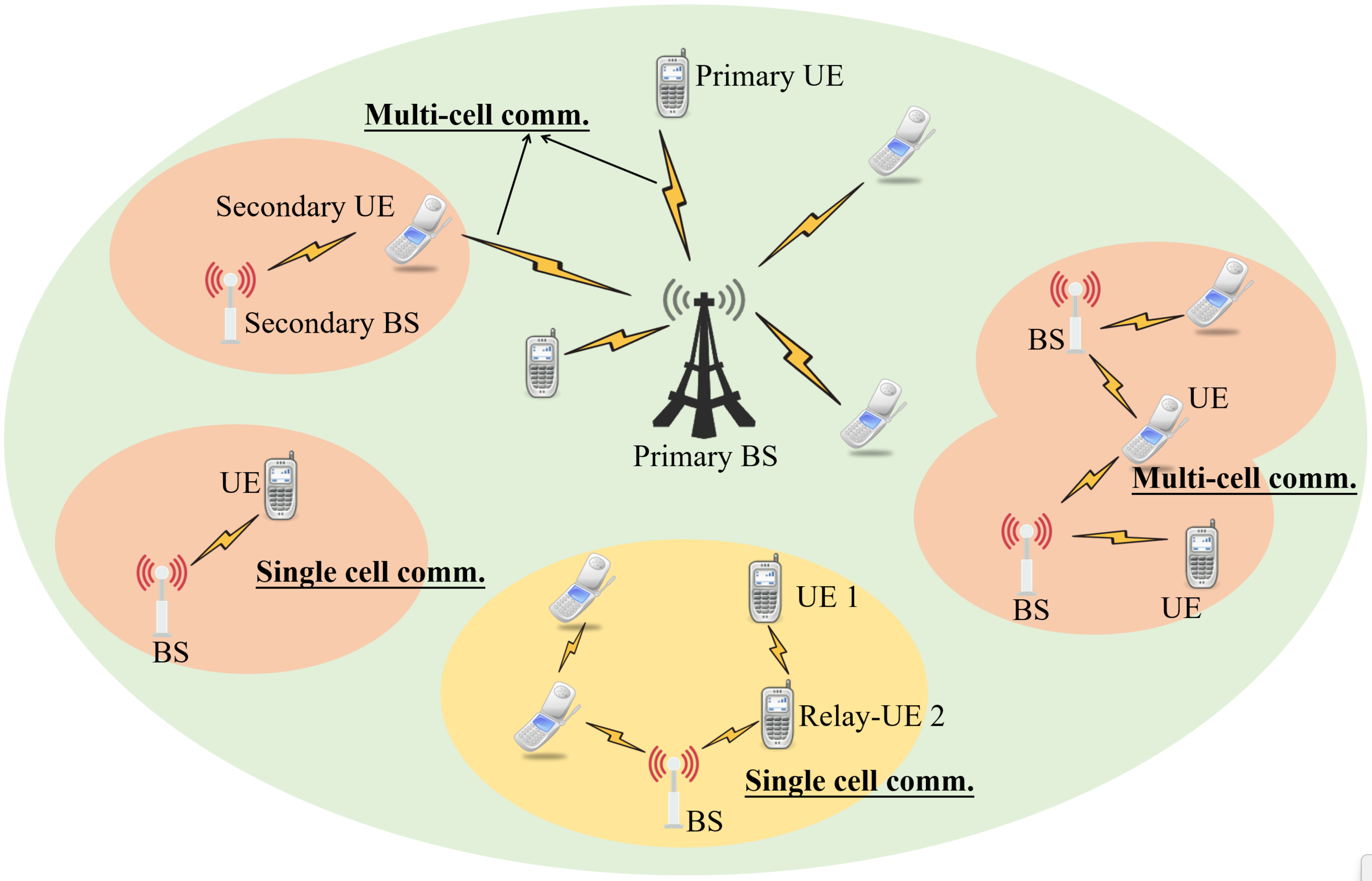}
    \caption{Different network scenarios in wireless communications. 
From the figure, single-cell communication can be easily identified in the bottom-left, while multi-cell communication is shown on the right. Cognitive networks within multi-cell communications, featuring primary and secondary base stations (BSs)/UEs, are displayed at the top of the figure. The relay networks are displayed at the bottom.}
    \label{figure-cell-communications}
\end{figure}

This section explores the applications of QC in various aspects of wireless communication, including power allocation, channel assignment, and user association, as illustrated in Fig. \ref{figure-cell-communications}. By leveraging the principles of quantum mechanics, researchers aim to address the challenges faced by traditional communication systems and unlock new opportunities for optimizing network performance and enhancing user experience.

\subsection{Power Allocation}\label{power allocation}

\vspace{5pt}
\subsubsection{Overview}
Power allocation within wireless communication \cite{lee2011distributed} involves the strategic distribution of available power resources among the diverse components of a communication system, such as BSs and user devices. The primary objective is to optimize power utilization to achieve communication that is both reliable and efficient, taking into account factors such as signal quality, network coverage, interference management, and energy consumption \cite{nasir2019multi}, which is also an important issue in 5G/6G cellular networks \cite{wei2023integrated}.
In more straightforward terms, power allocation revolves around determining the appropriate amount of power assigned to each transmitter or device in a wireless network to ensure effective communication.
Power allocation is usually implemented at the physical layer, involving only nodes within one same communication cell, optimizing the transmission power of signals to maximize transmission efficiency or minimize bit error rate. For example, power optimization with relay selection have been studied \cite{done-third-zhongrunhu-1-ok}. By selecting appropriate relay nodes and allocating corresponding power, the energy efficiency is maximized.

\begin{table*}[!t]
\centering
    \caption{Summary of QC Works on Power Allocations in Wireless Networks}
    \label{Summary-Power-Allocation}
    \scalebox{0.9}{
    \begin{tabular}{|>{\centering\arraybackslash}m{1.9cm}|>{\centering\arraybackslash}m{1.9cm}|>{\centering\arraybackslash}m{0.8cm}|>{\centering\arraybackslash}m{3.5cm}|>{\centering\arraybackslash}m{3.5cm}|>{\centering\arraybackslash}m{2.5cm}|>{\centering\arraybackslash}m{0.8cm}|>{\centering\arraybackslash}m{1.5cm}|} 
    \hline
    \textbf{QC category} & \textbf{Network structure} & \textbf{Work} & \textbf{Network details} & \textbf{Highlights} & \textbf{Optimization objectives} & \textbf{QC alg.} & \textbf{Control manner} \\ \hline
    
    \multirow{5}{*}{\centering  \makecell{\\ \\ \\ \\ \\ \\ \\ \\ \\ \\ \\Conventional QC}} & \multirow{3}{*}{ \centering  \makecell{\\ \\ \\ \\ Single cell \\communication}} &  \cite{done-third-zhongrunhu-9-ok} & K transmitter-receiver pairs with Gaussian Interference & Utilize a QC approach to solving power allocation & Maximize weighted-sum rate & QPSO & Centralized \\ \cline{3-8}
    
     &  &  \cite{done-gaopeng-14-ok-fine} & Primary users and secondary users competing for spectral resources & Employ Bernstein approximation to address probability constraints and co-evolutionary theory to handle large-scale problems. & Maximize network throughput & QPSO & Centralized \\ \cline{3-8}
     &  &  \cite{done-third-zhongrunhu-1-ok} & One-hop cooperative relay between BS and multiple users in a NOMA system & Apply QPSO to maximize QoS from the perspective of MIMO & Maximize computation efficiency and energy efficiency & QPSO & Centralized  \\ \cline{2-8}
     
     & \multirow{2}{*}{ \centering \makecell{\\ \\ \\Overlay \\networks}} &  \cite{done-second-gaopeng-18-ok} & Primary users and one primary BS; Cognitive users and one cognitive BS & Formulate resource scheduling in a two-layer wireless network, and propose a model to describe outage probability & Minimize system outage probability & QPSO & Centralized  \\ \cline{3-8}
     &  &  \cite{done-third-yangdongling-4-ok} & K users and N sub-carries in OFDMA system: two heterogeneous services, minimum-rate guarantee and best effort services & Divide the problem: channel assignment given power allocation, power allocation given channel assignment; transform it into a concave optimization and solved by Dinkelbach method & Maximize energy efficiency & QPSO & Centralized  \\ \hline
    \multirow{3}{*}{\centering  \makecell{\\ \\  \\ AI-driven QC}} & \multirow{3}{*}{ \centering  \makecell{\\   \\Single cell \\communication}} &  \cite{done-zhongrunhu-11-ok-fine} & Multiple users and a BS in a NOMA system & Propose federated teleportation based QNN and implement it in a case study & Maximize the weighted-sum rate & \multirow{2}{*}{ \centering  \makecell{\\ QNN}} & \multirow{2}{*}{\centering  \makecell{ \\ Distributed}} \\ \cline{3-6}
     &  &  \cite{done-zhongrunhu-15-ok-fine} & MIMO in a NOMA system & Present a QNN with parallel training & Maximize the weighted-sum rate &  &  \\ \cline{3-8} 
     &  &  \cite{done-zhongrunhu-20-ok} & Users and a BS in a NOMA system & Present quantum circuits of layerwise embedding with implementation in NISQ computers & Maximize cumulative energy efficiency over time & QDRL & Centralized (only one UAV) \\ \hline
\end{tabular}
}
\end{table*}

\begin{table*}[!t]
    \centering
    \caption{Details of QC Algorithms for Power Allocation}
    \label{QC-Power-Allocation}
    \begin{tabular}{|>{\centering\arraybackslash}m{1cm}|>{\centering\arraybackslash}m{2cm}|>{\centering\arraybackslash}m{2cm}|>{\centering\arraybackslash}m{4.5cm}|>{\centering\arraybackslash}m{3cm}|>{\centering\arraybackslash}m{2cm}|}
    \hline
    \textbf{Work} & \textbf{Algorithm} & \textbf{Orth. QC} & \textbf{Implemented in quantum computers} & \textbf{Platform} & \textbf{No. qubits} \\ \hline
    \cite{done-third-zhongrunhu-9-ok} & QPSO & \XSolidBrush & \XSolidBrush & Computer simulation & 0  \\ \hline
    \cite{done-gaopeng-14-ok-fine} & QPSO & \XSolidBrush & \XSolidBrush & Computer simulation & 0  \\ \hline
    \cite{done-third-zhongrunhu-1-ok} & QPSO & \XSolidBrush & \XSolidBrush & Computer simulation & 0   \\ \hline
    \cite{done-second-gaopeng-18-ok} & QPSO & \XSolidBrush & \XSolidBrush & Computer simulation & 0  \\ \hline
    \cite{done-third-yangdongling-4-ok} & QPSO & \XSolidBrush & \XSolidBrush & Computer simulation & 0  \\ \hline
    \cite{done-zhongrunhu-11-ok-fine} & QNN & \Checkmark & \Checkmark & IBM Qiskit & 16  \\ \hline
    \cite{done-zhongrunhu-15-ok-fine} & QNN & \Checkmark & \Checkmark & IBM Qiskit & 9  \\ \hline
    \cite{done-zhongrunhu-20-ok} & QDRL & \Checkmark & \Checkmark & IBM Qiskit & 12  \\ \hline
    \end{tabular}
\end{table*}

In this section, we undertake an investigation of existing QC-driven approaches along two fundamental dimensions: the complexity of network scenarios and the categories of QC algorithms. The former encompasses networks ranging from one-hop non-orthogonal channel multiple access (NOMA)/orthogonal frequency-division multiple access (OFDMA) systems to multiple-cell and overlay systems. Meanwhile, we classify the latter into conventional QC algorithms, such as classical QAOA and quantum-inspired QPSO, and AI-driven QC algorithms like QNN, with an emphasis on comprehending their suitability and effectiveness across various network landscapes.
\subsubsection{Applications of Conventional QC for Power Allocation}

\textbf{QC in a single-cell system}:

A single-cell communication system in wireless networks typically consists of a BS or access point and multiple user devices, responsible for handling wireless communications within a local area. User devices within this area communicate with the BS. A typical single-cell system is shown in Fig. \ref{figure-cell-communications}.
\cite{done-third-zhongrunhu-9-ok} and \cite{done-gaopeng-14-ok-fine}  consider power allocation within a straightforward context, focusing on a single cell of a multiple access (MA) system while applying conventional QC algorithms. 
$K$ transmitter-receiver communication pairs with Gaussian interference are involved in \cite{done-third-zhongrunhu-9-ok}. The problem is conceptualized as a non-convex and NP-hard programming formulation aimed at maximizing the weighted-sum rate. To address this problem, a QPSO based approach is employed, leveraging the characteristics of QC and biological behaviors. However, the network structure is simple and only Shannon capacity is considered without addressing interference.
Energy limitation issues are addressed in \cite{done-gaopeng-14-ok-fine} in  cognitive radio sensor networks by constructing a mathematical model for power allocation. Bernstein approximation technique is used to control the approximation error of function value to tackle probability constraints. In addition, it adopts the concept of co-evolution to deal with large scale problems. The proposed optimization problem is complex and NP-hard, which is corresponding to the co-evolutionary theory. A stochastic optimization algorithm based on polynomial approximation and QPSO is proposed to solve the problem. 

Though one-hop cooperative relay between the BS and multiple users, such as UE 1 communicated with BS via relay-UE 2 in Fig. \ref{figure-cell-communications}, is considered \cite{done-third-zhongrunhu-1-ok}, it is still within the area of a single-cell system. The relay link is introduced between the source node and destination node to reduce channel interference. Thus, there are two different links: the BS to relay nodes, and relay nodes to users. The strategies of power allocation for multiple relays and users significantly influence the interference among nodes in the local communication area. The situation becomes worse when a full-duplex operation of channels from the perspective of multiple input multiple output antennas (MIMO) is considered. Besides, another QoS metric, secrecy rates of end-users, is considered to represent the overhearing from eavesdroppers near end-users. The problem formulation is proved to be non-convex and solved by QPSO.

\textbf{QC in an overlay system}:

In contrast, more complex scenarios, spanning from multiple-cell networks to overlay networks, are investigated in \cite{done-second-gaopeng-18-ok}, and \cite{done-third-yangdongling-4-ok} while still utilizing conventional QC algorithms. An overlay system refers to an architecture in which additional network components or services are deployed over a single-cell system, like mesh networks, or virtualized network functions, and among others. 

The problem of scheduling multiple resources including power allocation and channel assignment in an overlay OFDMA network is studied in \cite{done-second-gaopeng-18-ok}. There is an additional constraint on the channel assignment for cognitive users, which complicates the problem comparing with resource scheduling in traditional wireless cognitive networks without primary networks. The work formulates the problem of resource scheduling in a two-layer wireless network, and proposes a model to describe system outage probability according to strategies of power allocation and channel assignment. 
\cite{done-third-yangdongling-4-ok} discusses the resource scheduling, including power allocation and channel assignment, in a single-cell OFDMA system with $K$ users and $N$ sub-carries, in which there are two heterogeneous service requirements from users: minimum-rate guarantee services and best effort services. It can be viewed as a virtual overlay network according to two heterogeneous service requirements. The work divides the problem formulation into two optimization parts: channel assignment for a given power allocation, which is solved stochastically by QPSO based solution; power allocation for a given channel assignment, which is transformed into a concave optimization problem and solved optimally by Dinkelbach method. Finally, an acceptable solution is obtained by iterating two optimization parts. 
Noting that there is no analysis to obtain the optimal solution by iterations even if the optimal solutions for two parts are obtained.
Specifically from the QC perspective, there are three modules same with the traditional PSO algorithm: particle position, fitness function, and evolution equation. The difference lies in that the iterative equation in the evolution operation for the mean best position bits, namely, decision variables of assigning frequency resource units to multiple users, is determined by the Hamming distance of two input binary bits. The $j$-th bit in the mean best solution from all particles is determined by the states of the corresponding $j$-th bit of each best particle.

\vspace{5pt}
\subsubsection{Applications of AI-driven QC for Power Allocation} \

The previous studies on power allocation rely on conventional QC algorithms like QPSO. However, it is worth noting that these conventional algorithms still demand significant computational resources and lead to high overhead, such as computation delays.
Recent research has turned towards AI-driven QC solutions for power allocation. These  approaches, such as QNN and QDRL, leverage advanced ML techniques to mitigate the computational complexities inherent in traditional QC algorithms. By harnessing the AI power, these methods offer promising avenues for more efficient and effective power allocation in wireless networks.
For example, \cite{done-zhongrunhu-11-ok-fine} proposes a distributed QNN framework for the purpose of optimizing wireless communication and gives a case study in non-orthogonal multiple access system for the framework implementation. The objective is also to maximize the weighted-sum rate under consideration of channel interference, and the proposed QNN algorithm is implemented in quantum circuits, which accelerates running time.

Other works implemented in quantum computers include \cite{done-zhongrunhu-15-ok-fine, done-zhongrunhu-20-ok} in the field of power allocation. 
\cite{done-zhongrunhu-15-ok-fine} presents a QNN with parallel training for optimizing wireless resource allocation. The proposal reduces the dimension of training data by sending only the statistical parameters of the dataset instead of the whole dataset. The proposed method is applied to optimize transmit precoding and power allocation in MIMO-NOMA. The analysis shows that the parallel training approach achieves a comparable sum rate with lower complexity compared to conventional methods. It highlights the potential benefits of QNNs in processing large dimensional solution spaces and in addressing the challenges of distributed datasets and training complexity. Specifically, similar to neurons in deep learning, the proposed QNNs for parallel training apply qubits as the basic units of computation. Each qubit, associated with a parameterized gate, is measured at the end of the QNN inference. The measurement yields the result for each individual qubit, and the corresponding loss gradient for the $i$-th qubit is calculated at the same time. This allows all weight parameters to be updated simultaneously, which significantly reduces the overall complexity of the for parallel training  process. There are three modules in the QNN operations: encoding, QNN inference, and decoding. The input encoding is performed to map an arbitrary classical value to its quantum state, in which encoding is implemented by a quantum-based operation. The QNN inference is similar to the forward propagation in deep learning. Afterwards, each quantum measurement gives result to either 1 or 0. The result of quantum measurements is then decoded as QNN output by means of assigning quantum states.

\cite{done-zhongrunhu-20-ok} proposes a layerwise QDRL method for optimizing continuous large space and time series problems using deep layer training. It leverages the advantages of QC to maximize reward and reduce training loss. The study focuses on jointly optimizing UAV trajectory planning, user grouping, and power allocation for energy efficiency. The main contributions of the study are the introduction of layerwise quantum embedding, and the presentation of quantum circuits for practical implementation. In quantum embedding, the encoding operation typically consists of multiple layers, with quantum measurements performed only after the final layer to obtain classical values. In this study, it adopts a layer-wise approach to training, where quantum measurements are conducted at the end of each layer. The classical values derived from these measurements are then used to compute the local loss for that layer. This localized loss is subsequently used to update the training parameters for the next layer, allowing for more granular optimization during the training process. However, the solution is assumed that users can perceive perfect CSI in the UAV downlink communication channel, which limits its application.

In conclusion, the integration of QC techniques in power allocation holds promise for optimizing various aspects of wireless networks. Table \ref{Summary-Power-Allocation} provides an overview of the main findings, optimization objectives, and QC algorithms utilized across different wireless networks. Noting that, there are relatively fewer studies employing QC algorithms for power allocation, largely due to the localized nature of power allocation in wireless networks, typically confined to a single-cell system. This observation is reflected in both the choice of QC algorithms and their control manners as depicted in Table \ref{QC-Power-Allocation}.  Note that the algorithms can be classified into orthodox QC algorithms and simulated ones. Orthodox QC algorithms refer to those that are directly implemented on quantum computers, leveraging quantum hardware for execution.
The majority of studies utilize QC simulated algorithms that are implemented on traditional computers.

\subsection{Channel Assignment}
\subsubsection{Overview}
Channel assignment is a fundamental aspect of wireless communication network management, where the task involves allocating communication channels to different users or devices within the network and well addressed in 5G networks \cite{wei2023integrated}, such as slicing solutions. In wireless networks, communication channels represent the available frequency bands or time slots that facilitate data transmission between transmitters and receivers. Efficient channel assignment is crucial for optimizing network performance, minimizing interference, and maximizing spectral efficiency.
The problem has been addressed by various algorithms and strategies, such as RL approaches \cite{9003500}. These methods aim to allocate channels in a way that minimizes interference between users while maximizing the utilization of available spectrum resources.

With the growing complexity and scale of wireless networks, traditional channel assignment approaches may face limitations in addressing the dynamic and heterogeneous communication environments. QC presents a promising avenue for tackling these challenges by offering novel algorithms and techniques that can potentially improve the spectral efficiency, reliability, and scalability of channel assignment in wireless communication networks.

\begin{table*}[!t]
\centering
\caption{Summary of QC Works on Channel Assignment in Wireless Networks}
\label{QC-Network-Optimization}
\scalebox{0.75}{
\begin{tabular}{|>{\centering\arraybackslash}m{2.5cm}|>{\centering\arraybackslash}m{3cm}|>{\centering\arraybackslash}m{1cm}|>{\centering\arraybackslash}m{2.5cm}|>{\centering\arraybackslash}m{4cm}|>{\centering\arraybackslash}m{3.5cm}|>{\centering\arraybackslash}m{2cm}|>{\centering\arraybackslash}m{2cm}|}
\hline
\textbf{QC category} & \textbf{Network structure} & \textbf{Work} & \textbf{Network details} & \textbf{Highlights} & \textbf{Optimization objectives} & \textbf{QC alg.} & \textbf{Control manner} \\ \hline
\multirow{6}{*}{\centering \makecell{\\ \\ \\ \\ \\ \\ \\ \\ \\ \\ \\ \\ Conventional QC}} & \multirow{4}{*}{\centering \makecell{\\  \\ \\ \\ \\ \\ \\Single-cell systems}} & \cite{done-second-chengyu-15-half-ok} & K users and a receiver over MIMO antennas & Encode the maximum likelihood detection problem into a Hamiltonian operator, and transform the problem into a level-p QAOA circuit & Maximize signal detection probability & QAOA & Centralized \\ \cline{3-8} 
 &  & \cite{done-third-zhongrunhu-13-ok} & N transmitter-receiver pairs in a cell & Transform sub-channel assignment into a capacitated Max k-Cut formulation, which is induced as a QUBO problem, and QA is applied to solve the problem & Maximize the sum rate of D2D transmitter-receiver pairs & QA & Centralized \\ \cline{3-8} 
 &  & \cite{done-third-yangdongling-12-ok} & N users competing for orthogonal channels in cognitive networks & Map channel assignment matrix to the chromosome of the algorithms to decrease search space & Maximize sum rates of users, the minimal rate of users, proportional fair rates of users & QGA & Centralized \\ \cline{3-8} 
 &  & \cite{done-second-shike-17-ok} & Mobile cellular networks & Present a case study on the assignment problem of root sequence index in LTE/NR physical random access channel configuration & Maximize orthogonality metric, the set cardinality of root sequence index, among mobile users & QA & Centralized \\ \cline{2-8} 
 & \multirow{2}{*}{\centering \makecell{\\  \\Multi-hop networks}} & \cite{done-chengyu-13-OK} & No two links within k hops is simultaneously activated. & Schedule air-links for maximum throughput while avoiding k-hop interference & Maximize network throughput & QA & Centralized \\ \cline{3-8} 
 &  & \cite{done-wucheng-16-ok} & Network graph: vertex denotes T/R nodes and edges denote links. & Schedule air-links for minimizing k-hop interference & Minimize k-hop interference & QA & Centralized \\ \hline
\multirow{3}{*}{\centering \makecell{\\ \\ \\  \\ \\ \\AI-driven QC}} & \multirow{2}{*}{\centering \makecell{\\ \\ \\Single-cell systems}} & \cite{done-wucheng-17-ok} & NOMA networks & Propose a QNN framework for resource allocation and a solution for non-orthogonal NOMA as a case study & Maximize the sum communication rate of users. & QNNs & Centralized \\ \cline{3-8} 
 &  & \cite{done-zuoqixue-14-ok} & Multiple UAVs a central server & Study the impact of channel characteristics, such as communication scattering.  Formulate UAV movement into MDP, and develop a quantum multi-agent AC algorithm & Maximize sum communication rates of ground users & Quantum multi-agent AC & Distributed \\ \cline{2-8} 
 & Overlay networks & \cite{done-third-gaopeng-13-ok} & Vehicle-to-everything: eNodeBs allocate channels to RSUs, RSUs allocate them to vehicles. & Formulate channel assignment as a discrete MDP and adopt QRL to find the optimal policy. & Minimize the transmission delay and energy consumption & QRL based on Grover Search Algorithm & Distributed \\ \hline

\multirow{5}{*}{\centering \makecell{\\ \\ \\ \\ \\ \\ \\ \\Conventional QC for \\IRS systems}} & \multirow{3}{*}{\centering \makecell{\\ \\ \\ \\Single-cell systems}} & \cite{done-second-shike-16-ok} & One BS and multiple users in a cell under  assistance of one IRS & Propose a scheduling method where the IRS is assigned to only a single user at a given time slot. & Maximize network throughput & QA & Centralized \\ \cline{3-8} 

 &  & \cite{done-second-gaopeng-10-ok} & One BS and users in a cell under assistance of one IRS & Propose a scheduling method where IRS is assigned to only one user at a given time slot. & Maximize secrecy energy efficiency & Quantum bald eagle search algorithm & Centralized \\ \cline{3-8} 
 
 &  & \cite{done-wucheng-20-OK} & One BS and users in a cell with assistance of multiple IRSs & Propose a scheduling method where one IRS is assigned to only one user at a given time slot & Maximize network throughput & QA & Centralized \\ \cline{2-8} 
 
 & \multirow{2}{*}{\centering \makecell{\\   \\Multi-cell systems}} & \cite{done-wucheng-9-ok} & Multiple BSs and users with assistance of multiple IRSs & Propose a scheduling method where each IRS is assigned to only one user at a given time slot & Maximize network throughput & QA & Centralized \\ \cline{3-8} 

 &  & \cite{done-zhongrunhu-10-ok-fine} & Multiple BSs and multiple users under assistance of one IRS & Share the IRS among BSs by periodically switching phase shifts to reduce hardware cost for multiple IRSs & Minimize the number of required time slots to complete the communication & QA & Centralized \\ \hline
\end{tabular}
}
\end{table*}
%


\begin{table*}[!t]
    \centering
    \caption{Details of QC Algorithms for Channel Assignment}
    \label{Channel Assignment Table2}
    \begin{tabular}{|>{\centering\arraybackslash}m{1cm}|>{\centering\arraybackslash}m{2cm}|>{\centering\arraybackslash}m{1.5cm}|>{\centering\arraybackslash}m{4.5cm}|>{\centering\arraybackslash}m{3cm}|>{\centering\arraybackslash}m{3.5cm}|}
    \hline
    \textbf{Work} & \textbf{Algorithm} & \textbf{Orth. QC} & \textbf{Implemented in quantum computers} & \textbf{Platform} & \textbf{No. qubits}  \\ \hline

    \centering\cite{done-second-chengyu-15-half-ok} & QAOA & \Checkmark &  \Checkmark & IBM Qiskit & 1, 2, 3  \\ \hline
    \centering\cite{done-third-zhongrunhu-13-ok} & QA &  \Checkmark &  \Checkmark  & D-Wave & 2000+  \\ \hline
    \centering\cite{done-chengyu-13-OK} & QA & \Checkmark & \Checkmark & D-Wave & 164, 405 \\ \hline
    \centering\cite{done-wucheng-16-ok} & QA & \Checkmark & \Checkmark &  D-Wave 2X & 164, 405, 759, 783 \\ \hline
    \centering\cite{done-second-shike-17-ok} & QA & \Checkmark & \Checkmark & D-Wave & No. optimization variables  \\ \hline
    \centering\cite{done-third-yangdongling-12-ok} & QGA & \XSolidBrush & \XSolidBrush & Computer simulation & 0  \\ \hline
    \centering\cite{done-third-gaopeng-13-ok}  & QRL & \XSolidBrush & \XSolidBrush & Computer simulation & 0  \\ \hline
    \centering\cite{done-zuoqixue-14-ok} & QDRL & \XSolidBrush & \XSolidBrush & Computer simulation & 0  \\ \hline
    \centering\cite{done-wucheng-17-ok}  &\makecell{QNNs,\\ RL-QNN} & \Checkmark & \Checkmark & IBM Qiskit &\makecell{$2(N_{\text{layer} } - 1)N_{\text{neuron}} + 1$, \\ $3 + 3(N_{\text{layer} } - 1)N_{\text{neuron}} + 1$}  \\ \hline
    \centering\cite{done-second-gaopeng-10-ok} & QGA & \XSolidBrush & \XSolidBrush & Computer simulation & 0  \\ \hline
    \centering\cite{done-wucheng-20-OK} & QA & \Checkmark & \Checkmark & D-Wave & $ N_{\text{user}} N_{\text{IRS}} N_{\text{time slots}}$  \\ \hline
    \centering\cite{done-second-shike-16-ok} & QA & unknown & unknown & unknown & unknown  \\ \hline
    \centering\cite{done-wucheng-9-ok} & QA & \Checkmark & \Checkmark & D-Wave & unknown  \\ \hline
    \end{tabular}
\end{table*}

\vspace{5pt}
\subsubsection{Applications of Conventional QC for Channel Assignment}\

\textbf{QC in a single-cell system}:

The application of QAOA is discussed to solve the problem of maximum likelihood detection of binary symbols transmitted over MIMO channels \cite{done-second-chengyu-15-half-ok}. The maximum likelihood detection problem is NP-hard and difficult to solve by classical computers. The study utilizes QAOA to encode maximum likelihood detection into a level-p QAOA circuit, which is solved by classical optimizers. The work provides an overview of QAOA and their applications in solving various problems.

Device to device (D2D) communications with $N$ transmitter-receiver pairs in a single-cell system are discussed in \cite{done-third-zhongrunhu-13-ok}. The authors highlight the challenge of mutual interference between D2D and nearby cellular communication and the need to reduce the interference through resource allocation including sub-channel assignment and power allocation. They formulate the problem as a capacitated max k-cut and applied quantum annealing (QA). Specifically, sub-channels are assigned to disjoint transmitter groupings, and the issue is formulated as a QUBO problem. The algorithm is implemented in a quantum computer. Though the paper considers a simple scenario of sub-channel allocation, it provides guidelines to apply optimization techniques of QC to solve similar problems.
Applying another QC algorithm QGA for cognitive radio spectrum allocation, illustrated in Fig. \ref{figure-cell-communications}, in a single-cell network is discussed in \cite{done-third-yangdongling-12-ok}. Specifically, the work proposes a mapping process between the channel assignment matrix and the chromosome of the algorithms to decrease the search space.  Thought it highlights the importance of dynamic spectrum access in order to improve spectrum utilization in the validation results, there is no consideration in the problem formulation and analysis.

The preceding works explore QC algorithms for channel assignment in local wireless networks. Their application in mobile cellular networks is also conducted in \cite{done-second-shike-17-ok}. It presents a case study on the root sequence index  assignment problem in LTE/NR physical random access channel configuration. They formulate the problem as a QUBO problem and solved it using a cloud-based QC platform. The results show that QA solver can successfully assign conflict-free root sequence index, but comparison with classical heuristics reveals that some classic algorithms are more effective in terms of solution quality and computation time. The reason lies in the semi-quantum algorithm and the type of quantum computer, which provides reference for the correct use of QC techniques. 

\textbf{QC in relay networks}:

The problem of channel assignment in relay networks is typically formulated as conflict graphs that are resolved by QA in \cite{done-chengyu-13-OK, done-wucheng-16-ok}.
A real-world application of QA, specifically the scheduling of air-links for maximum throughput while avoiding interference near network nodes is studied in \cite{done-chengyu-13-OK}. It introduces a novel gap expansion technique that benefits QA more than simulated annealing, leading to improved speedup and network queue occupancy. The proposed QA is implemented in quantum computers. However, it is assumed that overall network information is given in the scheduling algorithm. \cite{done-wucheng-16-ok} explored network performance in the same scenario with \cite{done-chengyu-13-OK}. Weighted maximum independent set is utilized to model the K-hop interference. It makes a comparison between simulated annealing and QA in solving the scheduling problem. The results show that the QA solution outperforms simulated annealing when using the gap expansion technique.

\vspace{5pt}
\subsubsection{Applications of AI-driven QC for Channel Assignment}\

\textbf{QC in a single-cell system}:

A framework of QNNs for resource allocation \cite{done-wucheng-17-ok} is designed in wireless communications, such as 5G/6G, to reduce time complexity while maintaining performance, and a reinforcement-learning-inspired QNN is proposed to further improve performance. Specifically, they present the quantum circuit design of the QNN to ensure practical implementation in NISQ computers. The proposed QNN architecture comprises three key components: an encoding stage, a multi-layer network, and a decoding stage. The encoding phase transforms the original input and weight values into quantum superposition states. The main operation of the QNN resembles that of a fully connected classical neural network, where the CCX(·) and CZ(·) gates are utilized as activation functions and inter-neuron connections, respectively. The decoding stage is responsible for extracting real-valued outputs from the QNN. Since the loss function is computed using real numbers, ratio decoding is applied to map qubit states to real-number outputs. The QNN is applied to group users in NOMA as a case study. The performance on channel assignment outperforms other solutions, but at the expense of a large number of qubits. 

RL based QC algorithms, quantum multi-agent AC networks, are performed in \cite{done-zuoqixue-14-ok} to construct a robust mobile access system with multiple UAVs. The proposed QDRL framework consists of a centralized quantum critic and multiple quantum actor networks. The $m$-th quantum actor selects the action with the highest probability among the available actions, based on its state and observation information. Meanwhile, the centralized critic evaluates the current state using a state-value function. The actor and critic networks are updated using the temporal difference actor-critic model, guided by the Bellman optimality equations.
Due to the unique QC properties, such as qubit rotation via unitary operations, multi-qubit entanglement, and complex conjugates in the Q and V functions, the traditional policy gradient derivative can not be computed directly. This is because the quantum state remains unknown prior to measurement. To address this challenge, the parameter-shift rule is applied to the Bellman optimality equations. This approach enables the calculation of gradients by linking classical derivatives with QC, thereby integrating classical RL with QC. The algorithm utilizes MARL to enable UAVs to learn collectively and optimize their actions within a shared environment. The study addresses scalability challenges by implementing a quantum centralized critic. One of its advantages lies in the design of a realistic environment with noise injection, the evaluation of quantum multiple agent reinforcement learning (QMARL) and noise reflection on training policies.

One of 6G scenarios, Metaverse, is discussed in \cite{done-third-gaopeng-13-ok}. The problem lies in insufficient sampling in model training for connected and autonomous vehicles. 
To efficiently achieve Metaverse, channel resources among road-side units (RSUs) and vehicles are significant for network performance, in which eNodeBs allocate channel resources to RSUs and RSUs allocate them to vehicles. It  formulates the channel resource assignment as a discrete Markov decision process (MDP) and adopts QRL to obtain the optimal policy. However, the study assumes that vehicle information is given without considering their dynamic, such as moving out communication ranges of RSUs. 

\vspace{5pt}
\subsubsection{Applications of QC Algorithms in IRS}
\begin{figure*}[ht] 
\centering
\subfloat[One BS and one IRS\label{a}]{
\includegraphics[width = 0.22\textwidth]{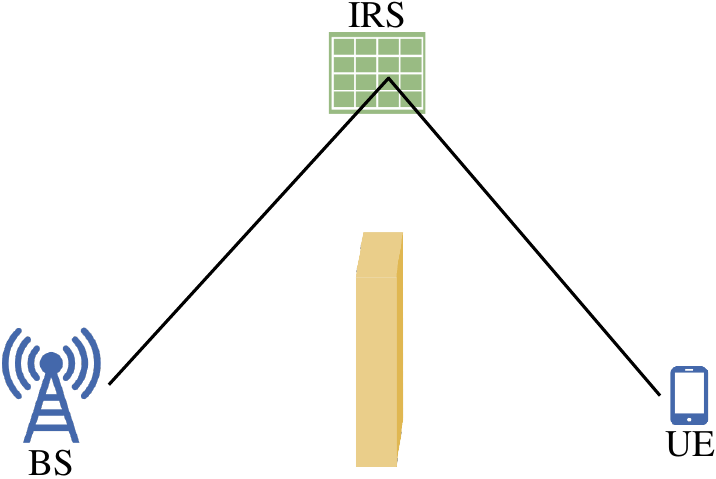}}
\hfill
\subfloat[One BS and multiple IRSs\label{b}]{
\includegraphics[width = 0.22\textwidth]{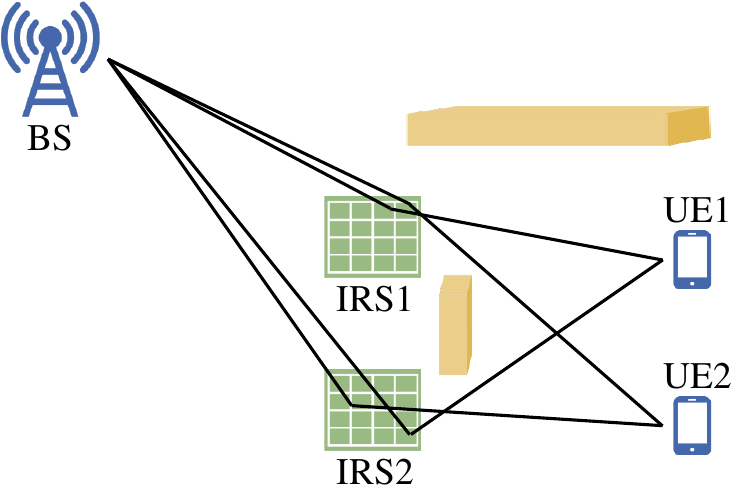}}
\hfill
\subfloat[Multiple BSs and multiple IRSs\label{c}]{
\includegraphics[width = 0.22\textwidth]{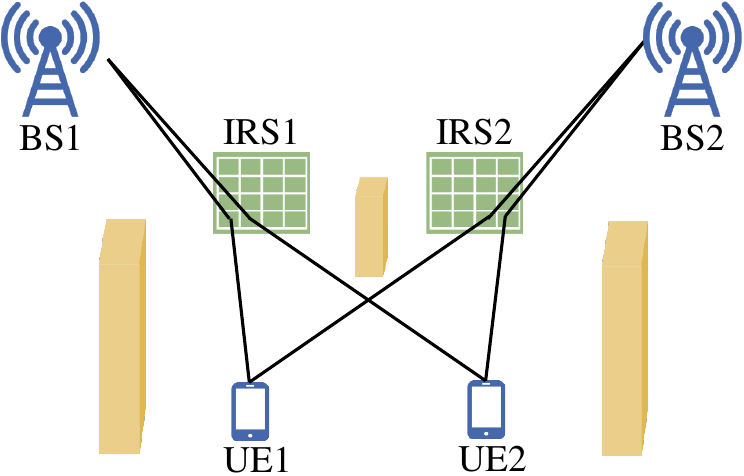}} 
\hfill
\subfloat[Multiple BSs and one IRS\label{d}]{
\includegraphics[width = 0.22\textwidth]{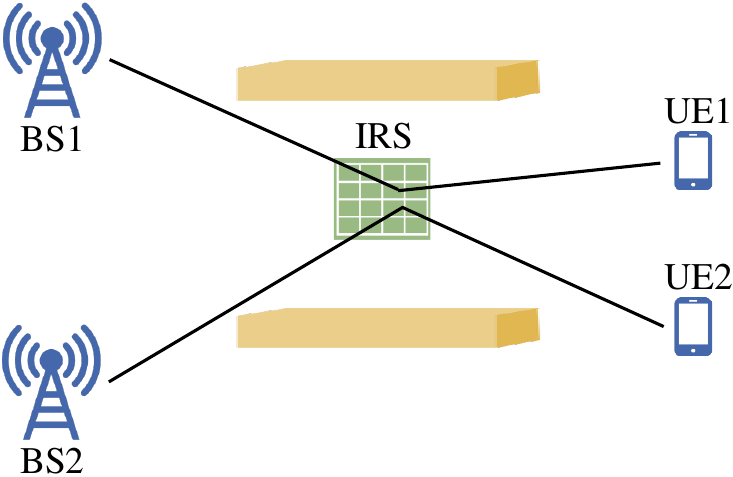}} 
\caption{Different IRS systems.}\label{IRS}
\end{figure*}

Intelligent reflective surfaces (IRSs) \cite{10443962} represent a revolutionary technology in 6G communications that leverages passive elements to manipulate electromagnetic waves. These surfaces are composed of programmable materials capable of adjusting the amplitude, phase, and polarization of incident electromagnetic signals as illustrated in Fig. \ref{a}. By intelligently controlling these properties, IRSs can effectively shape and redirect wireless signals to enhance communication performance.  IRS systems fundamentally enhance wireless communication by reflecting and manipulating signals, which makes them totally different in technique from channel or spectrum assignment. This independent treatment allows us to highlight its unique role in optimizing communication, which is distinct from traditional techniques in spectrum assignment.

\cite{done-wucheng-20-OK} proposes an approach to controlling the propagation path from the transmitter to the receiver by appropriately according the reflection phase of IRS elements in a communication cell consisting of one BS and multiple users under assistance of multiple IRSs as illustrated in Fig. \ref{b}. The core problem is to schedule the assignment of IRSs to a given users for a unit time. The authors first formulate the scheduling problem as a QUBO problem and then develop an algorithm based on QA to solve the issue. The solution is validated in quantum computers with the platform D-Wave.
There are other works \cite{done-second-shike-16-ok, done-second-gaopeng-10-ok} for the same scenario as \cite{done-wucheng-20-OK}. Taking \cite{done-second-shike-16-ok} as an example, it compares its tensor contraction and regression with PSO and factorization machine based QA. However, there is no explanation for its advantage over others.

\cite{done-wucheng-20-OK} and \cite{done-second-shike-16-ok} consider the scheduling problem with only one BS in a communication cell. To improve system scalability, \cite{done-wucheng-9-ok} tackles sub-channel assignment with multiple BSs as illustrated in Fig. \ref{c}. The network structure is different from the scenario in which there is only one BS, when multiple BSs are introduced into the network. In this work, IRSs and users are coordinated by a BS and a server (such as a cellular station), but are not necessarily in the same coverage cell of the BS. When there are multiple BSs in different cells under assistance of multiple IRSs, the channel scheduling problem is formulated as a combinatorial optimization, which is also in the area of QUBO. And a QA based solution is proposed to solve the problem. However, it is a centralized scheduling method for multiple IRSs. In addition, there is no detailed discussion on QA implementation though the simulations validated its effectiveness. 
  
\cite{done-wucheng-20-OK} and \cite{done-second-shike-16-ok} consider  network performance in a communication cell with multiple users, an IRS and one BS. The network is easily extended to a communication cell with multiple users, multiple IRSs, and multiple BSs in different cells. To reduce the network overhead such as hardware costs, it obtains the similar network performance with \cite{done-wucheng-9-ok} by sharing an ISR of switching phase shift as illustrated in Fig. \ref{d}, which is studied in \cite{done-zhongrunhu-10-ok-fine}. Specifically, when a particular phase shift is set at a given time, the corresponding time resources and BSs are assigned to UEs existing in the reflection direction of the radio waves transmitted from each BS, which enables efficient communication. In the case of multiple operators, each operator performs such radio resource-allocation control in parallel, enabling communications with the IRS without coordination required between the operators. The process is formulated as QUBO and solved by classical QC algorithm. However, it can not achieve the optimal performance since multiple operators work independently. And thus, it is still viewed as a centralized approach. 
Other works related to IRS scheduling for channel assignment includes \cite{done-second-gaopeng-10-ok} that considers both power allocation and communication security. Its scenario is similar with \cite{done-wucheng-20-OK, done-second-shike-16-ok} in a communication cell. However, the solution for channel assignment of scheduling phase shift is independent with other optimization variables. A quantum-inspired bald eagle search algorithm is developed to solve the problem. 

In summary, QC-enabled algorithms offer a promising solution to enhance performance in channel assignment, as the overview of these studies presented in Table \ref{QC-Network-Optimization} and QC details in Table \ref{Channel Assignment Table2}.

\subsection{User Association}\label{user-association}
\subsubsection{Overview}
User association in wireless networks refers to the process of determining which access point, BS, or sensor sink that a user device connects to for communication. This decision is crucial for optimizing network performance, ensuring efficient resource utilization and service quality \cite{10416899}.
The primary goal of user association is to provide seamless connectivity while maximizing network capacity and throughput. Various factors influence user association decision, including signal strength, available bandwidth, network load, user mobility, and QoS requirements.
Note that network dynamics, particularly, sink (such as UAVs) mobility, are one of key features in user association, which is explored in numerous studies \cite{zhao2019deep}, such as in mobile sensor networks and UAV networks. Solutions encounter computational constraints like with a substantial number of devices and dynamic environment. The challenge can be addressed through the utilization of QC techniques.
Thus, in this section, we primarily categorize our discussion according to two key criteria: QC algorithm category and mobility of BSs or sinks.

\vspace{5pt}
\subsubsection{Applications of Conventional QC for User Association} \ 
\textbf{Without consideration of mobility:}

\begin{table*}[!t]
\centering
\caption{Summary of QC Works on User Association in Wireless Networks}
\label{user-asso-table1}
\scalebox{0.9}{
\begin{tabular}{|>{\centering\arraybackslash}m{1.5cm}|>{\centering\arraybackslash}m{1.5cm}|>{\centering\arraybackslash}m{1cm}|>{\centering\arraybackslash}m{2.5cm}|>{\centering\arraybackslash}m{4cm}|>{\centering\arraybackslash}m{2cm}|>{\centering\arraybackslash}m{2cm}|>{\centering\arraybackslash}m{2cm}|}
\hline
\textbf{QC category} & \textbf{Network dynamic} & \textbf{Work} & \textbf{Network details} & \textbf{Highlights} & \textbf{Optimization objectives} & \textbf{QC alg.} & \textbf{Control manner} \\ \hline
\multirow{13}{*}{\centering \makecell{\\ \\ \\ \\ \\ \\ \\ \\ \\ \\ \\ \\ \\ \\ \\ \\ \\ \\ \\ \\ \\ \\ \\ \\ \\ \\ \\ \\ \\ Conventional \\QC}} & \multirow{10}{*}{\centering \makecell{\\ \\ \\ \\ \\ \\ \\ \\ \\ \\ \\ \\ \\ \\ \\ \\ \\ \\Without \\mobility}} & \cite{done-third-zhongrunhu-17-ok} & Multiple sensors communicate with a BS under assistance of cluster head & Propose a quantum-inspired evolutionary algorithm to select clusters from multiple sensors & Maximize network lifespan & Quantum-inspired evolutionary algorithm & Centralized \\ \cline{3-8} 
 &  & \cite{done-second-gaopeng-3-ok} & Multiple sensors communicate with a BS under assistance of cluster head & Propose a quantum-inspired gravitational search algorithm to select clusters from multiple sensors & Maximize network lifespan & Quantum-inspired gravitational search algorithm & Centralized \\ \cline{3-8} 
 &  & \cite{done-wucheng-15-ok} & Multiple sensors communicate with a BS under assistance of cluster head & Propose a quantum-inspired gravitational search algorithm to assign operational states for multiple sensors & Maximize network lifespan & Quantum-inspired gravitational search algorithm & Centralized \\ \cline{3-8} 
 &  & \cite{done-gaopeng-11-ok-fine} & Multiple sensors communicate with a BS under assistance of cluster head & Propose a quantum elite gray wolf optimization to assign operational states for multiple sensors & Maximize network multiple QoS & Quantum elite gray wolf optimization & Centralized \\ \cline{3-8} 
 &  & \cite{done-second-gaopeng-15-ok} & Multiple sensors communicate with a BS under assistance of cluster head & Propose a Quantum Clone Whale Optimization Algorithm to assign clusters from multiple sensors & Maximize network multiple QoS & Quantum clone whale optimization algorithm & Centralized \\ \cline{3-8} 
 &  & \cite{done-forth-shike-3-ok} & Multiple sensors communicate with a BS under assistance of cluster head & Propose a Quantum Artificial Bee Colony to assign clusters from multiple sensors & Maximize network lifespan & Quantum artificial bee colony & Centralized \\ \cline{3-8} 
 &  & \cite{done-fourth-chengyu-17-ok} & Multiple sensors communicate with a BS under assistance of cluster head & Propose a QPSO to assign clusters from multiple sensors & Maximize network lifespan & QPSO & Centralized \\ \cline{3-8} 
 &  & \cite{done-third-shike-11-ok} & Multiple sensors communicate with a BS under assistance of cluster head & Propose a QGA to assign clusters from multiple sensors under assistance of cluster head & Maximize network lifespan & QGA & Centralized \\ \cline{3-8} 
 &  & \cite{done-chengyu-4-ok} & Multiple sensors communicate with a BS under assistance of cluster head & Propose a QPSO to assign clusters from multiple sensors & Maximize network lifespan & QPSO & Centralized \\ \cline{3-8} 
 &  & \cite{done-second-chengyu-19-ok} & Online social networks & Predict online social connectivity to recommend user connection& Predict link connectivity & Quantum ant colony optimization & Centralized \\ \cline{2-8} 
 & \multirow{3}{*}{\centering \makecell{\\ \\ \\ \\ \\ \\ \\ \\ Mobility}} & \cite{done-chengyu-14-ok} & Wireless sensor networks with a mobile sink & Analyze path planning of a mobile sink and transform the problem to a travelling salesman problem & Find the shortest path of a mobile sink & QA & Centralized \\ \cline{3-8} 
 &  & \cite{done-second-wucheng-8-ok} & Multiple-UAV networks & Organize multiple-UAV into clusters while considering three metrics namely average distance, distance to UAVs and the number of UAV neighbors & Optimize the fitness function with average distance, distance to UAVs and the number of UAV neighbors & QGA & Centralized \\ \cline{3-8} 
 &  & \cite{done-zhongrunhu-19-ok-fine} & Multiple-UAV networks to provide computation resources dynamically to ground users & Propose a distributed leader-follower coalition formation algorithm in a multiple-UAV networks to accomplish service provision dynamically in an unknown environment & Multiple objectives of minimizing coalition cost, maximizing reliability and reputation & QGA & Distributed \\ \hline
\multirow{2}{*}{\centering \makecell{\\ \\ \\ \\ AI-driven QC}} & \multirow{2}{*}{\centering \makecell{\\ \\ \\ \\ Mobility}} & \cite{done-zuoqixue-15-ok} & Multiple UAVs communicate with multiple BSs, and there is only communication between a UAV and a BS & Propose a distributed leader-follower coalition formation algorithm in a multiple-UAV networks to accomplish service provision dynamically in an unknown environment & Minimize the movement penalty and connection outage duration & QDRL & Distributed \\ \cline{3-8} 
 &  & \cite{done-second-zuoqixue-6-ok} & Multiple ground users communicate with a UAV & Optimize UAV trajectory planning, NOMA user grouping, and power allocation, solved by proposed QDRL that can tackle issues of a large and continuous action and state spaces & Maximize communications throughput per unit of energy consumed by the UAV & QDRL & Centralized \\ \hline
\end{tabular}
}
\end{table*}

The study examines basic wireless sensor network \cite{done-third-zhongrunhu-17-ok}, consisting of sensors and a BS. It introduces a clustering method based on quantum evolution algorithm (QEA) to extend the network lifespan. However, it does not account for network dynamics, such as traffic patterns and device mobility.
In \cite{done-second-gaopeng-3-ok}, a comparable scenario to  \cite{done-third-zhongrunhu-17-ok} is explored, with the introduction of the quantum-inspired gravitational search to enhance energy efficiency and prolong network lifespan. The study considers communication constraints and application-specific criteria, encompassing parameters such as mean relative deviation, spatial density error, and sensor-out-of-range error. Nonetheless, the paper lacks an in-depth analysis of the optimality of the proposed solution.
Meanwhile, \cite{done-wucheng-15-ok} presents an alternative quantum-inspired approach to the issue addressed in \cite{done-third-zhongrunhu-17-ok} and \cite{done-second-gaopeng-3-ok}. Named the quantum-inspired gravitational search algorithm, this solution aims to determine the operational states of sensors while preserving network connectivity. It introduces a weighted fitness function, integrating connectivity, energy, and application-specific parameters. However, accurately determining the appropriate weights for these factors poses a challenge.

Several additional quantum-inspired approaches have been proposed in the literature, including works such as \cite{done-gaopeng-11-ok-fine, done-second-gaopeng-15-ok, done-forth-shike-3-ok, done-fourth-chengyu-17-ok, done-third-shike-11-ok, done-chengyu-4-ok}. In the proposed quantum-inspired elite gray wolf approach \cite{done-gaopeng-11-ok-fine}, each individual in the population is represented by a set of quantum probability amplitudes corresponding to the states of all qubits. The initialization of the quantum probability amplitude matrix is performed by a logistic chaotic map, and the matrix is subsequently updated using quantum rotation gates based on the individual fitness value at each iteration.
To enhance the search for optimal clustering, the method introduces an elite pool that stores the best-performing individuals and their corresponding quantum probability amplitude matrices from previous iterations. During the $g$-th iteration, top individuals are selected from the elite pool based on their fitness and are inherited to guide the population evolution in the subsequent $(g+1)$-th iteration. Additionally, a quantum NOT gate is applied to prevent the algorithm from converging prematurely to local optima.

These contributions introduce techniques such as quantum elite gray wolf optimization, quantum clone whale optimization, quantum artificial bee colony, QPSO, and QGA, all aimed at addressing the same problem outlined in \cite{done-third-zhongrunhu-17-ok}.
Additionally, other studies explore link connectivity within QC algorithms. For example, \cite{done-second-chengyu-19-ok} investigates online social networks like facebook, weibo, and twitter. The goal is to predict link connectivity, facilitating recommendations for registered users to connect with others.

\textbf{With consideration of mobility:}

Subsequent studies consider more complex scenarios involving UAV networks and wireless sensor networks featuring mobile sinks.
For example, \cite{done-chengyu-14-ok} focuses on path planning of a mobile sink to enhance energy efficiency. In this scenario, the mobile sink starts its path from a designated location, traverses through the communication range of each sensor along the predetermined path, conducts a round of data collection, and returns to its initial starting point. They employ mathematical transformations to align the problem characteristics with those of a traveling salesman (TSP) problem. An optimal TSP path is obtained by using quantum tunneling effect and quantum circuit to achieve parallelism.

Regarding UAV networks, \cite{done-second-wucheng-8-ok} considers a scenario with multiple UAVs but does not address data relay among UAVs. The research designs an objective function incorporating parameters like average distance, distance to neighboring UAVs, and the number of UAV neighbors to select cluster heads. While the proposal accounts for UAV mobility, it lacks mobility analysis within the model. Additionally, the paper explores UAV clustering application in image classification, but the model lacks a direct correlation between UAV clustering and image classification.
In contrast, \cite{done-zhongrunhu-19-ok-fine} tackles the challenge of UAV coalition formation for providing computation resources in a large-scale heterogeneous UAV network. The study operates in an unknown and dynamic environment, where UAVs dynamically acquire environmental knowledge. Leader UAVs assess follower UAV' contributions in resource availability and share this information during coalition formation. QGA guides the inclusion of idle UAVs into coalitions, considering UAV mobility characterized by randomness and lacking correlation with ground points of interest (PoI). However, the study lacks an analysis of PoI impact on system performance, differing significantly from edge computing research, particularly in task offloading investigations \cite{zhao2023drl}.

\vspace{5pt}
\subsubsection{Applications of AI-driven QC for User Association}\

Several other studies \cite{done-zuoqixue-15-ok,done-second-zuoqixue-6-ok} integrate AI techniques with QC algorithms within their solutions while considering network dynamics. 
For example, \cite{done-zuoqixue-15-ok} discusses the path planning for cellular-connected UAVs to minimize the weighted sum of time cost and expected outage duration. Traditional offline optimization techniques face challenges due to practical considerations like local building distribution and directional antenna radiation patterns. To overcome this, the authors propose a QDRL solution that maps the problem into MDP, enabling the UAV to find the optimal flying direction within each time slot. 
Unlike approaches focusing on designing multi-layer quantum networks or updating network parameters, this study proposes a quantum-inspired representation of experience prioritization to improve data efficiency. In practical applications, certain buffer experiences are sampled for training with undesirably high frequency, leading to over-fitting. Additionally, the limited buffer size exacerbates this issue, resulting in biased and unfair sampling performance.
To address these challenges, experience prioritization is represented in terms of qubits and determined by linear coefficient quantum ground states. This quantum representation creates a connection between quantum eigenstates and the selection or rejection of specific experiences. By leveraging quantum amplitude amplification, the method enables the manipulation of quantum collapse to better manage experience sampling and prevent over-fitting.
However, the analysis does not establish direct correlations among multiple UAVs, simplifying the model into individual path planning for each UAV.

Due to robust capacity of QDRL for addressing challenges including those characterized by a large state and action spaces, it can also be effectively applied in the optimization of multiple objectives. \cite{done-second-zuoqixue-6-ok} addresses multiple interconnected challenges, encompassing UAV trajectory planning, NOMA user grouping, and power allocation. Within this framework, users within a group maintain communication with the UAV within their communication ranges along the UAV trajectory. The decision regarding energy consumption by the UAV directly influences the objective. Given the extensive and continuous nature of both action and state spaces, the paper introduces a QDRL approach to effectively tackling the complexity of the problem.

In conclusion, QC algorithms present a compelling approach for user association in wireless networks. Table \ref{user-asso-table1} offers a consolidated overview of these studies. However, there are two open issues in these QC algorithms addressing user association. First, these works typically relied on computer simulations for QC algorithm implementation. Second, few studies account for the impact of mobility on performance, even in UAV networks.

\section{QC in Networking} \label{Traffic Routing}

\subsection{Overview}

\begin{figure}
    \centering
    \includegraphics[width=0.9\linewidth]{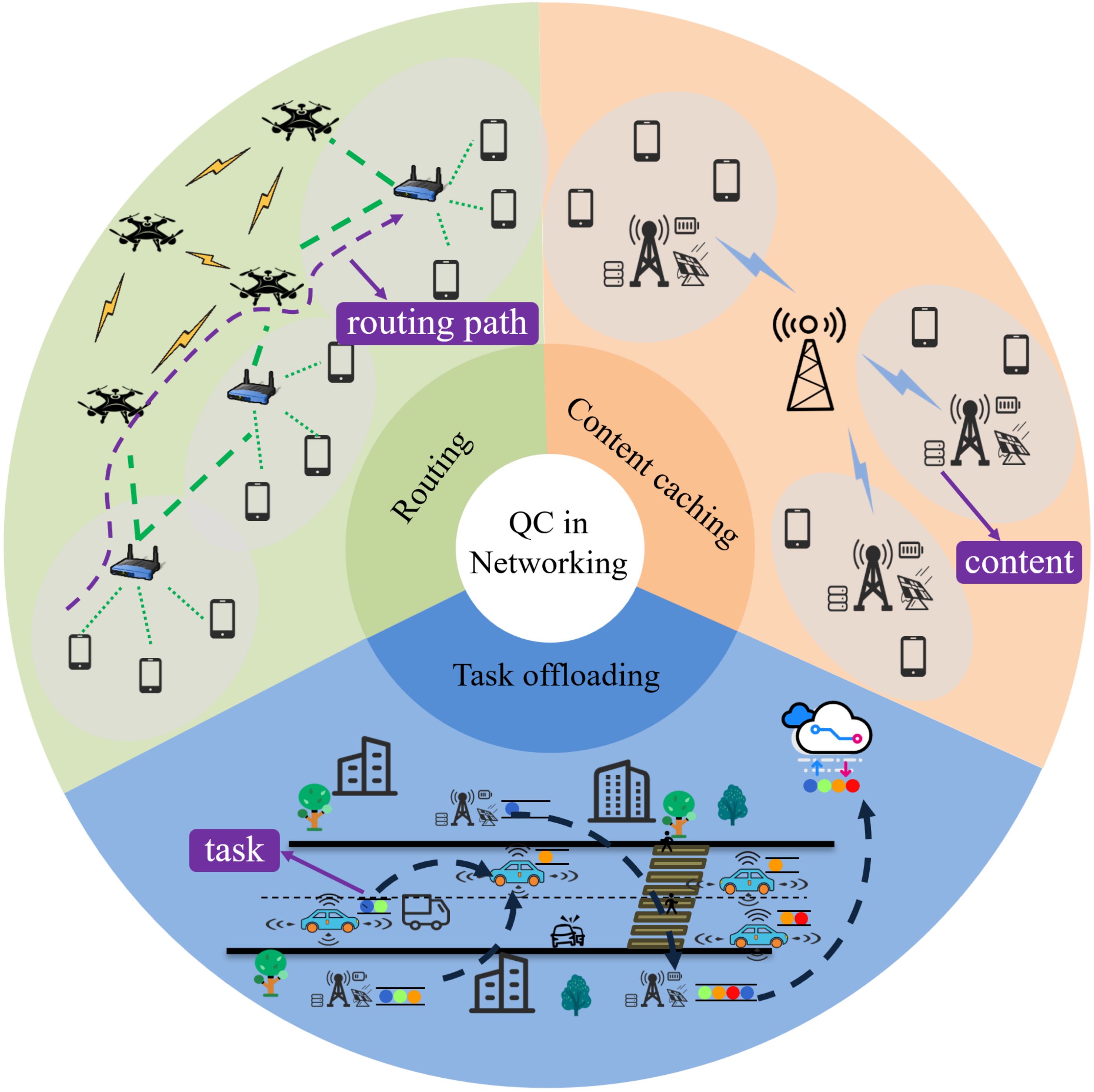}
    \caption{Different network scenarios in wireless networking.}
    \label{networking}
\end{figure}

This section explores how QC principles can revolutionize various aspects of networking, including routing, task offloading, and content caching as illustrated in Fig. \ref{networking}. 
These topics are crucial for optimizing network performance in scenarios with high mobility, dynamic environments, and complex resource management that are becoming increasingly important in the 6G era and beyond.
Routing is a fundamental issue in wireless networking, as it forms the backbone for ensuring efficient data transmission. Task offloading and content caching, in turn, are often built upon effective routing strategies.
By focusing on these areas, the paper addresses key research challenges that are vital for improving network efficiency, scalability, and reliability. Additionally, this focus emphasizes the potential of QC to tackle networking issues that are difficult to address with classical methods.
\subsection{Routing}

\subsubsection{Overview}\

Routing presents fundamental challenges in network systems, encompassing the task of determining optimal paths for data packets to travel from a source to a destination \cite{10542397}. In wireless networks, where devices may be mobile and network conditions dynamically change, routing becomes even more complex. The primary goal of routing is to ensure efficient and reliable data transmission while minimizing latency, packet loss, and resource consumption.
Existing approaches to routing encompass a variety of algorithms and protocols designed to address different aspects of routing optimization. Traditional methods like optimized link state routing protocol (OLSR) and ad hoc on-demand distance vector (AODV) attempt to dynamically establish and update routes based on network topology changes. However, they suffer from overhead due to frequent updates and incur latency in route discovery. Additionally, traditional routing algorithms may not adequately address challenges such as mobility, variable channel conditions, and energy constraints in wireless networks \cite{mao2024intelligent}.

QC-driven approaches offer promising solutions to overcome the limitations of traditional routing methods. By leveraging the principles of quantum mechanics, they can explore vast solution spaces more efficiently and potentially find optimal or near-optimal routes in significantly less time.
In addition, QC enables the development of AI-driven routing algorithms that can adapt and learn from network dynamics, optimizing routing decisions in real-time. These quantum-enhanced approaches have the potential to improve routing efficiency, minimize latency, and enhance network reliability.

\begin{table*}[!t]
\centering
\caption{Summary of QC Works on Routing in Wireless Networks}
\label{routing1}
\scalebox{0.7}{
\begin{tabular}{|>{\centering\arraybackslash}m{2cm}|>{\centering\arraybackslash}m{2cm}|>{\centering\arraybackslash}m{2cm}|>{\centering\arraybackslash}m{4cm}|>{\centering\arraybackslash}m{4cm}|>{\centering\arraybackslash}m{4cm}|>{\centering\arraybackslash}m{2cm}|>{\centering\arraybackslash}m{2cm}|}
\hline
\textbf{QC category} & \textbf{Network feature} & \textbf{Work} & \textbf{Network structure} & \textbf{Highlights} & \textbf{Optimization objectives} & \textbf{QC alg.} & \textbf{Control manner} \\
\hline

\multirow{15}{*}{\centering \makecell{\\ \\ \\ \\ \\ \\ \\ \\ \\ \\ \\ \\ \\ \\ \\ \\ \\ \\ \\ \\ \\ \\ \\Conventional \\QC}} & \multirow{11}{*}{\centering \makecell{\\ \\ \\ \\ \\ \\ \\ \\ \\ \\ \\ \\ \\ \\ \\Without \\consideration \\of \\ mobility}} & \cite{done-chengyu-11-OK} & Wireless mesh networks with multi-hop relays in a 6G scenario & The paper gives a detailed example of applying QAOA in solving a single-objective formulation. & Maximize the network utility including battery cost, sum throughput for all nodes, and the number of hops & QAOA & Centralized \\ \cline{3-8} 
 &  & \cite{10325636} & A fleet of vehicles with a central depot is routed to visit a set of customers &Propose a multi-phase method using Ising machines to minimize travel distance while balancing depot load and vehicle load. & Find the shortest travel route while balancing loads. & QA & Centralized \\ \cline{3-8} 
 &  & \cite{done-zhongrunhu-4-ok-fine}& A fully connected multi-hop network with a single source node, a single destination node and a cloud of relay nodes & Develop a multi-objective dynamic programming framework for generating Pareto-optimal routes relying on the correlations of the specific links & Identify the entire set of weakly Pareto-optimal routes with multiple objectives & Boyer-Brassard-Hoyer-Tapp Quantum Search Algorithm & Centralized \\ \cline{3-8} 
 &  & \cite{done-shike-8-ok} & Self-organizing network in which data are relayed to destinations by intermediate nodes & Propose an optimal multi-objective quantum algorithm to evaluate paths using Pareto optimality & Maximize network utility: overall bit error rates, linear-domain path losses and delays & Boyer-Brassard-Hoyer-Tapp Quantum Search Algorithm & Centralized \\ \cline{3-8} 
 &  & \cite{done-fourth-wucheng-2-ok} & Wireless sensor networks with multi-hop relays & Propose a QC solution focusing on quantum representation, measurement, and rotation angle & Minimize the total energy cost over all relaying paths  & Quantum ant colony optimization & Centralized \\ \cline{3-8} 
 &  & \cite{done-third-gaopeng-20-ok} & Wireless mesh networks & Propose a quantum ant colony algorithm to predict pheromone values to measure paths & Predict pheromone values & Quantum ant colony optimization & Centralized \\ \cline{3-8} 
 &  & \cite{done-yangdongling-3-ok} & Mesh networks in IoT & Improve the robustness of network connectivity for IoT topology & Sum of the number of nodes in maximum connected sub-graph & QGA & Centralized \\ \cline{3-8} 
 &  & \cite{done-chengyu-7-ok} & Wireless sensor networks relaying data by multiple hops to sinks & Propose a QGA for multi-sink deployment while balancing energy consumption among sensors& Maximize network utility including the number of hops, energy consumption, and sink dispersion & QGA & Centralized \\ \cline{3-8} 
 &  & \cite{done-second-gaopeng-16-ok} & A weighted digraph wireless mesh network:  source and destination nodes as multicast trees & QoS multicast routing is converted into ILP with QoS constraints, solved by QPSO combined with loop deletion operation & Minimize  network delay & QPSO & Centralized \\ \cline{3-8} 
 &  & \cite{done-zuoqixue-10-ok} & A directed network graph in industrial IoT & Study the impact of p-degrees of network connectivity on network performance & Minimize energy consumption and network delay & QPSO & Centralized \\ \cline{3-8} 
 &  & \cite{done-third-yangdongling-7-ok} & Cooperative data transmission from a cluster head to BS & Propose a QPSO algorithm to assign relay-cluster heads & Maximize network lifespan & QPSO & Centralized \\ \cline{2-8} 
 & \multirow{4}{*}{\centering \makecell{\\ \\ \\ \\Consideration \\of \\mobility}} & \cite{done-second-gaopeng-6-ok} & Mobile ad-hoc networks & Embed Q-learning of RL in quantum genetic strategy & Minimize local energy consumption & QGA & Centralized \\ \cline{3-8} 
 &  & \cite{done-second-gaopeng-8-ok} & Mobile ad-hoc networks & Embed augmented Q-Learning algorithm and combine the OLSR to optimize the selection of multi-point delay sets & Maximize the network utility:  overhead of topology control, delivery packet rates and time delay of packet transmission & QGA & Centralized \\ \cline{3-8} 
 &  & \cite{done-fourth-wucheng-1-ok} & UAV path planning & Combine the differential evolution algorithm with QPSO & Sum of the required time to finish the tasks on each target & QPSO & Centralized \\ \cline{3-8} 
 &  & \cite{done-second-wucheng-11-ok} & Mobile multi-hop networks with communications from UAVs to BSs under relay assistance & Design a method for neuro-fuzzy cluster based topology construction and find   routes for inter-cluster UAV communication & Minimize delay and power consumption & Quantum ant lion optimization & Centralized \\ \hline
\multirow{3}{*}{\centering \makecell{\\ \\ \\AI-driven QC}} & \multirow{2}{*}{\centering \makecell{\\  Without \\consideration \\of \\mobility}} & \cite{done-zuoqixue-2-ok} & Wireless sensor networks with multi-hop relays from the source nodes and destination nodes & A QRL algorithm is developed based on Grover iteration for efficient decision-making & Maximize the network utility of energy, channel and data buffer & QRL & Centralized \\ \cline{3-8} 
 &  & \cite{done-third-zuoqixue-17-ok} & Multi-robot network: autonomous mobile robots to identify locations and detect surrounding static and dynamic objects & Propose a centralized training and decentralized execution QMARL for multi-robot control and coordination & Maximize network utility of transportation delay, load balancing, and the ratio of positive delivering goods & QMARL & Distributed \\ \cline{2-8} 
 & Consideration of mobility & \cite{done-second-shike-6-ok} & Multiple UAVs as mobile BSs to gather data from other devices & Propose a QMARL for massive agents & Reward maximization & QMARL & Distributed \\ \hline
\end{tabular}
}
\end{table*}

\begin{table*}[!t]
    \centering
    \caption{Details of QC Algorithms in Routing}
    \label{routing2}
    \begin{tabular}{|>{\centering\arraybackslash}m{1cm}|>{\centering\arraybackslash}m{4cm}|>{\centering\arraybackslash}m{1.5cm}|>{\centering\arraybackslash}m{2.5cm}|>{\centering\arraybackslash}m{3cm}|>{\centering\arraybackslash}m{2cm}|>{\centering\arraybackslash}m{1cm}|}
    \hline
    \textbf{Work} & \textbf{Alg.} & \textbf{Orth. QC} & \textbf{Quantum computer} & \textbf{Platform} & \textbf{No. qubits}  & \textbf{Mobility} \\ \hline
    \cite{done-chengyu-11-OK} & QAOA & \Checkmark & \Checkmark & IBM QC & 65  & \XSolidBrush \\ \hline
    \cite{10325636} & QA & \Checkmark & \Checkmark & Fixstars Amplify AE, GUROBI, SA, Christofides, and Feld & Unknown  & \XSolidBrush \\ \hline
    \cite{done-zhongrunhu-4-ok-fine} & Boyer-Brassard-Hoyer-Tapp Quantum Search Algorithm & \Checkmark & \Checkmark & Unknown & Unknown  & \XSolidBrush \\ \hline
    \cite{done-shike-8-ok} & Boyer-Brassard-Hoyer-Tapp Quantum Search Algorithm & \Checkmark & \Checkmark & Unknown & Unknown  & \XSolidBrush \\ \hline
    \cite{done-fourth-wucheng-2-ok} & Quantum ant colony optimization & \XSolidBrush & \XSolidBrush & Computer simulation & 0  & \XSolidBrush \\ \hline
    \cite{done-third-gaopeng-20-ok} & Quantum ant colony optimization & \XSolidBrush & \XSolidBrush & Computer simulation & 0  & \XSolidBrush \\ \hline
    \cite{done-yangdongling-3-ok} & QGA & \Checkmark & \Checkmark & Unknown & Unknown  & \XSolidBrush \\ \hline
    \cite{done-chengyu-7-ok} & QGA & \XSolidBrush & \XSolidBrush & Computer simulation & 0  & \XSolidBrush \\ \hline
    \cite{done-second-gaopeng-16-ok} & QPSO & \XSolidBrush & \XSolidBrush & Computer simulation & 0  & \XSolidBrush \\ \hline
    \cite{done-zuoqixue-10-ok} & QPSO & \XSolidBrush & \XSolidBrush & Computer simulation & 0  & \XSolidBrush \\ \hline
    \cite{done-third-yangdongling-7-ok} & QPSO & \XSolidBrush & \XSolidBrush & Computer simulation & 0  & \XSolidBrush \\ \hline
    \cite{done-second-gaopeng-6-ok} & QGA & \XSolidBrush & \XSolidBrush & Computer simulation & 0  & \Checkmark \\ \hline
    \cite{done-second-gaopeng-8-ok} & QGA & \XSolidBrush & \XSolidBrush & Computer simulation & 0  & \XSolidBrush \\ \hline
    \cite{done-fourth-wucheng-1-ok} & QPSO & \XSolidBrush & \XSolidBrush & Computer simulation & 0  & \Checkmark \\ \hline
    \cite{done-second-wucheng-11-ok} & Evolutionary Quantum Pareto Optimization & \XSolidBrush & \XSolidBrush & Computer simulation & 0  & \Checkmark \\ \hline
    \cite{done-zuoqixue-2-ok} & QRL & \XSolidBrush & \XSolidBrush & Quantum virtual machine & 4  & \XSolidBrush \\ \hline
    \cite{done-third-zuoqixue-17-ok} & QMARL & \XSolidBrush & \XSolidBrush & Computer simulation & 0  & \XSolidBrush \\ \hline
    \cite{done-second-shike-6-ok} & QMARL & \Checkmark & \Checkmark & VQC & Number of UAVs  & \Checkmark \\ \hline
    \end{tabular}
\end{table*}

\vspace{5pt}
\subsubsection{Applications of Conventional QC for Routing} \

\textbf{Without consideration of mobility}

The paper \cite{10325636} introduces a multi-phase method to address vehicle routing with backhauls, mapping key elements of each phase onto QUBO models suitable for Ising machines. The two-phase method comprises a clustering phase followed by a routing phase, while the three-phase method includes a depot clustering phase, a vehicle clustering phase, and a routing phase. The clustering-related phases are expansions of the knapsack problem, while the routing phases are tackled as TSP.  
Its modeling process of QUBO formation can be regarded as a tutorial for applying QC in the field of networking.

The paper \cite{done-zhongrunhu-4-ok-fine} presents a multi-objective dynamic programming framework aimed at generating potentially Pareto-optimal routes by leveraging correlations among specific links, thereby significantly reducing the total number of routes. This framework extends the widely single-objective Bellman-Ford algorithm to handle multiple objectives. Additionally, the paper introduces a quantum-assisted algorithm that capitalizes on both the dynamic programming framework and the synergies between quantum programming and hybrid programming to effectively solve the multi-objective routing problem. Similar works from the authors with different objectives or different QC algorithms are presented in \cite{done-shike-9-ok, done-second-gaopeng-19-ok}. The study \cite{done-shike-8-ok}, similar to \cite{done-zhongrunhu-4-ok-fine}, utilizes the Boyer-Brassard-Hoyer-Tapp quantum search algorithm, but in a different scenario within self-organizing networks. They introduce an optimal multi-objective quantum-assisted algorithm aimed at assessing legitimate routes while incorporating the concept of Pareto optimality to reduce complexity. 
However, the study overlooks the influence of node mobility, a critical characteristic in self-organizing networks like Ad-hoc networks.

The energy optimization model introduced in \cite{done-fourth-wucheng-2-ok} targets sensor-enabled IoT environments. A QC-driven solution is proposed for optimizing the problem, emphasizing quantum representation, measurement, and rotation angle. However, a notable drawback of this approach is that minimizing the total energy cost across all relaying paths does not necessarily equate to prolonging the network lifetime. There are other works \cite{done-third-gaopeng-20-ok} based on quantum ant colony algorithms to solve routing problems. It utilizes the metric, pheromone value inspired by ant colony algorithm, to measure a routing path. Due to  dynamic network changes, it requires a large computation time to track pheromone traits, especially in a large wireless sensor network.  

Additional popular quantum evolution based proposals for routing issues include QGA solutions \cite{done-yangdongling-3-ok, done-chengyu-7-ok} and QPSO solutions \cite{done-second-gaopeng-16-ok, done-zuoqixue-10-ok, done-third-yangdongling-7-ok}. \cite{done-yangdongling-3-ok} improves the network robustness of connectivity for IoT topology duo to node failures caused by malicious attacks. It proposes a QGA algorithm to maximize the cumulative sum of the number of nodes in the maximum connected sub-graph of the network. Additionally, \cite{done-chengyu-7-ok} is designed to be resilient, which adapts to any changes in a planned topology of deployed sinks in anticipation of sensor failures. Also it achieves energy consumption balance among sensors by equalizing the path length and brings down the cost of wireless sensor networks by minimizing the number of sinks. Multicast routing in wireless mesh networks are considered with the objective of minimizing network delay \cite{done-second-gaopeng-16-ok}. The routing problem is converted into an integer programming problem with QoS constraints and is solved by QPSO combined with loop deletion operation. 
Connectivity in routing is explored in \cite{done-zuoqixue-10-ok, done-third-yangdongling-7-ok}. \cite{done-zuoqixue-10-ok} considers heterogeneous scenarios of network topology for the optimal path configuration by exploring and exploiting the hunts. It applies multiple inputs from heterogeneous industrial IoT into QPSO techniques. The differential evolution operator and crossover operations are used for information interchange among the nodes to avoid trapping into local minima. Specifically, it studies the impact of p-degrees of network connectivity on network performance. The same scenario of IoT is also studied in \cite{done-third-yangdongling-7-ok} to optimize selection of cluster heads for multi-hop inter-cluster communication, which is different from user association in a single-cell communication in subsection \ref{user-association}. It proposes a cooperative multiple-input-multiple-output connectivity scheme between sensors and cluster heads to extend network lifetime. QPSO is applied to select the optimum cooperative coalitions of each hop in the routing path.

\textbf{With consideration of mobility}

The preceding studies overlook node mobility, a critical aspect in multi-hop wireless networks where devices and nodes dynamically move in a network environment. Several works have employed QC algorithms to address the challenge. 
For instance, the optimized link state routing protocol, designed for mobile ad-hoc networks, is explored in \cite{done-second-gaopeng-6-ok, done-second-gaopeng-8-ok} from the same researchers, both leveraging QGA approaches. 
The dynamic topology inherent in mobile ad-hoc networks pose significant challenges in finding and sustaining an optimal end-to-end path. However, the heuristic Q-learning strategy in RL enables dynamic adjustment of routing paths through interactions with environments. An enhanced QEA strategy has incorporated a novel Q-learning strategy, optimizing the selection of multi-point relays \cite{done-second-gaopeng-6-ok}. This improvement addresses the limitations of traditional protocols and demonstrates properties of convergence and global optimization. Other network aspects are additionally considered in \cite{done-second-gaopeng-8-ok}, such as the delivery rate of data packets, and overhead of network topology control.

Additional networks, including UAV networks, are examined with a focus on node mobility in \cite{done-fourth-wucheng-1-ok, done-second-wucheng-11-ok}, utilizing QPSO and quantum ant lion algorithm, respectively. Specifically, \cite{done-fourth-wucheng-1-ok} combines the differential evolution algorithm with QPSO to further enhance the performance of both algorithms. The primary objective is to describe a
general route planning strategy. A route planner for UAVs is designed to generate a safe and flyable path in the presence of different threat environments. \cite{done-second-wucheng-11-ok} presents an approach for designing an energy-efficient neuro-fuzzy cluster-based topology construction with a metaheuristic route planning algorithm. The model utilizes three input parameters: residual energy in UAVs, average distance to nearby UAVs, and UAV degree, for selecting cluster heads and constructing the topology. Additionally, quantum ant lion based metaheuristic route planning is employed to select an optimal set of routes for inter-cluster UAV communication. However, the proposal divides the problem into two independent sub-problems, potentially leading to sub-optimal solutions.

\vspace{5pt}
\subsubsection{Applications of AI-driven QC for Routing} \

In recent years, AI techniques, such as RL and NNs, can be integrated with QC to enhance decision-making. AI-driven QC has emerged as a promising paradigm for addressing routing challenges in complex wireless environments.

\textbf{Without consideration of mobility}

A framework \cite{done-zuoqixue-2-ok} is introduced aiming at leveraging QRL to enhance the design of next-generation wireless systems. The framework, termed QRL for relay and transmit power selection, enables sensor nodes to make optimal decisions regarding relay and transmit power selection by leveraging both present and past local state knowledge. Initially, the relay and transmit power selection problem is formulated as MDP. Subsequently, the QRL optimization aspect of the MDP is addressed, focusing on jointly optimizing energy consumption and throughput to maximize network utility. To solve this MDP problem, a QRL algorithm is developed, drawing upon Grover's iteration. The algorithm enables efficient decision-making within the network, facilitating enhanced performance and resource utilization. Especially, the simulation is implemented using quantum virtual machine hosted on Rigetti's forest platform, which provides an additional approach to validating QC algorithms. QMARL is also applied in tackling routing problems. The work \cite{done-third-zuoqixue-17-ok} introduces a quantum-based solution for MARL tailored for autonomous control and coordination of multiple robots within smart factory environments. The delivery problem shares similarities with the routing problem. It presents an innovative QMARL framework, featuring centralized training and decentralized execution. The framework leverages parameter sharing on policy, design of variational quantum circuits, and 2-variables dense encoding.
In particular, the proposed system enables the QDRL server to make distributed and sequential decisions for each autonomous mobile robot. These decisions revolve around determining the optimal allocation of goods, including the number of goods to carry in each autonomous mobile robot and when to request quality control. The objective is to prevent both overflow and underflow scenarios during the delivery of goods by each autonomous mobile robot.

\textbf{With consideration of mobility}

The previous studies focusing on AI-driven QC overlook the consideration of node mobility in wireless networks. This aspect is addressed by a limited number of studies, such as \cite{done-second-shike-6-ok}. 
Multiple mobile UAV platforms act as mobile BSs to gather data from other devices. Note that trajectory planning of multiple UAVs falls within the domain of routing, as data from devices are relayed by multiple UAVs to reach the destination.
The proposed QMARL-based algorithm presents a significant advantage in reducing action dimensionality. By utilizing projective value measures, it efficiently compresses the dimensionality to a logarithmic scale. This dimensionality reduction facilitates improved convergence during training, particularly in large-scale MARL scenarios.

\subsection{Task Offloading} 

\subsubsection{Overview} \

Research in task offloading encompasses a spectrum of crucial topics \cite{li2022resource, 10638123}. Efficient task partitioning and scheduling, ensuring optimal allocation of tasks between edge devices and the cloud, form a foundational aspect. The effective management of limited edge resources from computing, storage, to communication, is a critical consideration \cite{tang2022blockchain}.

\begin{table*}[!t]
\centering
\caption{Summary of QC Works on Task Offloading in Wireless Networks}
\label{task-offloading1}
\scalebox{0.7}{
\begin{tabular}{|>{\centering\arraybackslash}m{2cm}|>{\centering\arraybackslash}m{2cm}|>{\centering\arraybackslash}m{2cm}|>{\centering\arraybackslash}m{4cm}|>{\centering\arraybackslash}m{4cm}|>{\centering\arraybackslash}m{4cm}|>{\centering\arraybackslash}m{1.5cm}|>{\centering\arraybackslash}m{2cm}|}
\hline
\textbf{QC category} & \textbf{Network feature} & \textbf{Work} & \textbf{Network structure} & \textbf{Highlights} & \textbf{Optimization objectives} & \textbf{QC alg.} & \textbf{Control manner} \\
\hline

\multirow{13}{*}{\centering \makecell{\\ \\ \\ \\  \\ \\ \\ \\ \\ \\ \\ \\ \\ \\ \\ \\ \\ \\ \\ \\ \\ \\ Conventional QC}} & \multirow{11}{*}{\centering \makecell{\\ \\ \\ \\ \\ \\ \\ \\ \\ \\ \\ \\ \\ \\ \\ Without \\Consideration \\of Mobility}} & \cite{done-shike-7-ok} & The network offers $K$ types of network function virtualization service. Virtual machines are divided into groups 
 serving network function virtualizations. & Formulate scheduling network function virtualization as ILP, and transform it into QUBO that is solved by QA & Minimize network delay & QA & Centralized \\ \cline{3-8} 
 &  & \cite{done-third-shike-10-ok} & Multiple virtual network function instances deployed into multiple physical hosts & Provide a detailed implementation, discuss the process of embedding ILP into QUBO & Minimize overall resource utilization & QA & Centralized \\ \cline{3-8} 
 &  & \cite{done-shike-15-ok-fine} & A computing graph network with vertices as computing servers, and edges as vertex connection links & Leverage linearization and Benders’ decomposition to convert mixed ILP problem into an ILP master problem and a linear programming sub-problem, present a hybrid Quantum Benders’ decomposition algorithm & Minimize total learning costs of CPU iterations & QA & Centralized \\ \cline{3-8} 
 &  & \cite{done-fourth-zuoqixue-20-ok} & IoV: RSUs, edge servers close to RSUs, and a cloud server. RSUs perceive environment. Edge servers and the cloud process data. & Introduce a quantified strategy to tackle the edge server placement problem. Employ Pareto QGA to obtain appropriate solutions & Maximize the ratio of the amount of data sent to edge servers, minimize load variance of edge servers and task waiting time & QGA & Centralized  \\ \cline{3-8} 
 &  & \cite{done-forth-shike-1-ok} & Three network layers: sensor layer, edge layer and cloud layer & Introduce task offloading to improve task execution speed for irrigation and fertilization & Minimize total task completion time in a long term & QGA & Centralized \\ \cline{3-8} 
 &  & \cite{done-chengyu-9-ok-fine} & Three layers: devices, edge nodes, and the cloud & Utilize a double-hashing technique for decoding process of quantum particles & Maximize 
 throughput, minimize delay and energy consumption, and balance system load & QPSO & Centralized \\ \cline{3-8} 
 &  & \cite{done-second-chengyu-1-ok} & Three layers: devices, edge nodes, and the cloud & Map task scheduling strategies to positions in the search space. The value in each position denotes computing resource. & Maximize user satisfaction & QPSO & Centralized \\ \cline{3-8} 
 &  & \cite{done-second-chengyu-12-ok} & Three layers: devices, edge nodes, and the cloud & Propose an improved QPSO to solve the optimization problem for optimal task offloading & Minimize energy consumption & QPSO & Centralized \\ \cline{3-8} 
 &  & \cite{done-third-yangdongling-14-ok} & IoV: offload tasks from vehicles to clouds, edge nodes, or other vehicles & Establish a multi-task offloading model based on dynamic edge–end collaboration & Minimize task offloading delay & QPSO & Centralized \\ \cline{3-8} 
 &  & \cite{done-second-chengyu-2-ok} & Two layers: devices (tasks), edge nodes (agents) & Accomplish task offloading and scheduling to agents aimed at finding the optimal coalition & Maximize coalition stability and user satisfaction & QPSO & Centralized \\ \cline{2-8} 
 & \multirow{2}{*}{\centering \makecell{ \\ With \\consideration \\of Mobility}} & \cite{done-gaopeng-9-ok} & IoV: R2I and V2V communications  & Maintain population diversity; Introduce crossover operator to exchange information among individuals & Minimize delay cost and energy consumption & QPSO & Centralized \\ \cline{3-8} 
 &  & \cite{done-fourth-yangdongling-2-ok} & Three-tier MEC based heterogeneous 5G networks with macro, pico and femto & Propose an uplink resource allocation with RF energy harvesting in a mobility scenario & Maximize network throughput & QPSO & Centralized \\ \hline
\multirow{8}{*}{\centering \makecell{\\ \\ \\  \\ \\ \\ \\ \\ \\ \\ \\ \\ \\ AI-driven QC}} & \multirow{8}{*}{\centering \makecell{\\ \\ \\ \\ \\ \\ \\ \\ \\ \\ \\ \\  Without \\consideration \\of Mobility}} & \cite{done-second-gaopeng-7-ok} & WiFi/5G inter-working vehicle network: RSUs, one BS, and vehicles & Design a QRL method to obtain real-time optimal resource management policy & Minimize system latency & QRL & Centralized \\ \cline{3-8} 
 &  & \cite{done-second-shike-4-ok} & M IoT devices connecting to a cloud server equipped with CPUs, GPUs and QPUs & Design a dynamic computation offloading strategy in which both classical and QC resources coexist & Minimizes total task-offloading latency & QDRL & Centralized \\ \cline{3-8} 
 &  & \cite{done-second-zhongrunhu-7} & An edge computing system: mobile users and quantum edge servers & Formulate a classical and quantum computation offloading problem as a mixed ILP, solved by hybrid continuous-discrete MDRL & Minimize total offloading and execution cost & Quantum multiple-agent DRL & Distributed \\ \cline{3-8} 
 &  & \cite{done-gaopeng-12-ok-fine} & MEC networks: multiple IoT devices and edge servers &  Employ quantum state representations to transform MDP into a QC problem & Maximize system energy efficiency & QDRL & Centralized \\ \cline{3-8} 
 &  & \cite{done-third-zhongrunhu-8-ok} & MEC networks: multiple IoT devices and edge servers & Combine QC and ML to achieve exploration and exploitation trade-off via quantum parallelism & Maximize system energy efficiency while guaranteeing transmission latency requirements & QDRL & Centralized \\ \cline{3-8} 
 &  & \cite{done-zhongrunhu-6-ok-fine} & VRs have computing capacity to perform immersive viewpoint rendering, associated with an edge AP to access VR servers & Integrate wireless VR QoE and content correlation together with computation offloading & Minimize the total energy consumption & QRL & Centralized \\ \cline{3-8} 
 &  & \cite{done-fourth-wucheng-13-ok} & VRs associated with multiple edge AP to receive videos and feedback from the server & Integrate content correlation, block consensus, fluctuating channel conditions, QoE, and the blockchain to offload tasks for security & Minimize energy consumption & QDRL & Centralized \\ \cline{3-8} 
 &  & \cite{done-zhongrunhu_16-ok-fine} & UAVs offloading tasks to BS, with limited resources, to construct Metaverse & Propose a QDRL algorithm to render UAV images to construct Metaverse & Maximize the benefits for Metaverse-service provider & QDRL & Centralized \\ \hline
\end{tabular}
}
\end{table*}

\begin{table*}[!t]
    \centering
    \caption{Details of QC Algorithms in Task Offloading}
    \label{task-offloading2}
    \begin{tabular}{|>{\centering\arraybackslash}m{1cm}|>{\centering\arraybackslash}m{2cm}|>{\centering\arraybackslash}m{1.5cm}|>{\centering\arraybackslash}m{2.5cm}|>{\centering\arraybackslash}m{2.5cm}|>{\centering\arraybackslash}m{3.5cm}|>{\centering\arraybackslash}m{1cm}|}
    \hline
    \textbf{Work} & \textbf{Alg.} & \textbf{Orth. QC} & \textbf{Quantum computer} & \textbf{Platform} & \textbf{No. qubits}  & \textbf{Mobility} \\ \hline
    
    \cite{done-shike-7-ok} & QA & \Checkmark & \Checkmark & D-wave & Unknown  & \XSolidBrush \\ \hline
    \cite{done-third-shike-10-ok} & QA & \Checkmark & \Checkmark & D-wave & Unknown  & \XSolidBrush \\ \hline
    \cite{done-shike-15-ok-fine} & QA & \Checkmark & \Checkmark & D-wave & Unknown  & \XSolidBrush \\ \hline
    \cite{done-fourth-zuoqixue-20-ok} & QGA & \XSolidBrush & \XSolidBrush & Computer simulation & 0  & \XSolidBrush \\ \hline
    \cite{done-forth-shike-1-ok} & QGA & \XSolidBrush & \XSolidBrush & Computer simulation & 0  & \XSolidBrush \\ \hline
    \cite{done-chengyu-9-ok-fine} & QPSO & \XSolidBrush & \XSolidBrush & Computer simulation & 0   & \XSolidBrush \\ \hline
    \cite{done-second-chengyu-1-ok} & QPSO & \XSolidBrush & \XSolidBrush & Computer simulation & 0  & \XSolidBrush \\ \hline
    \cite{done-second-chengyu-12-ok} &QPSO & \XSolidBrush & \XSolidBrush & Computer simulation & 0  & \XSolidBrush \\ \hline
    \cite{done-third-yangdongling-14-ok} & QPSO & \XSolidBrush & \XSolidBrush & Computer simulation & 0  & \XSolidBrush \\ \hline
    \cite{done-second-chengyu-2-ok} & QPSO & \XSolidBrush & \XSolidBrush & Computer simulation & 0 & \XSolidBrush \\ \hline
    \cite{done-gaopeng-9-ok} & QPSO & \XSolidBrush & \XSolidBrush & Computer simulation & 0  & \Checkmark \\ \hline
    \cite{done-fourth-yangdongling-2-ok} & QPSO & \XSolidBrush & \XSolidBrush & Computer simulation & 0 & \Checkmark \\ \hline
    \cite{done-second-gaopeng-7-ok} & QRL & \XSolidBrush & \XSolidBrush & Computer simulation & 0 & \XSolidBrush \\ \hline
    \cite{done-second-shike-4-ok} & QDRL & \XSolidBrush & \XSolidBrush & Computer simulation & 0  & \XSolidBrush \\ \hline
    \cite{done-second-zhongrunhu-7} & QMADRL & \Checkmark & \Checkmark & Unknown & [1000, 5000] per edge server  & \XSolidBrush \\ \hline
    \cite{done-gaopeng-12-ok-fine} & QDRL & \XSolidBrush & \XSolidBrush & Computer simulation & 0  & \XSolidBrush \\ \hline
    \cite{done-third-zhongrunhu-8-ok} & QDRL & \Checkmark & \Checkmark & Variational quantum circuits & Unknown  & \XSolidBrush \\ \hline
    \cite{done-zhongrunhu-6-ok-fine} & QRL & \XSolidBrush & \XSolidBrush & Computer simulation & 0  & \XSolidBrush \\ \hline
    \cite{done-fourth-wucheng-13-ok} & QDRL & \XSolidBrush & \XSolidBrush & Computer simulation & 0  & \XSolidBrush \\ \hline
    \cite{done-zhongrunhu_16-ok-fine} & QDRL & \XSolidBrush & \XSolidBrush & Computer simulation & 0  & \XSolidBrush \\ \hline
    \end{tabular}
\end{table*}

\vspace{5pt}
\subsubsection{Applications of Conventional QC for Task Offloading} \

\textbf{Without consideration of mobility}

Traditional task offloading, which involves scheduling function virtualization services to users, is explored in \cite{done-shike-7-ok, done-third-shike-10-ok}. Specifically, in \cite{done-shike-7-ok}, the network provides $K$ types of network function virtualization services, with virtual machines organized into groups dedicated to serving specific network functions. The scheduling problem for network function virtualization is initially framed as an integer linear programming (ILP) challenge, subsequently converted into a QUBO model, and addressed by QA.
Similarly, \cite{done-third-shike-10-ok} addresses the task of offloading virtual network functions within a simple network setup. It provides a detailed implementation introduction, discussing the process of embedding the ILP problem into a QUBO format. It includes the formation of the Q-matrix, crucial for solvability by QA, and the mapping of quantum hardware to the problem domain.

With the evolution of 5G/6G and beyond, edge computing emerges as a compelling paradigm aimed at delivering services to users by offloading computation tasks from users and cloud infrastructure to edge nodes situated in close proximity to users.
In response to the challenge posed by limited edge server placement in IoV networks, a method called quantum edge server placement is developed in \cite{done-fourth-zuoqixue-20-ok}. The method employs dual encoding, utilizing both binary and quantum encoding techniques, to represent the locations of edge servers. Leveraging the non-dominated sorting genetic algorithm, it effectively explores potential placement solutions. By employing an assessment function, it derives optimal placement schemes, thereby improving the efficiency of edge server deployment in IoV scenarios. However, it is worth noting that it does not account for network dynamics, such as the impact of vehicle mobility. QGA-based task offloading  also finds applications in intelligent agriculture, including irrigation and fertilization tasks \cite{done-forth-shike-1-ok}. Quantum operators utilizes rotation gates and non-gates to update the matrix, mitigating the algorithm complexity by eliminating query operations from the rotation angle table during quantum rotation gate updates.
However, it is worth noting that in practice, the latency requirements for drip irrigation and fertilization tasks are not stringent, and the demand for computational resources in these applications tends to be low. Consequently, these factors result in  less attention being given to finding optimized solutions for such scenarios.

Another conventional technique for task offloading is QPSO, widely applied in service placement within IoT networks \cite{done-chengyu-9-ok-fine}, as well as in multi-task applications \cite{done-second-chengyu-12-ok, done-third-yangdongling-14-ok, done-second-chengyu-2-ok}. 
A QPSO method is proposed specifically for IoT task offloading in edge computing systems in \cite{done-chengyu-9-ok-fine}. The problem is formulated mathematically, with quantum particles efficiently designed to handle the entire IoT task offloading process. These particles guarantee the generation of complete and valid solutions to the problem. Notably, the decoding process of quantum particles employs a novel double-hashing technique, ensuring uniform task offloading to the edge nodes. Furthermore, the fitness function is designed considering four distinct objectives: maximizing throughput, minimizing delay, reducing energy consumption, and balancing system load. 

Federated learning, a method of distributing computational tasks across multiple decentralized servers \cite{done-shike-15-ok-fine}, can be considered a variant of task offloading. The approach addresses the challenge of training ML models across a distributed network without centralized data aggregation. The process involves formulating a joint participant selection and learning scheduling problem as a mixed ILP problem, aimed at minimizing the total learning costs. To tackle this optimization problem, a hybrid quantum-classical benders' decomposition algorithm is proposed. By combining classical and QC methodologies, the approach aims to efficiently solve the optimization problem associated with multi-modal federated learning in distributed networks.

The previous studies fail to account for the impact of multiple tasks on system performance. For instance, in  \cite{done-chengyu-9-ok-fine}, while mathematical models involve multiple users with multiple tasks, there is no mutual influence among them. The task offloading problem is mathematically formulated as a mixed-integer nonlinear programming model in \cite{done-second-chengyu-1-ok}. The formulation takes into account constraints such as deadlines and resource capacities, with the objective of maximizing user satisfaction. Within the proposed QPSO algorithm, task scheduling solutions are mapped to positions in the search space using real number coding. Here, the value in each position denotes the computing resource to which the corresponding task is assigned. The algorithm adopts a population evolutionary approach of QPSO to iteratively search for the global best solution.

The study presented in \cite{done-second-chengyu-12-ok} introduces an innovative solution to task offloading within mobile edge computing (MEC) environments characterized by users and servers. In this study, the task offloading strategy is reformulated as an energy optimization problem. To address this challenge, an enhanced QPSO algorithm is proposed. Conventional PSO approaches constrain the search range of particles, restricting them from exploring positions that are far from the swarm, even if those positions yield better solutions than the current best. This study introduces a novel approach by utilizing quantum state representation to address this limitation. By representing multiple potential solutions as quantum states before measurement, the method enables particles to explore a broader solution space, thereby reducing the likelihood of getting trapped in local optima. The effectiveness of the proposed QPSO-driven offloading strategy is demonstrated through its convergence. A bound state described by the probability density function, appears at any interval of the completely searchable space with a certain probability. However, it is noteworthy that the impact of different tasks remains relatively weak, as multiple tasks from various users compete for the limited resources available at edge nodes. Additionally, the study lacks analysis regarding the impact of mobility within MEC environments. Similar limitations are observed in the study conducted by \cite{done-third-yangdongling-14-ok} within the IoV environment. Additionally, the assumption that the next hop is predetermined in scenarios where tasks are offloaded from vehicles to other vehicles lacks practicality. 
The first limitation is addressed in \cite{done-second-chengyu-2-ok}. It identifies two interrelated impacts among tasks: firstly, when a complex task arrives, it is offloaded to a coalition of edge nodes for collaborative resolution; secondly, the complex task is divided into multiple sub-tasks, each offloaded to distinct coalitions. The proposal calculates the Shapley value of the optimal solution by employing a similarity metric based on task rewards of each combined edge group.

\textbf{With consideration of mobility}

The previous works in this subsection overlook the mobility of end-uses or edge nodes even though in MEC or IoV scenarios. 
The channel resource allocation is presented in \cite{done-fourth-yangdongling-2-ok} in the background of task offloading in a three-tier, such as macro, pico and femto, MEC based heterogeneous 5G networks. User equipments offload computation tasks to the local server, which significantly reduces the transmission latency. In the process of data transmission, multiple nodes compete for the channel resource. However, a random mobility model is designed to describe the movement of the user equipment.
One approach in \cite{done-gaopeng-9-ok} introduces a joint offloading strategy based on QPSO for MEC-enabled vehicular networks. The strategy focuses on minimizing delay costs and energy consumption by offloading tasks from vehicles to available service nodes, including service vehicles and nearby RSUs. To facilitate task offloading via V2V communication, a vehicle selection algorithm is introduced to determine an optimal offloading decision sequence. Moreover, to maintain population diversity in the QPSO algorithm, a crossover operator is employed to exchange information among individuals. However, it is assumed in the mobility model that vehicle speeds and coordinates are given in advance, presenting a practical limitation. 

\vspace{5pt}
\subsubsection{Applications of AI-driven QC for Task Offloading} \

The emergence and rapid development of AI have transformed QC techniques for task offloading into a higher level of smartness and efficiency through many scenarios, e.g., IoV \cite{done-second-gaopeng-7-ok} and smart healthcare \cite{done-second-shike-4-ok}. Specifically, in \cite{done-second-gaopeng-7-ok}, a resource management strategy is introduced to achieve the balance between communication and computation requirements by continuously monitoring the status of local vehicles, MEC-assisted WiFi links, and cloud-assisted 5G cellular networks. It empowers vehicles to make adaptive and flexible decisions regarding task offloading and transmission. 
To tackle the optimization problem in the face of time-varying channel conditions, the problem is reformulated as a MDP problem, solved by QRL to derive the real-time optimal resource management policy. The approach assumes a virtual quantum server for performing optimization, whereas works \cite{done-second-shike-4-ok, done-second-zhongrunhu-7} explicitly introduce quantum servers.

In \cite{done-second-shike-4-ok}, a dynamic computation offloading schedule is designed to minimize total latency within a hybrid setup, accommodating both classical (CPUs/GPUs) and quantum (QPUs) computing resources in cloud servers. This strategy ensures sustainable computation offloading while meeting individual latency constraints and the requisite success ratio for each task. Initially, the latency minimization problem is reformulated as a step-wise mixed-integer non-convex optimization task by utilizing Lyapunov techniques. Subsequently, DQN is employed for computation offloading mode selection, CPU/GPU mode or QPU mode, for cloud servers. 
Additionally, quantum edge servers are established in \cite{done-second-zhongrunhu-7}. It formulates a classical and quantum computation offloading problem as a mixed-integer programming model, which is tackled using a multi-agent hybrid continuous-discrete DRL algorithm.

QC coupled with DRL techniques also finds extensive applications in addressing task offloading challenges across various scenarios, including IoT \cite{done-gaopeng-12-ok-fine, done-third-zhongrunhu-8-ok}, VR \cite{done-zhongrunhu-6-ok-fine, done-fourth-wucheng-13-ok}, and Metaverse environments \cite{done-zhongrunhu_16-ok-fine}. 
In \cite{done-gaopeng-12-ok-fine}, a MEC-enabled IoT framework is proposed, integrating QC capabilities. The architecture enables IoT users to make sequential offloading and transmission decisions in an adaptive and flexible manner. The study formulates a joint optimal stochastic computation task offloading and dynamic resource allocation problem aimed at maximizing system energy efficiency. The optimization considers factors such as transmission latency, energy consumption, time-varying channel conditions, and data buffer constraints. 
However, in highly dynamic and complex IoT environments,  a Q-table in classical DRL algorithms face significant challenges. The stochastic and time-varying nature of channel conditions, combined with a large search space, leads to high time and space complexities. The discretization of continuous state and action spaces in the Q-table further hinders learning efficiency, primarily due to the high dimensionality. Additionally, the slow learning speed and limited storage capacity of classical DRL models are inadequate to meet the stringent offloading decision-making and processing demands of IoT systems.
Quantum state representations are employed to transform the continuous-time MDP problem into QC optimization, which is effectively addressed by the proposed QDRL algorithm. 
All potential quantum representation states can be modified in magnitude through a set of orthonormal eigenstates. Leveraging quantum parallelism, a unitary matrix operates on the superposition of mixed qubits, simultaneously affecting all states.
The research team also extends their solution into 6G MEC based IoT system \cite{done-third-zhongrunhu-8-ok}, but applying variational quantum circuits.

Representative applications in 6G, VR and Metaverse, presented challenges in transmitting resources for large-volume streaming data, excessive computing resources for encoding and decoding operations, and extreme energy consumption in transmitting and receiving devices. QC coupled with DRL techniques emerges as a promising approach to address these challenges \cite{done-zhongrunhu-6-ok-fine, done-fourth-wucheng-13-ok, done-zhongrunhu_16-ok-fine}. The study \cite{done-zhongrunhu-6-ok-fine} presents an energy-aware resource management strategy for VR-supported industrial IoT. It addresses rendering task offloading and resource allocation challenges considering factors like content correlation, fluctuating channel conditions, and VR QoE. To handle this problem effectively, the problem is transformed into a sequential decision-making scenario under varying channel conditions over time. By defining a reward function that encapsulates energy consumption and VR QoE, the optimization problem is formulated as an MDP, which is then tackled using QRL to maximize the long-term system reward. The researchers extend their work to enhance resilience against malicious attacks by integrating the blockchain technique into the system, particularly in the context of VR-enabled medical treatment applications in 6G \cite{done-fourth-wucheng-13-ok}. Additionally, they apply QDRL instead of QRL to enhance optimization efficiency.
Similar to the work \cite{done-gaopeng-12-ok-fine}, the study transforms the joint resource allocation based on RL into QDRL to overcome the challenge of "curse of dimensionality". It chooses an $n$-qubit quantum system and uses $n$ qubits to represent the eigen actions. An entangled quantum action is constructed by an orthogonal set of base vectors that are called eigen actions. Each eigen action corresponds to a classic action in RL. Then, the classic actions can be transformed into an entangled quantum action using the base vectors. The states and actions in RL is transformed into a quantum-collapse-based mapping according to the probability amplitude. Finally, the quantum searching for the optimal actions is performed according to the Grover iteration based on Hadamard gate.

Another innovative application of 6G, Metaverse, leverages DRL-driven QC algorithms to facilitate seamless immersion between reality and virtuality. Service subscribers access a real-time virtual world, constructed from images captured by on-site UAVs, within the Metaverse framework \cite{done-zhongrunhu_16-ok-fine}. A key challenge arises in determining the optimal BSs to which UAVs offloaded their tasks to maximize benefits, particularly when BSs have limited bandwidth and computation resources. Specifically, to tackle the challenge posed by new environments or re-deployments, where traditional RL methods often encountered model collapse, collective learning is introduced. This approach enables UAVs to leverage shared intelligence, facilitating rapid adaptation to new environments. The work can be improved by considering UAV mobility to achieve a more efficient system.

In conclusion, QC offers valuable insights into task offloading, particularly in the context of evolving networking technologies like 6G, IoV, and UAV networks. Detailed summaries of these works are presented in Table \ref{task-offloading1}, with further elaboration on their QC algorithms available in Table \ref{task-offloading2}. 
It is worth noting that task offloading within these scenarios typically involves distributed control and node mobility. However, only a few works provide solutions to address these aspects, showed in Table \ref{task-offloading2}. The primary reason for the first characteristic is current high overhead of deploying quantum machines. To the best of our knowledge, mobility remains an open challenge for task offloading integrated with QC.

\subsection{Content Caching}
\label{sec:contentcaching}

\subsubsection{Overview}\

In the fast-paced world of digital content consumption,
content caching, a technique widely employed by content delivery networks (CDNs), web servers, information centric network, etc., plays a pivotal role in optimizing the delivery of online content, profoundly influencing the overall user experience \cite{mao2023ai}.
It involves the strategic storage of frequently accessed content or computation results on servers distributed geographically, ensuring that users can access information with minimal latency, cost or other preferred objective. This practice significantly increases responsiveness and network bandwidth utilization.

Content caching scenarios are prevalent across various applications, spanning from edge computing \cite{song2021energy}, 5G cellular networks \cite{xie2019energy},  IoT \cite{done-gaopeng-1-ok-fine} to video streaming \cite{xie2022joint}. With the development of QC algorithms, the solutions leverage QGA \cite{song2021energy}, QDL \cite{done-wucheng-10-ok}, and QRL \cite{lin2020edge} to tackle problems. 
The inherent technical challenges in content caching stem from several key factors. First, ensuring effective caching cooperation among different network entities is crucial for optimizing the distribution of cached content and maximizing overall network performance. Second, resource limitations, such as limited storage capacity and bandwidth availability, pose significant challenges in efficiently managing cached content and satisfying user demands. Third, the dynamic nature of network traffic and user preferences requires adaptive caching strategies that can adjust in real-time to changing conditions. Finally, the distributed control manner of content caching systems necessitates robust algorithms and protocols for coordinating caching activities and ensuring seamless integration within the network architecture.

\begin{table*}[!t]
\centering
\caption{Summary of QC Works on Content Caching in Wireless Networks}
\label{Content-Caching1}
\begin{tabular}{|>{\centering\arraybackslash}m{2cm}|>{\centering\arraybackslash}m{1cm}|>{\centering\arraybackslash}m{3cm}|>{\centering\arraybackslash}m{3cm}|>{\centering\arraybackslash}m{3cm}|>{\centering\arraybackslash}m{1.5cm}|>{\centering\arraybackslash}m{1.5cm}|}
\hline
\textbf{QC category} & \textbf{Work} & \textbf{Network structure} & \textbf{Highlights} & \textbf{Optimization objectives} & \textbf{QC alg.} & \textbf{Control manner} \\ \hline

\multirow{4}{*}{\centering \makecell{\\ \\ \\ \\ \\ \\ \\ Conventional QC}} & \cite{song2021energy} & 5G cellular networks: a macro, picos and users & Introduce an adaptive weighted cost function to combine delay and energy cost & Minimize delay and energy cost & QGA & Centralized \\ \cline{2-7} 
 & \cite{xie2017energy} & 5G: a macro BS, multiple small-cell BSs that can be activated as either a pico  or a femto & Design a joint caching and activation policy to improve energy efficiency & Maximize system energy savings & QEA & Centralized \\ \cline{2-7} 
 & \cite{xie2019energy} & 5G: a core gateway, multiple eNodeB and users & Propose a hierarchical cooperative caching policy among a core gateway and eNodeBs to realise energy efficiency & Minimize total transmission energy consumption & QEA & Centralized \\ \cline{2-7} 
 & \cite{xie2022joint} & 5G: Cache and relay bit-rate video streaming with assistance of UAVs & Propose a QEA algorithm to optimize joint caching and user association for adaptive  video streaming & Minimize  total content delivery delay & QEA & Centralized \\ \hline
\multirow{4}{*}{\centering \makecell{\\ \\ \\ \\ \\ \\ \\AI-driven QC}} & \cite{done-wucheng-10-ok} & Fog-computing assisted networks: a centralized cloud, fogs and users &  Integrated DL into QC to address  video caching with a detail process of training a QDL model & Update QDL model parameters according to the dataset & QDL & Centralized \\ \cline{2-7} 
 & \cite{done-gaopeng-1-ok-fine} & Edge computing assisted networks: a BS, edge servers and IoT devices & Optimize resource allocation, content caching, and task offloading in dynamic large-scale IoT networks & Maximize energy efficiency & QRL & Centralized \\ \cline{2-7} 
 & \cite{lin2020edge} & Edge computing assisted networks: a BS, edge servers and users & Optimize content caching and spectrum allocation, considering dynamic content popularity, transmission method, and requirements on QoE & Minimize content delivery delay & QRL & Centralized \\ \cline{2-7} 
 & \cite{done-gaopeng-2-ok-fine} & Edge computing assisted networks:  a BS, edge servers and users & Integrate DRL into QC to address task offloading and content caching. & Maximize the caching benefit & QDRL & Centralized \\ \hline

\end{tabular}
\end{table*}

\subsubsection{Applications of Conventional QC for Content Caching}

To alleviate backhaul link congestion and improve resource utilization in 5G ultra-dense cellular networks consisting of a macro, multiple picos and users, it is a traditional issue to cache content cooperatively in picos with resource limitations as illustrated in \cite{song2021energy}. This study focuses on an energy–delay tradeoff problem in collaborative edge computing-assisted and energy harvesting-powered ultra-dense networks. It regards delay and energy as two types of cost and introduces a weighted cost function to transform the tradeoff problem into a cost minimization problem. An alternating optimization based on an improved QGA is proposed to solve the cost minimization problem. The proposal divides the cost minimization problem into two sub-problems, adaptive tuning weight sub-problem and caching decision sub-problem. It improves QGA in terms of repairing unfeasible solutions and adaptively updating quantum genes. The problem arises from the fact the solutions derived from the individual sub-problems can not ensure the optimality of the original problem.

One of important features in 5G cellular networks is heterogeneity, i.e., small-cell BSs can be activated when necessary to be either a pico-cell BS or femto-cell BS \cite{xie2017energy}, and cooperative caching performs in core gateways \cite{xie2019energy}. A joint caching and activation mechanism in \cite{xie2017energy} is designed to maximize system energy savings, while a cooperative caching policy is proposed in \cite{xie2019energy} to reduce the duplicate transmission and improve the green content delivery. Both two methods formulate the issue as an ILP problem, which is solved by QEA. Another heterogeneity comes from assistance of UAVs for edge caching in 5G cellular networks, which has also received extensive attention from both industry and academia \cite{xie2022joint}. UAV-assisted edge caching is a promising technique to improve content transmission and enhance delivery efficiency by caching video closer to users. The study \cite{xie2022joint} adaptively adjusts video quality based on time-varying network conditions and different user preferences according to proposed QEA. 
However, the work can be further enhanced by leveraging UAV mobility to dynamically adapt to user mobility.

\subsubsection{Applications of AI-driven QC for Content Caching} \

AI-driven QC can significantly enhance the efficiency of content caching by leveraging its advantages in computing \cite{done-wucheng-10-ok, done-gaopeng-1-ok-fine, lin2020edge, done-gaopeng-2-ok-fine}. In \cite{done-wucheng-10-ok}, DRL is integrated into QC to address the traditional challenge of video caching with a detail process of training a QDL model. Specifically, a DRL agent utilizes the self-organizing map algorithm to prioritize caching contents. The prioritized contents are stored in a quantum memory module using a two-level spin quantum phenomenon. Following the selection of the most suitable lattice map via self-organizing maps, data points below a specified threshold are mapped onto the data frame to retrieve the videos. These videos are considered high-priority for training based on the input data provided in the dataset. Similarly, the models for the other level prioritized content can be trained accordingly.

Embedding RL into QC \cite{done-gaopeng-1-ok-fine, lin2020edge} is another effective approach to solving complex problems in content caching, especially for the problems in sequential decision processes. In \cite{done-gaopeng-1-ok-fine}, there are two sub-problems of task offloading and content caching with limited resources in a traditional edge computing assisted network. 
The network architecture comprises a BS, edge servers, and IoT devices. The proposed approach aims to optimize dynamic resource allocation, content caching strategy, and computation offloading policy to enhance energy efficiency in dynamic large-scale IoT networks. A QRL based algorithm is introduced to address the challenges posed by complex and dynamic environments, varying content popularity, and user preferences. However, there are two shortcomings in the proposal. First, the rationale behind jointly addressing the offloading and caching problem is unclear, lacking practical justification. Secondly, the solution relies on integer-based decisions for offloading and caching, potentially limiting its flexibility and scalability.  QRL is also applied in the application of caching video streaming to improve QoE \cite{lin2020edge}. Its proposal jointly optimizes content caching and spectrum allocation to minimize content delivery delay, taking into account time-varying content popularity, transmission method selection, and different requirements on QoE in a similar wireless network structure with \cite{done-gaopeng-1-ok-fine}. The second shortcoming in \cite{done-gaopeng-1-ok-fine} also arises from the binary caching and offloading decisions in its proposal.
To address the same problem as \cite{done-gaopeng-1-ok-fine}, the study \cite{done-gaopeng-2-ok-fine} proposes a solution by integrating DQN into QC, which involves the joint optimization of task offloading and content caching. However, a limitation arises as the two sub-problems are optimized independently, potentially leading to sub-optimal solutions for the original problem.

In conclusion, QC algorithms are widely applied in the area of content caching, as presented in Table \ref{Content-Caching1}. There are still several issues in existing literature. Many related works implement their QC algorithms in computer simulations duo to the current  developing stage of quantum computers. Additionally, current solutions often employ centralized control methods, primarily because computation services typically take the form of quantum clouds. Lastly, similar to the challenge in task offloading, node mobility remains an open issue when considering content caching.

\section{Miscellaneous Issues} \label{Miscellaneous Issues}

The first section explores QC impact on wireless security, focusing on encryption, authentication, and privacy techniques like federated learning and blockchain. The second part shows the QC and other emerging methods for pinpointing wireless device locations. The third section discusses techniques for optimizing video content delivery over wireless networks, addressing challenges such as the optimization to improve streaming performance.

\subsection{Security and Privacy}
On the one hand, quantum communication stands out as a promising frontier in securing sensitive information against eavesdropping and unauthorized access. Unlike classical cryptographic techniques, quantum communication leverages the principles of quantum mechanics to ensure the security of transmitted data. For example, quantum communication protocols like quantum key distribution \cite{second-zuoqixue-19-wrong}, exploit the inherent properties of quantum particles, such as superposition and entanglement, to establish secure communication channels. On the other hand, QC enables quantum computers to perform complex computations at exponentially faster rates than classical computers, making them well-suited for cryptographic applications, such as secure coding such as secure URLLC in 5G and beyond cellular networks \cite{done-third-zhongrunhu-3-ok, done-second-yangdongling-17-ok}. 

Attacks on wireless pilot signals pose a significant threat to URLLC services.  
In grant-free URLLC scenarios, accurate allocation of distinct pilot resources to multiple users sharing the same time-frequency resource is crucial for the next-generation eNodeB to accurately identify these users under access collisions. This precise allocation is essential for maintaining accurate channel estimation, which is necessary for ensuring reliable data transmission. However, this procedure is susceptible to attacks targeting pilots. 
The research in \cite{done-third-zhongrunhu-3-ok} introduces an approach utilizing quantum learning to implement nonrandom superimposed coding for encoding and decoding pilots across multidimensional resources. The method enables rapid learning and precise elimination of attack uncertainties. Specifically, pilots for uplink access by multiple users are encoded as distinct sub-carrier activation patterns. The enodeB then decodes the desired pilots from observed sub-carrier activation patterns that represent a superposition. This is achieved through a joint design involving attack mode detection and user activity detection via a quantum learning network. 
The study in \cite{done-second-yangdongling-17-ok} solves the same security problem as \cite{done-third-zhongrunhu-3-ok}  but in the area of random structure coding. It introduces an approach to encoding and decoding pilot signals using a random structured code across multidimensional physical resources. The proposed code employs  random encoding with minimal structure to mitigate the impact of attacks. By leveraging group spatial channel features, the decoding process is modeled as a computational trap, leading to a security challenge parallel with random computing with redundancy. To address this, the study utilizes a quantum algorithm to learn the computational trap model, enabling the removal of computational redundancy and elimination of dispersed attacks. The proposal demonstrates the existence of a quantum black-box model corresponding to the computational trap and derives a precise expression for computational performance.

QC can also integrate with existing popular security and privacy techniques like blockchain \cite{done-second-chengyu-10-ok} and federated learning \cite{done-second-chengyu-16-ok, mao2023security, done-second-yangdongling-18-ok} to enhance network performance.
Inspired by classical federated learning techniques, the work in \cite{done-second-chengyu-16-ok} aims to enhance data privacy by aggregating updates from local computations to share model parameters. To obtain approximate optima in the parameter landscape, it introduces an extension of the traditional variational quantum algorithm. The variational quantum algorithms are also applied in future 6G space-air-ground integrated networks to address the complexity emerging from vast datasets and ML models aiming at data security and privacy \cite{done-second-yangdongling-18-ok}.  

\subsection{Localization and Tracking}
Existing localization and tracking solutions often require continuous communication and processing, leading to significant energy consumption, particularly in battery-constrained devices \cite{li2019af}. Additionally, it faces challenges in scaling to support a large number of devices simultaneously, leading to increased computational complexity and resource consumption. 
QC facilitates simultaneous processing of multiple possibilities, which accelerates and enhances localization and tracking tasks. This capability finds applications in various scenarios, including tracking mobile user trajectories \cite{done-second-shike-12-ok}, managing wireless sensor networks \cite{done-chengyu-18-ok, done-third-yangdongling-5-ok}, and coordinating multiple UAV networks \cite{done-second-wucheng-7-ok}. 
User trajectory prediction in \cite{done-second-shike-12-ok} involves two components: reservoir computing and QC. Compared to training simple recurrent neural networks, reservoir computing is computationally more efficient because only the output layer weights are trainable. However, achieving optimal performance in reservoir computing requires careful selection of reservoir weights to create complex and nonlinear dynamics.
QC is employed to generate such complex reservoir dynamics, mapping input time series into a higher-dimensional computational space consisting of dynamical states. Once high-dimensional dynamical states are obtained, a simple linear regression is applied to train the output weights, enabling efficient prediction of mobile users' trajectories.

QC also finds extensive applications in localization and tracking within sensor networks, particularly addressing scalability challenges. 
The approach in \cite{done-chengyu-18-ok} enhances the differential crossover in QPSO to address nonlinear localization problems influenced by node density and coverage in IoT. By utilizing the differential evolution operator, it mitigates group movements within narrow ranges, thus preventing convergence to local optima and enhancing global search capabilities. The cross operator facilitates information exchange among individuals within a group, while exceptional genes are moderately retained to sustain the evolutionary process, thereby incorporating both proactive and reactive features. In contrast to traditional location scenarios as presented in \cite{done-chengyu-18-ok}, \cite{done-third-yangdongling-5-ok} addresses the localization with multiple mobile robots equipped with sensors aiming to locate a specific source node. 
By modeling each robot as a double integrator, \cite{done-third-yangdongling-5-ok} converts the source location challenge into a multi-robot path planning task, considering constraints such as collisions between robots and obstacles, as well as collisions among robots. A framework for the mobile policy of the robots is devised. And an adaptive weight strategy, a swarm timely update mechanism, and a leading-following behavior are integrated into QPSO. These enhancements aims to enable multiple robots to adhere to the mobile policy effectively, ensuring robustness and efficiency in their search.
  
QC, paired with AI, presents new ways to tackle location and tracking tasks in UAV networks \cite{done-second-wucheng-7-ok}. 
In \cite{done-second-wucheng-7-ok}, a DRL approach is proposed for controlling vertical take-off and landing aircraft in the presence of wind disturbances. The tracking control problem is framed as MDP. To enhance control accuracy despite wind disturbances, various wind fields are incorporated into the learning environment to broaden exploration and mimic the impact of winds on flight control. To ensure both tracking accuracy and practical implementation, a quantum-inspired experience replay strategy is introduced based on QC principles. This strategy features a preparation operation scheme to foster exploration and accelerate convergence, along with a depreciation operation method to enhance sample diversity and bolster system robustness. 

Previous efforts to locate and track devices relies on sensors or GPS, which is impractical for scenarios like indoor person tracking with only a cellphone. Existing solutions for such an application typically utilize received signal strength (RSS) to analyze the location, but face challenges of scalability to worldwide and algorithm complexity. The work \cite{10274098} provides a fast matching quantum fingerprinting algorithm by calculating the quantum cosine similarity between online RSS vector at an unknown location and each RSS vector in a cellular fingerprint. Its efficiency is validated in the testbed IBM quantum machine simulator and cellular BSs. While it is an innovative approach, the matching accuracy can be further enhanced by utilizing channel state information (CSI) instead of RSS. However, acquiring CSI presents a practical challenge due to hardware limitations. Moreover, RSS, while widely used, is susceptible to noise factors like pedestrian interference, which is not accounted for in the solution.

\subsection{Video Streaming}

Video serves as a primary means of conveying information through visual and auditory media and finds widespread applications in entertainment, education, and information dissemination.
A typical video system includes several components such as video encoding/decoding, network transmission, and buffer, differing significantly from conventional data transmission. Additionally, emerging technologies \cite{li2023spherical,liu2021point} such as multi-view videos, free-viewpoint videos, VR/360 videos, and point cloud videos, have surfaced in recent years. These innovations allow users to choose viewing angles, enabling them to selectively watch specific video parts during viewing processes, for example, in the case of 360 videos/point cloud videos, where users can switch perspectives over time.

In wireless communications and networking, one classical issue is adaptive bitrate control in order to ensure video streaming quality. The work \cite{video-streaming-wei2023qudash} maximizes user QoE by increasing average bitrate and decreasing video rebuffering events. Specifically, the proposal formulates the control problem into a QUBO model from three aspects: video quality, bitrate change, and buffer condition, which is solved in the platform Digital Annealer. Representative applications, VR and Metaverse in 6G and beyond, also employ QC algorithms to improve performance. Ultrahigh viewport rendering demands and excessive terminal energy consumption restrict the application of wireless VR. The work \cite{lin2021task} considers content correlation between VR equipments, fluctuating channel conditions, and VR QoE, by formulating the joint problem of viewport rendering offloading, computing, and spectrum resource allocation. The formulation is solved by proposed RL algorithm and quantum parallelism is integrated into the RL to overcome the problem of high data dimension.

\section{Challenges and Open Issues}
\label{Challenges and Open Issues}

\subsection{QC in Wireless Communications and Networking with Multiple Variables}
QC has been applied in wireless communications and networking, primarily focusing on individual variables. However, tackling optimization problems involving multiple variables poses a more formidable challenge while considering network characteristics, such as high dimensionality, conflicting objectives, resource constraints and so forth. Addressing multiple variables in wireless systems presents significant hurdles for optimization and control. The exponential increase in dimensionality of optimization problems renders them computationally intensive and impractical for real-time decision-making. Balancing conflicting objectives, such as maximizing throughput while minimizing latency and interference, becomes intricate. Moreover, dynamic environments exacerbate the complexity, introducing uncertainty in predicting optimal configurations. Complex interactions among variables, compounded by resource constraints, further compound the challenge. Efficient algorithms that integrate optimization theory, signal processing, and ML is essential to effectively tackle these challenges.
On the other hand, AI-driven quantum algorithms sometimes require datasets for training, which is currently hindered by insufficient available datasets.

The challenge of handling multiple variables in wireless networks using QC can be addressed by integrating hybrid quantum-classical optimization frameworks. These frameworks  utilize classical techniques such as ML and optimization theory to manage simpler or well-understood parts of the problem, while quantum algorithms like QAOA focus on the most computationally intensive tasks. Additionally, techniques like dimensionality reduction and adaptive quantum algorithms can help mitigate the exponential growth of variables and provide real-time solutions. Developing quantum-inspired algorithms that efficiently handle conflicting objectives, such as balancing throughput, latency, and interference, will also be critical for the challenge.

\subsection{QC in Wireless Communications and Networking for Particular Services}
Introducing QC into wireless communications and networking for specific services poses several challenges while considering stringent service requirements, complex trade-offs, dynamic traffic patterns, and so forth. For instance, services like video streaming demand stringent requirements beyond conventional wireless systems, such as low latency, high throughput, and reliable transmission. Optimizing these parameters using quantum algorithms becomes more challenging due to the complexity introduced by the additional service requirements. Quantum algorithms must navigate the trade-offs between different objectives, such as maximizing video quality while minimizing latency and packet loss. Furthermore, the dynamic nature of video traffic adds an extra layer of complexity, requiring adaptive optimization strategies that can efficiently allocate resources in real-time. Additionally, ensuring the security and privacy of video data transmission becomes more critical with the integration of QC, necessitating robust encryption schemes and secure communication protocols that can withstand quantum attacks. Overall, integrating QC into wireless communications and networking or specific services like video streaming presents challenges in balancing competing objectives, adapting to dynamic traffic patterns, and ensuring robust security measures, requiring innovative solutions tailored to the unique requirements of each service.

To optimize QC in wireless networks for specific services like video streaming, hybrid quantum-classical algorithms can be developed to address the trade-offs between quality, latency, and packet loss. QML can predict traffic patterns, allowing for adaptive resource allocation in real-time. To ensure robust security in these systems, quantum-resistant encryption protocols must be integrated to safeguard data from quantum attacks. Additionally, early-stage quantum simulations can be used to test and refine these solutions, paving the way for practical implementations in complex service scenarios.

\subsection{Understanding and Utilizing QC in Communications and Networking}
While QC has demonstrated remarkable efficiency and achieved excellent results in some systems, their widespread application still faces numerous challenges compared to techniques like convex optimization and DRL. Particularly, QC, as a novel approach distinct from conventional methods, presents difficulties in understanding its operational principles. Unlike traditional algorithms whose operations are more easily comprehensible, the workings of QC involve intricate quantum mechanics concepts, making them less intuitive to grasp. Bridging this comprehension gap and facilitating the integration of QC into wireless systems pose significant challenges. Educating individuals about this technology and its applications in wireless communications and networking requires simplifying complex quantum concepts into more understandable terms and providing accessible learning resources. Additionally, developing user-friendly tools and interfaces that abstract away the complexity of quantum operations can aid in democratizing the utilization of QC in wireless systems. Overcoming these challenges demands collaborative efforts from researchers, educators, and industry professionals to demystify QC and foster its practical implementation in wireless communications and networking  effectively.

Addressing the complexity and understanding of QC in wireless communications requires the creation of educational tools and resources that break down quantum concepts into more accessible terms for professionals in the field. By developing user-friendly interfaces and software that abstract complex quantum operations, wireless engineers can more easily experiment with and adopt QC techniques. Collaborative efforts between academia and industry, along with standardized training programs, will help bridge the knowledge gap and accelerate the practical use of QC in real-world wireless systems.

\subsection{QC Cloud Services in Wireless Communications and Networking}

As QC becomes increasingly available to the public through cloud services, there is a growing need to ensure these services can be effectively supported in  wireless networks in the near future. Factors such as communication latency and stability will be critical considerations in meeting network requirements. Furthermore, the integration of QC into the classical computing pipeline will involve determining which components are most suitable for quantum processing to maximize benefits. This includes addressing the overhead associated with transitioning between classical and QC, a crucial aspect for optimizing overall efficiency and performance.

To support QC cloud services over wireless networks, a focus on optimizing communication latency and stability is key. One potential solution is to deploy quantum edge computing to localize quantum processing tasks closer to users, reducing network overhead and latency. Additionally, integrating QC with classical systems should prioritize tasks most suited to quantum processing, thereby maximizing performance while minimizing the overhead associated with switching between classical and quantum systems. Improved network protocols that efficiently manage these transitions will be necessary for enhancing the performance of QC cloud services in wireless networks.

\subsection{Integration of DRL with QC}
While many studies have explored the integration of DRL with QC to address issues in wireless networks, expanding on the challenges that QC faces in DRL requires an exploration of the key areas where these two fields intersect.
1) Measurement and information loss:
In QC, a quantum state collapses upon measurement, leading to a loss of information. Since DRL relies heavily on learning from observed states and rewards, this presents a challenge when attempting to optimize quantum systems in real-time without collapsing the states.
2) Noise and decoherence:
Quantum systems are highly sensitive to environmental noise and decoherence, which degrades the quality of quantum information and calculations. DRL algorithms that attempt to optimize quantum circuits may struggle with learning stable policies in the presence of such noise.
3) Quantum reward structures:
DRL relies on reward signals to guide the learning process. In quantum systems, the nature of the reward structures may be less intuitive than classical systems due to the probabilistic nature of quantum operations, leading to challenges in defining meaningful rewards.

To overcome the challenges of integrating DRL with QC in wireless networks, hybrid frameworks can be designed that separate quantum and classical tasks. Classical DRL algorithms can handle observation and learning, while quantum algorithms focus on optimizing decision-making processes in high-dimensional spaces. Developing quantum error correction techniques and robust algorithms that account for quantum noise and decoherence will also help stabilize learning. Quantum-inspired reward structures that align with DRL principles must be created to enhance the learning process while ensuring the stability and reliability of quantum systems.

\section{Conclusions} \label{Conclusions}

This paper presents a comprehensive survey of QC in wireless communications and networking, providing an in-depth overview of the field's fundamentals, advancements, and application potentials. By integrating QC into the optimization processes of wireless networks, significant improvements can be made in areas such as resource allocation, power management, edge computing, and security.

The primary contributions of this work lie in its interdisciplinary coverage, bridging quantum mechanics and wireless communications, and in providing a clear roadmap for applying QC algorithms to various aspects of wireless networking. The survey not only highlights the theoretical frameworks of QC but also examines its practical implications in real-world scenarios, such as edge computing and content caching. Additionally, it identifies key challenges and potential solutions for leveraging quantum algorithms in next-generation wireless networks.

The ongoing development of QC platforms, along with their application in wireless communications, opens up promising avenues for future research. Specific areas such as QRL, edge intelligence with QC, and advanced security protocols powered by quantum technologies will likely evolve further. Future research should focus on minimizing hardware noise and gate errors to improve algorithmic efficiency, as well as exploring hybrid classical-quantum frameworks for large-scale implementations. Expected outcomes from these advancements include faster optimization processes, more robust security mechanisms, and enhanced communication systems capable of meeting the demands of the upcoming 6G era.

\bibliographystyle{IEEEtran}
\bibliography{reference}

\begin{IEEEbiography}
[{\includegraphics[width=1in,height=1.25in,clip,keepaspectratio]{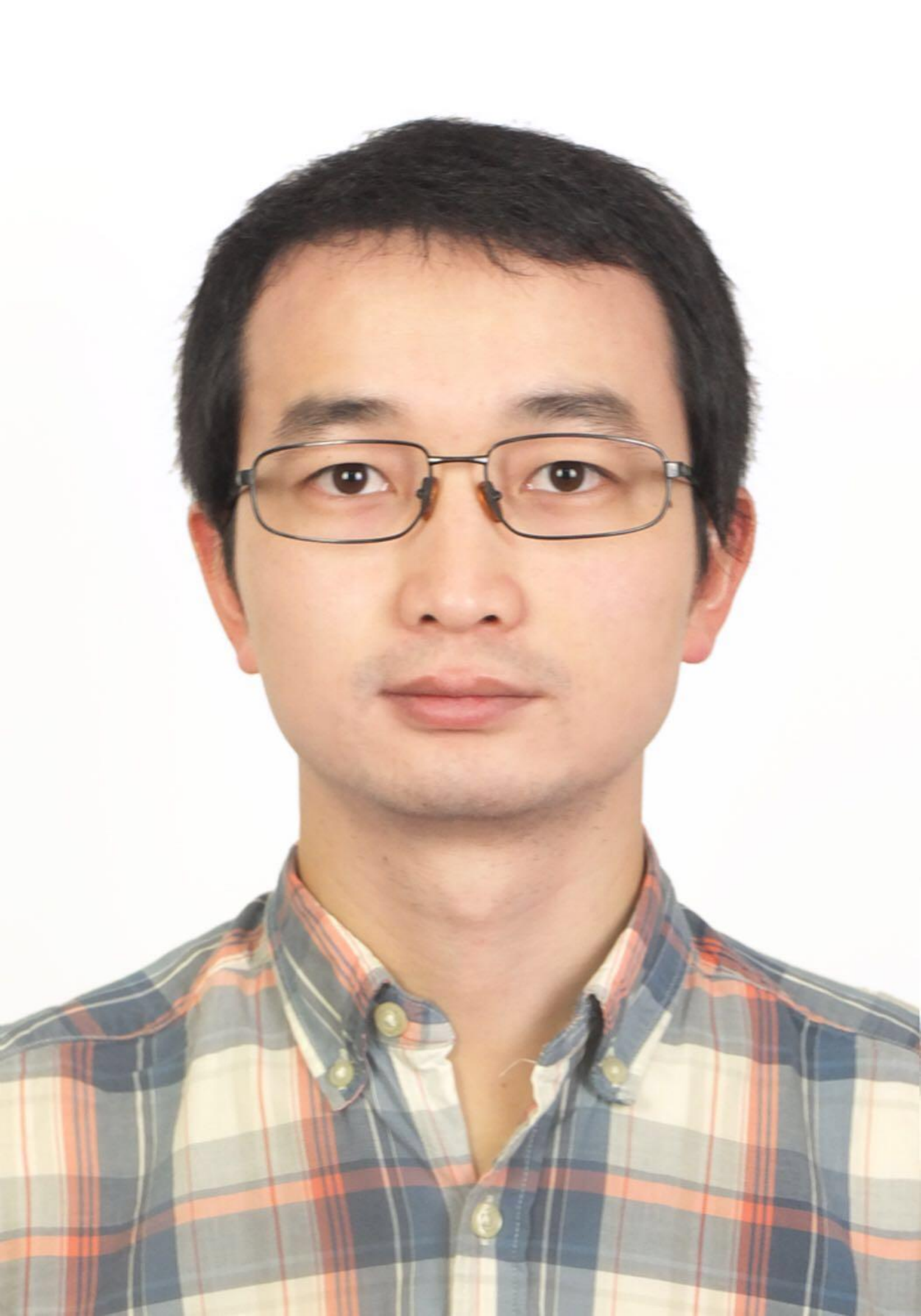}}]
{Wei Zhao} (S’12-M’16) received his Ph.D. degree in the Graduate School of Information Sciences, Tohoku University. He is currently an Associate Professor at the School of Computer Science and Technology, Anhui University of Technology. His research interests include deep reinforcement learning, edge computing, and resource allocation in wireless networks. He was the recipient of the IEEE WCSP-2014 Best Paper Award, and IEEE GLOBECOM-2014 Best Paper Award. He is a member of IEEE.
\end{IEEEbiography}

\vspace{-40pt}

\begin{IEEEbiography}
[{\includegraphics[width=1in,height=1.25in,clip,keepaspectratio]{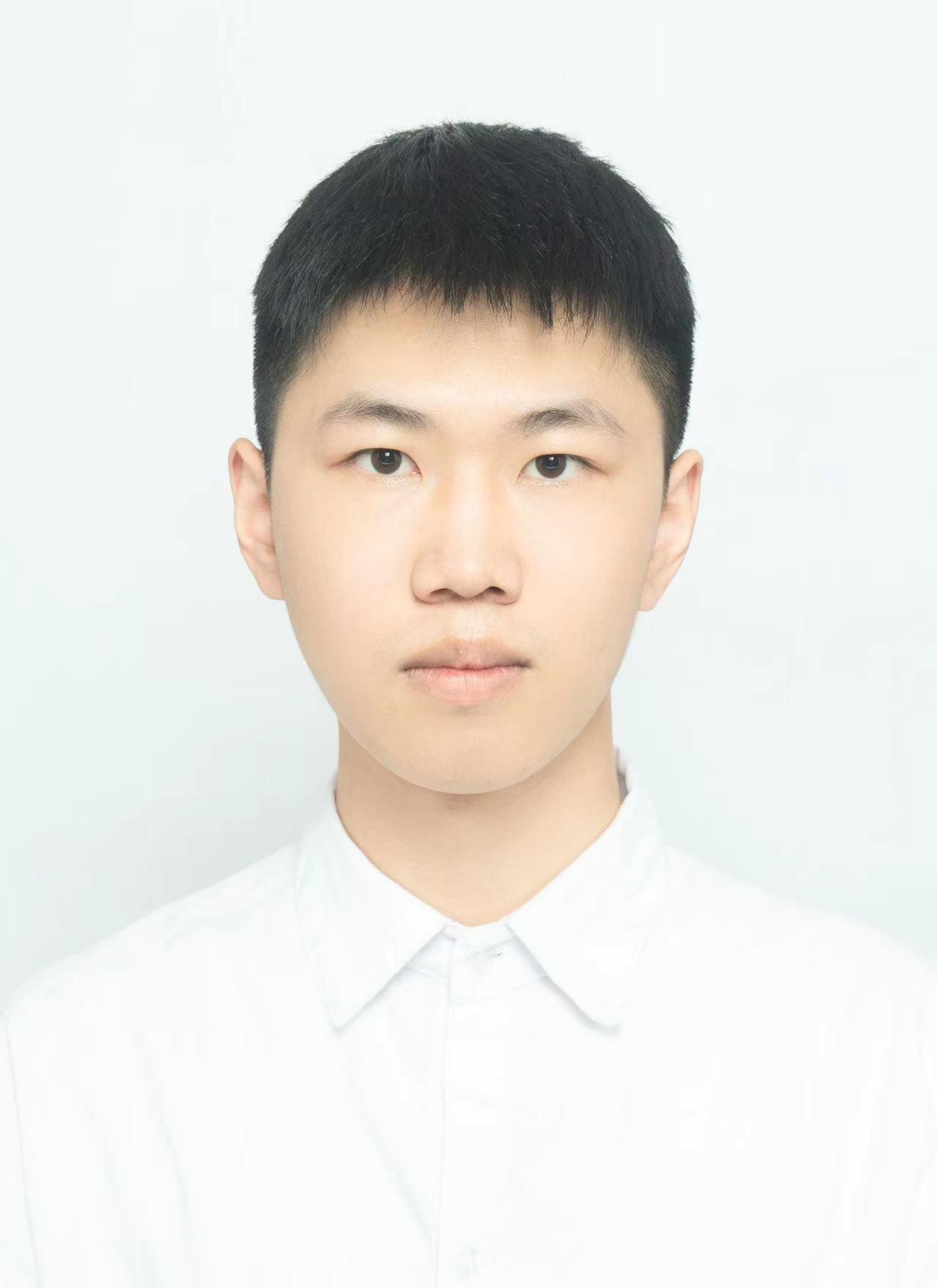}}]
{Tangjie Weng} is currently pursuing a Master's degree in Engineering at Anhui University of Technology. His research interests center around deep reinforcement learning and edge computing, specifically in their application to unmanned aerial vehicles (UAVs). By leveraging advanced computing techniques, He aims to enhance UAV decision-making processes and overall performance for various applications.
\end{IEEEbiography}

\vspace{-40pt}

\begin{IEEEbiography}
[{\includegraphics[width=1in,height=1.25in,clip,keepaspectratio]{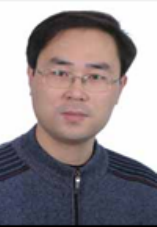}}]
{Yue Ruan} (M’17) received the Ph.D. degree in Computer Software and Theory from the School of Computer Science and Engineering, Southeast University, China, in 2017. After that, he was a visiting scholar at the Physics Department, Western Australia University, Australia. Currently, he is a lecturer at the School of Computer Science and Technology, Anhui University of Technology, China.His research interests include QC and ML, esp. quantum ML for solving optimization problems in communication scenarios.
\end{IEEEbiography}

\vspace{-35pt}

\begin{IEEEbiography}
[{\includegraphics[width=1in,height=1.25in,clip,keepaspectratio]{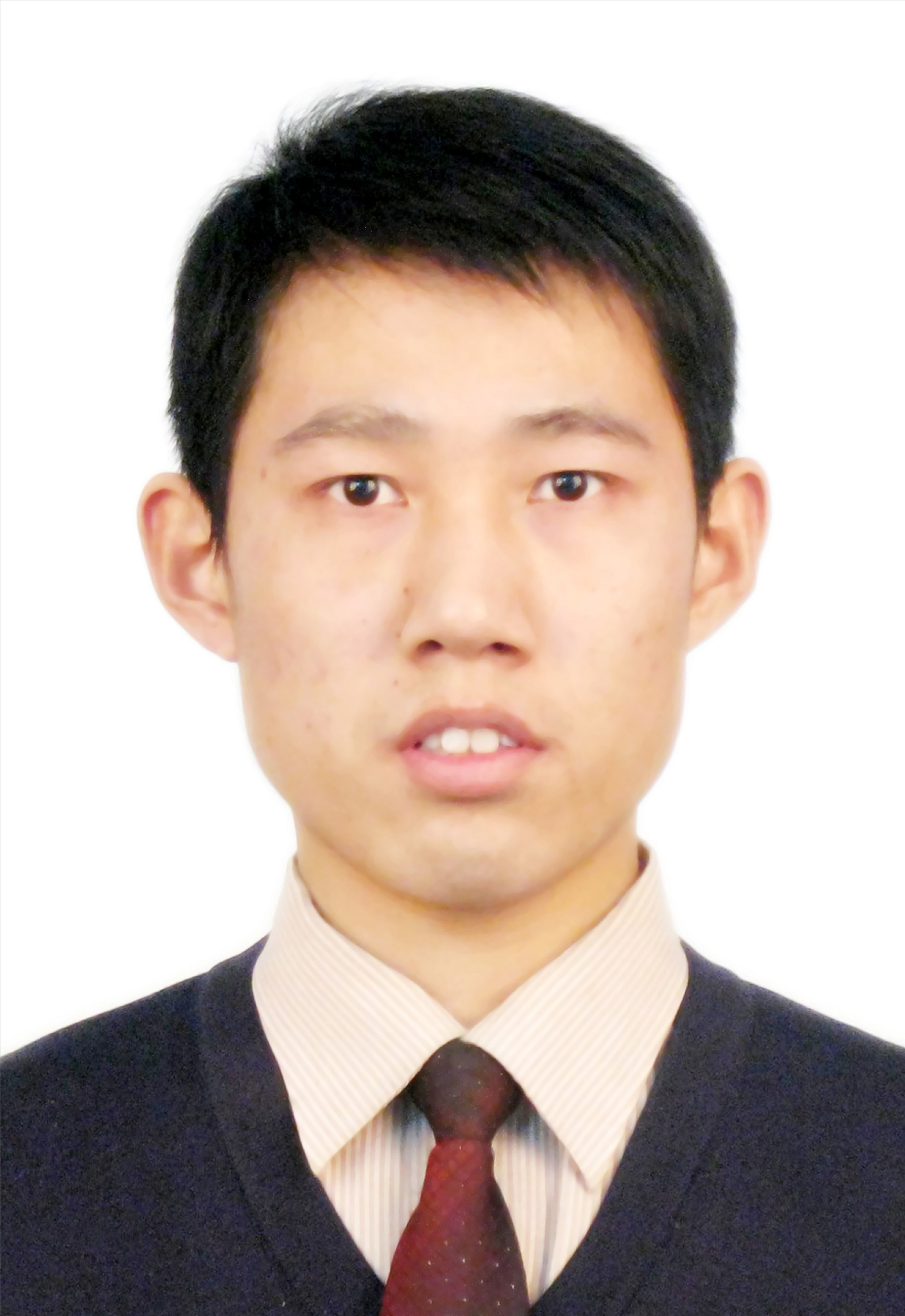}}]
{Zhi Liu} (S’11-M’14-SM’19) received the Ph.D. degree in informatics in National Institute of Informatics. He is currently an Associate Professor at the University of Electro-Communications. His research interest includes video network transmission and MEC. He is now an editorial board member of IEEE Transactions on Multimedia, IEEE Networks and Internet of Things Journal. He is a senior member of IEEE.
\end{IEEEbiography}

\vspace{-40pt}

\begin{IEEEbiography}
[{\includegraphics[width=1in,height=1.25in,clip,keepaspectratio]{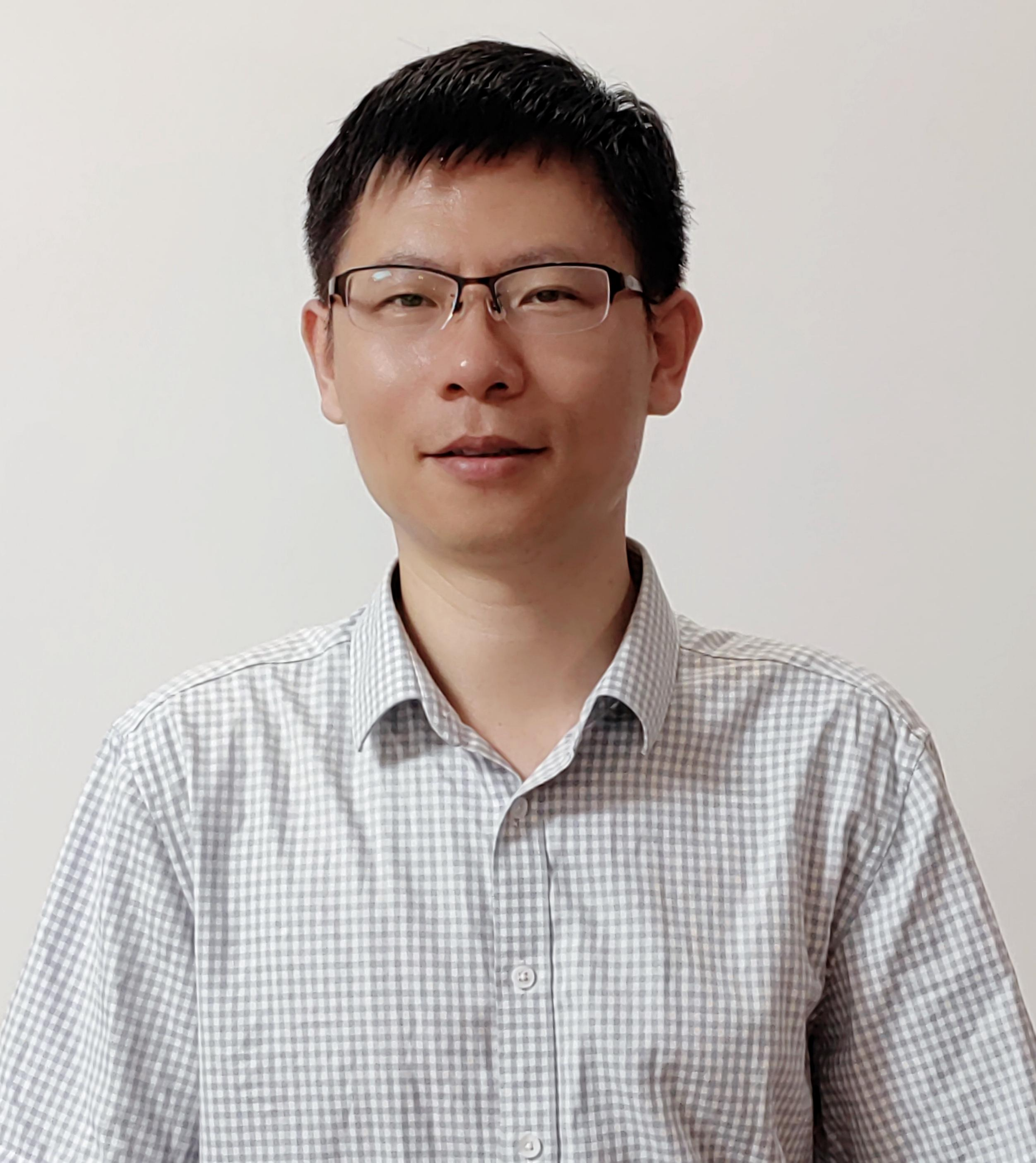}}]
{Xuangou Wu} received the Ph.D. degree in computer software and theory from Anhui University of Science and Technology of China, Hefei, China, in 2013. From August 2018 to September 2019, he was a visiting scholar with Department of Computer Science \& Engineering, Hong Kong University of Science \& Technology. He is currently a full Professor with the School of Computer Science and Technology, Anhui University of Technology, China. His research interests include Internet of Things, network security, and privacy protection.
\end{IEEEbiography}

\vspace{-35pt}

\begin{IEEEbiography}
[{\includegraphics[width=1in,height=1.25in,clip,keepaspectratio]{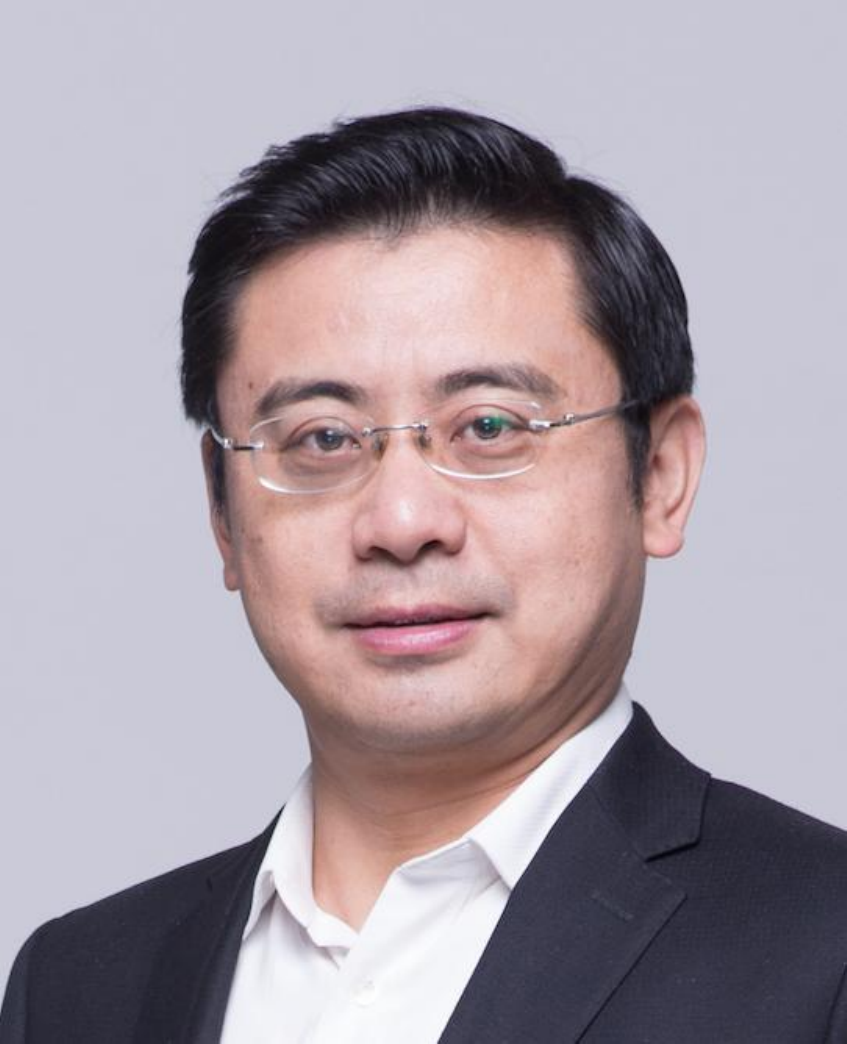}}]
{Xiao Zheng} received the Ph.D. degree in computer science and technology from Southeast University, China, in 2014. He is currently a professor with the School of Computer Science and Technology, Anhui University of Technology, Anhui, China. His research interests include service computing, mobile cloud computing and privacy protection. He has been a guest editor of IEICE Transactions on Communications. He is a senior member of CCF, and a member of the ACM.
\end{IEEEbiography}

\vspace{-40pt}

\begin{IEEEbiography}
[{\includegraphics[width=1in,height=1.25in,clip,keepaspectratio]{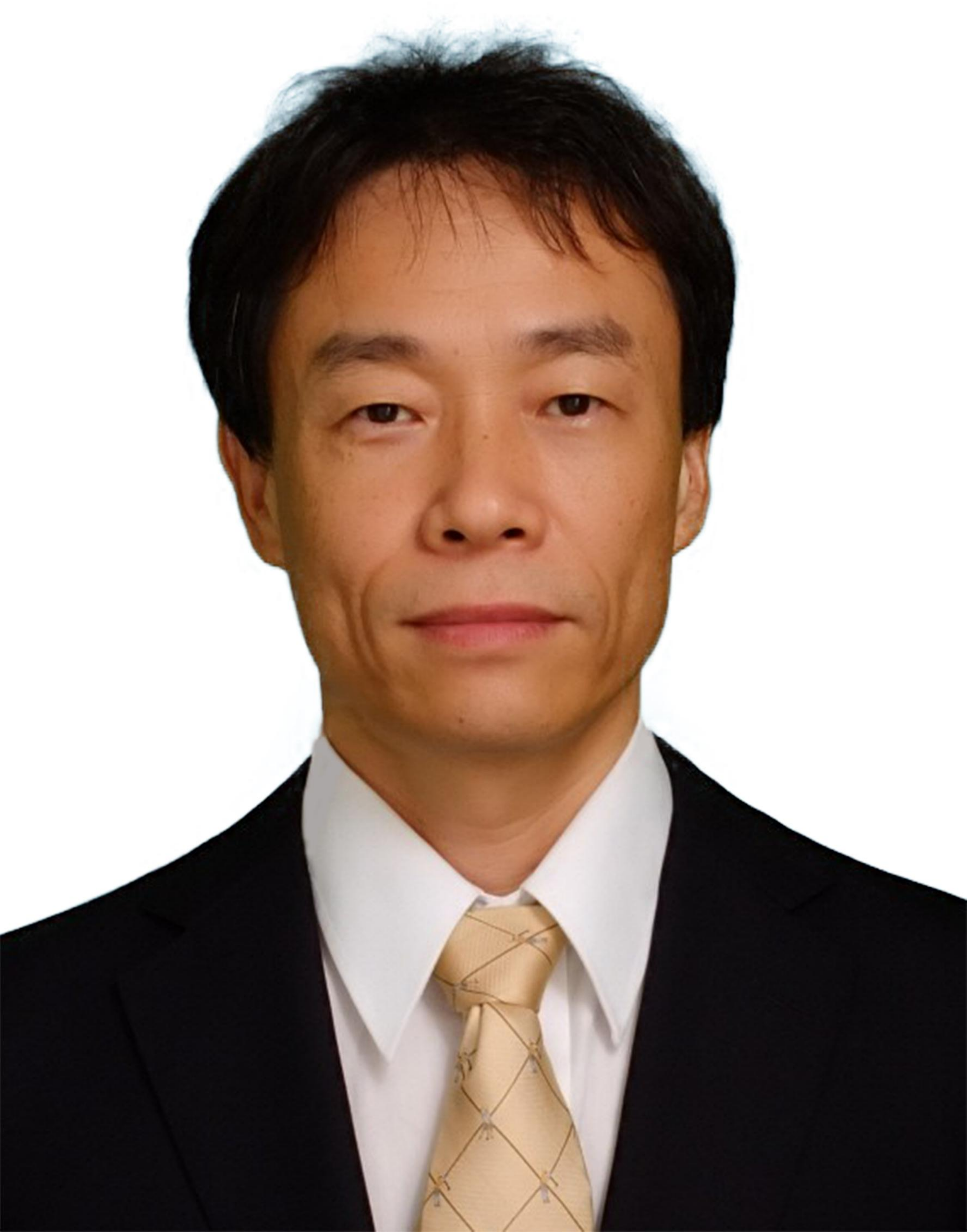}}]
{Nei Kato} (M’04, SM’05, F’13) is a full professor and the Dean of the Graduate School of Information Sciences, Tohoku University. He has been engaged in research on computer networking, wireless mobile communications, satellite communications, ad hoc and sensor and mesh networks, smart grid, AI, IoT, big data, and pattern recognition. He has published more than 500 papers in prestigious peer-reviewed journals and conferences. He was the Vice-President (Member \& Global Activities) of IEEE Communications Society (2018-2019), the Editor-in-Chief of IEEE Transactions on Vehicular Technology (2017-2020), and the Editor-in-Chief of IEEE Network (2015-2017). He is a Fellow of the Engineering Academy of Japan, IEEE and IEICE.
\end{IEEEbiography}

\end{document}